\documentclass[a4paper,11pt]{article}

\UseRawInputEncoding

\usepackage[utf8]{inputenc}
\usepackage[T1]{fontenc}
\usepackage{epigraph}

\usepackage{graphicx}
\usepackage{array}
\usepackage[normalem]{ulem}
\usepackage{slashed}
\usepackage[dvipsnames,table]{xcolor}
\usepackage[bottom]{footmisc}
\usepackage{booktabs}
\usepackage{xspace}
\usepackage{graphics}
\usepackage{longtable}
\usepackage{subfig}
\usepackage{jheppub} 

\catcode`~=11 
\newcommand{\urltilde}{\kern -.15em\lower .7ex\hbox{~}\kern .04em}
\catcode`~=13 

\usepackage[nottoc,notlot,notlof]{tocbibind}

\usepackage{breqn}

\usepackage{relsize}

\usepackage{marvosym}
\usepackage{amssymb}
\usepackage{bbold}
\usepackage{comment}
\usepackage{color}
\usepackage{tikz}
\usetikzlibrary{plotmarks}

\usepackage{amsmath}

\DeclareMathOperator{\sgn}{sgn}

\usepackage{mathtools}
\DeclarePairedDelimiter\ceil{\lceil}{\rceil}
\DeclarePairedDelimiter\floor{\lfloor}{\rfloor}

\usepackage[capitalise]{cleveref}

\def\spose#1{\hbox to 0pt{#1\hss}}
\def\ltapprox{\mathrel{\spose{\lower 3pt\hbox{$\mathchar"218$}}
 \raise 2.0pt\hbox{$\mathchar"13C$}}}
\def\gtapprox{\mathrel{\spose{\lower 3pt\hbox{$\mathchar"218$}}
 \raise 2.0pt\hbox{$\mathchar"13E$}}}

\DeclareMathOperator{\Tr}{Tr}
\newcommand{\1}{1\!\!\!\bot}

\usepackage{fancyhdr}
\def\vecx{\vec{x}\mkern 1mu}
\def\vecy{\vec{y}\mkern 1mu}
\def\vecz{\vec{z}\mkern 2mu}
\def\vecr{\vec{r}\mkern 2mu}
\def\vecR{\vec{R}\mkern 1mu}
\def\vecRpp{\vec{R}^{\mkern 1mu\prime}\mkern 1mu}
\def\vecRp{\vec{R}^{\mkern 1mu\prime}}
\def\veczero{\vec{0}\mkern 2mu}
\def\veczeropp{\vec{0}^{\mkern 2mu\prime} \mkern 1mu}
\def\veck{\vec{k}\mkern 2mu}
\def\vecK{\vec{K}\mkern 2mu}
\def\veckpp{\vec{k}^{\mkern 1mu \prime} \mkern 1mu}
\def\veckp{\vec{k}^{\mkern 1mu \prime}}
\def\cdots{\!\cdot\!}
\def\scriptU{\scriptscriptstyle U}
\def\scriptC{\scriptscriptstyle C}
\def\scriptT{\scriptscriptstyle \Theta}
\def\mk{\mkern 1mu}
\def\mkh{\mkern 1.5mu}
\allowdisplaybreaks

\title{Bloch Waves, Magnetization and
Domain Walls: \\[2mm]
The Case of the Gluon Propagator}
\author[a]{Attilio Cucchieri,}
\author[a]{and Tereza Mendes}
\affiliation[a]{Instituto de F\'\i sica de São Carlos,
                Universidade de São Paulo, IFSC--USP, \\[2mm]
                13566-590, São Carlos, SP, Brazil}
\emailAdd{attilio@ifsc.usp.br}
\emailAdd{mendes@ifsc.usp.br}

\date{\today}

\abstract{
We expand our previous study \cite{Cucchieri:2016qyc} of
replicated gauge configurations in lattice SU($N_c$) Yang-Mills theory
--- employing Bloch's theorem, from condensed-matter physics --- to 
construct gauge-fixed field configurations on significantly larger
lattices than the original, or primitive, one.
We present a comprehensive discussion of the general gauge-fixing problem,
identifying advantages of the replicated-lattice approach.
In particular, the consideration of Bloch waves leads us to a 
visualization of the extended gauge-fixed configurations
in terms of (color) magnetization domains.
Moreover, we are able to explore features of the method to
optimize the evaluation of gauge fields in momentum space,
furthering our knowledge of the ``allowed momenta'', an issue that has 
hindered wider applications of this approach up to now.
Interestingly, our analysis yields
both a better conceptual understanding of the problem and a more
efficient way to compute the desired large-volume observables.
}

\begin{document}

\epigraph{In memory and recognition of Daniel Zwanziger}

\maketitle
\flushbottom


\section{Introduction}

We study the problem of fixing the so-called minimal Landau
gauge in Yang-Mills theory using a replicated gauge-field configuration 
on an extended
lattice $\Lambda_z$, obtained by copying --- $m$ times along
each direction --- the link configuration defined on the
original lattice $\Lambda_x$ \cite{Cucchieri:2016qyc}.
We employ periodic boundary conditions (PBCs),
both for the original and for the extended lattice.
This setup is then used for the evaluation of the gluon propagator
in pure gauge theory, aiming at explaining and understanding the results
obtained from numerical simulations of the propagator in the infrared regime
for two, three and four space-time dimensions \cite{Cucchieri:2009xxr,
Cucchieri:2012cb,Cucchieri:2011ig}.
We recall that this method was first proposed by D. Zwanziger in
ref.\ \cite{Zwanziger:1993dh}, as a way to take the infinite-volume limit
in lattice gauge theory in two steps.

In a previous study \cite{Cucchieri:2016qyc}, we have worked out the
numerical implementation of the method and conducted a feasibility test
--- in two and three dimensions --- applying it 
to the gauge-fixing problem and evaluating the lattice gluon 
propagator in momentum space $D(\veck)$ for the SU(2) case.
We thus obtained results for two- and three-dimensional lattices of sides 
up to 16 times larger than the original one, corresponding to lattice
volumes respectively up to a few hundred and a thousand times larger than
the starting one.
We also verified
good agreement when comparing $D(\veck)$ with numerical data
obtained by working directly on a large lattice of the same size
as the extended lattice $\Lambda_z$.
This exercise proved very promising, since the computational 
cost could be greatly reduced, but there were a few unresolved issues,
which we now address.
More specifically, we found that a nonzero gluon propagator could be 
obtained only for certain values of momenta.
These {\em allowed} momenta
included the ones given by the discretization on the original (small) 
lattice $\Lambda_x$, but it was not clear to us if (and which) other momenta 
could also produce a nonzero value for $D(\veck)$.
We also obtained that the gluon propagator at zero momentum was strongly
suppressed when evaluated on $\Lambda_z$, a result that we interpreted just
qualitatively, as a peculiar effect due to the extended gauge transformations.
At that time, we could not offer, for either of these two results, a robust 
analytic explanation, which would complete our conceptual description of 
the proposed approach.
In point of fact, achieving such a comprehension is essential also for
a more efficient application of the method.
Indeed, in ref.\ \cite{Cucchieri:2016qyc}, while
thermalization and gauge fixing were carried out --- in the numerical
code --- using only the original lattice $\Lambda_x$, we still needed
to use the (gauge-fixed) gauge field defined on the extended lattice 
$\Lambda_z$ for the evaluation of the gluon propagator.
Clearly, a better understanding of the setup and its properties must allow
the entire numerical implementation of the method to be based on variables defined 
solely on $\Lambda_x$, in order for the computational cost to be independent 
of the replica factor $m$.
This is the main goal of the present work.

\vskip 3mm
The manuscript is organized as follows.
In section \ref{sec:mlg-PBCs}, we review --- for general SU($N_c$) 
gauge theory in the $d$-dimensional case ---
the numerical problem of imposing the minimal-Landau-gauge condition 
for a thermalized link configuration $\{ U_{\mu}(\vecx) \}$ on a 
lattice $\Lambda_x$, with PBCs, as well as the definition of 
the (lattice) gluon propagator in momentum space $D(\veck)$.
Even though most of the topics discussed in this section are well
known, the presentation is useful in order to set the notation and 
prepare the ground for our analysis of the replicated-lattice case.
In particular, we explicitly address the invariance of the lattice 
formulation under translations and global gauge transformations,
which will be important for our later discussion.
Then, in section \ref{sec:bloch-PBCs}, we extend the analysis to the 
case of a replicated field configuration, i.e.\ we discuss the minimal
Landau gauge on the extended lattice $\Lambda_z$ with PBCs, providing 
a more detailed description than the one presented in 
ref.\ \cite{Cucchieri:2016qyc}.
Specifically, after recalling the usual demonstration of Bloch's theorem for
a crystalline solid and its more relevant consequences, we review
the proof presented in refs.\ \cite{Cucchieri:2016qyc, Zwanziger:1993dh}, 
highlighting the properties of the translation operator ${\cal T}$ and
the role played by global transformations. 
This analysis naturally suggests a new interpretation of the
gauge-fixing condition for the extended lattice $\Lambda_z$,
which is presented in section \ref{sec:minimizing-revisited}.
It also permits the visualization of the gauge-fixed configurations
in terms of {\em domains}, which will be later identified with different
values of an effective (color) magnetization.
Afterwards, we show, in section \ref{sec:linkNP}, which gauge-fixed
link variables are nonzero on the extended lattice when
evaluated in momentum space.
This is, of course, the essential ingredient to predict which
momenta have a nonzero gluon propagator $D(\veck)$.
From our presentation it will be clear that, for the majority of
the momenta $\veck$, 
the gluon propagator is indeed equal to zero.
On the other hand, the allowed momenta, i.e.\ the
momenta for which a nonzero $D(\veck)$ is obtained,
include, but are not limited to, the momenta determined
by the discretization on the original (small) lattice
$\Lambda_x$.
However, as we will see, allowed momenta that are not
defined on $\Lambda_x$ depend on the outcome of the numerical
gauge fixing, i.e.\ they are usually different for different
gauge-fixed configurations.
Hence, in a numerical simulation, they usually show very poor
statistics.
We also carefully analyze, in section \ref{sec:gluon-extended}, 
the evaluation of the gluon propagator at zero momentum, 
as well as its limit for large values of the parameter $m$.
Some of the analytic results presented in section \ref{sec:linkNP}
are tested numerically in section \ref{sec:numerical-PBCs}, where
we also illustrate the color-magnetization domains for the different 
lattice replicas and we present our conclusions.
Finally, details about the Cartan sub-algebra for the SU($N_c$) group
are reported in appendix \ref{sec:Cartan}.


\section{Minimal Landau gauge with PBCs}
\label{sec:mlg-PBCs}

Let us first consider the usual minimal-Landau-gauge condition for 
Yang-Mills theory in the $d$-dimensional case and for the SU($N_c$) 
gauge group, on a lattice $\Lambda_x$ with volume $V=N^d$ and PBCs
(see for example ref.\ \cite{Giusti:2001xf}).
The gauge-fixing condition 
is imposed by minimizing --- with respect to the gauge transformation 
$\{ h(\vecx) \}$ --- the functional \cite{Zwanziger:1993dh}
\begin{eqnarray}
{\cal E}_{\scriptU}[h] \,& \equiv &\,
\frac{\Tr}{2\,N_c\,d\,V} \,
\sum_{\mu = 1}^d \, \sum_{\vec{x}\in\Lambda_x}
\, \left[ \, \1 \, - \, U_{\mu}(h;\vecx) \, \right] \,
\left[ \, \1 \, - \, U_{\mu}(h;\vecx) \,
\right]^{\dagger} \label{eq:minimizingV} \\[2mm]
              \,& = &\, \frac{\Re \, \Tr}{N_c\,d\,V} \,
\sum_{\mu = 1}^d \, \sum_{\vec{x}\in\Lambda_x} \, \left[ \,
   \1 \, - \, U_{\mu}(h;\vecx) \, \right]\,.
\label{eq:minimizing}
\end{eqnarray}
Here, $\Tr$ is the trace (in color space), $^{\dagger}$ stands for the 
Hermitian conjugate, $\Re$ selects the real part, the vector $\vec{x}$ has
integer components $x_{\mu}$ from $1$ to $N$, the transformed
gauge link is given by
\begin{equation}
U_{\mu}(h;\vecx)  \,\equiv\, h(\vecx)
\, U_{\mu}(\vecx) \, h(\vec{x} + \hat{e}_{\mu})^{\dagger}\,,
\label{eq:Uofh}
\end{equation}
where the (thermalized) link configuration $\{ U_{\mu}(\vecx) \}$ is
kept fixed, and $\hat{e}_{\mu}$
is the unit vector in the positive $\mu$ direction.
Both $U_{\mu}(\vecx)$ and the gauge-transformation variable $h(\vecx)$ are 
SU($N_c$) matrices in the fundamental $N_c \times N_c$ representation and 
we denote by $\1$ the $N_c \times N_c$ identity matrix.
As discussed below, this ensures a lattice implementation of the familiar 
Landau-gauge condition in the continuum, i.e.\ the condition of null 
divergence for the gauge field.

Let us impose periodicity, by requiring that
\begin{equation}
U_{\mu}(\vec{x} + N \hat{e}_{\nu}) \, = \, U_{\mu}(\vecx)
\label{eq:Uperiodicity-noh}
\end{equation}
and
\begin{equation}
h(\vec{x} + N \hat{e}_{\nu}) \, = \, h(\vecx)
\label{eq:hperiodicity}
\end{equation}
for $\mu, \nu = 1, \ldots, d$.
Combining these two conditions in (\ref{eq:Uofh}), we obtain that
\begin{equation}
U_{\mu}(h;\vec{x} + N \hat{e}_{\nu}) \, = \, U_{\mu}(h;\vecx) \; ,
\label{eq:Uperiodicity}
\end{equation}
i.e.\ the gauge-transformed link variables $U_{\mu}(h;\vecx)$ are
also periodic\footnote{We stress that, in order to have periodicity
for the original and for the gauge-transformed link configurations, we
only need the gauge transformation to be periodic up to a {\em global}
center element $z_{\mu}$ per direction
(see, e.g., \cite{Greensite:2003bk}).
We do not consider this possibility here.} on $\Lambda_x$.

Note that the minimizing functional ${\cal E}_{\scriptU}[h]$
is non-negative\footnote{One can easily show that the minimum value of
${\cal E}_{\scriptU}[h]$ is equal to zero, corresponding to
$U_{\mu}(h;\vecx) = \1$.}
and, due to cyclicity of the trace, it is invariant under {\em global}
gauge transformations $h(\vecx) = v \in$ SU($N_c$).
At the same time, eq.\ (\ref{eq:minimizingV}) tells us that
the minimal-Landau-gauge condition selects on each gauge
orbit --- defined by the original link configuration 
$\{ U_{\mu}(\vecx) \}$ --- the configuration whose distance from the trivial
vacuum $U_{\mu}(\vecx) = \1$ is minimal \cite{Zwanziger:1993dh}.
Of course, there may be more than one minimum $\{ U_{\mu}(h;\vecx)\}$ 
of ${\cal E}_{\scriptU}[h]$ for a given $\{ U_{\mu}(\vecx) \}$, corresponding
to different solutions of the minimization problem.
Indeed, it is well-known that --- both on the lattice and in the
continuum formulation --- there are multiple solutions to the
general Landau-gauge-fixing problem along each gauge orbit,
i.e.\ multiple configurations $\{ U_{\mu}(h;\vecx)\}$ corresponding
to a null divergence of the gauge field
\cite{Gribov:1977wm,Zwanziger:1982na,DellAntonio:1989wae,Vandersickel:2012tz}.
These are called Gribov copies.
Let us remark that not all such copies
will be also (local, or relative) minima of ${\cal E}_{\scriptU}[h]$, since 
the minimal-Landau-gauge condition is more restrictive than the general one.
The set of all local minima of the functional ${\cal E}_{\scriptU}[h]$
defines the first Gribov region $\Omega$.
It includes representative configurations of all gauge orbits, as well
as some of their Gribov copies, while the remaining ones lie outside of
$\Omega$.

Clearly, if the configuration\footnote{Usually, in order to simplify the 
notation, the gauge-fixed link configuration $\{ U_{\mu}(h;\vecx)\}$ is
redefined simply as $\{ U_{\mu}(\vecx) \}$.
Here, however, we prefer to keep the dependence on the
gauge transformation $\{ h(\vecx) \}$ explicit, for better comparison of 
the setup on the original lattice $\Lambda_x$ with that attained on the 
extended lattice $\Lambda_z$ (see sections \ref{sec:bloch-PBCs},
\ref{sec:minimizing-revisited} and \ref{sec:linkNP}).}
$\{ U_{\mu}(h;\vecx) \}$ is a local minimum
of the functional ${\cal E}_{\scriptU}[h]$, the stationarity
condition implies that its first variation with respect to the
matrices $\{ h(\vecx) \}$ be zero.
This variation may be conveniently obtained 
\cite{Zwanziger:1993dh,Cucchieri:2005yr} from a gauge transformation 
$ h(\vecx) \to R(\tau; \vecx) \, h(\vecx)$, with $R(\tau; \vecx)$
close to the identity and taken in a one-parameter subgroup of the
gauge group SU($N_c$). 
We thus write
\begin{equation}
R(\tau; \vecx) \, \equiv \, \exp{ \Bigl[
   \, i\, \tau \! \sum_{b = 1}^{N_c^2 -1}
         \! \gamma^b(\vecx) \, t^b \, \Bigr] }
         \, \approx\, \1 \,+\,i\, \tau \! \sum_{b = 1}^{N_c^2 -1}
         \! \gamma^b(\vecx) \, t^b\,,
\label{eq:hofgamma}
\end{equation}
where the parameter $\tau$ is real and small.
Here, $t^b$ are the $N_c^2 - 1$ traceless Hermitian generators
of the SU($N_c$) gauge group and the factors $\gamma^b(\vecx)$ are
also real.
About this, we recall that SU($N_c$) is a real Lie group and that its Lie
algebra $su(N_c)$ is also real \cite{Cornwell}.
Then, we can write any element $g\in$ SU($N_c$) as $g=\exp{\left(
i\, \sum_b \gamma^b \,t^b\right)}$, with $\gamma^b \in \mathbb{R}$, ensuring
that $g^{\dagger}=g^{-1}$.
At the same time, the condition $\Tr\,(t^b) = 0$ implies that
$\det(g) = 1$.
We consider generators $t^b$ normalized such that
\begin{equation}
\Tr \, ( t^b t^c ) \, = \, 2 \, \delta^{bc} \; ,
\label{eq:Trlambda}
\end{equation}
which is the usual normalization condition satisfied by the Pauli
matrices, in the SU($2$) case, and by the Gell-Mann matrices, in the
SU($3$) case.

Using this one-parameter subgroup, we may regard ${\cal E}_{\scriptU}[h]$ 
as a function ${\cal E}_{\scriptU}[h](\tau)$ of $\tau$.
Its first derivative with respect to $\tau$ is then given, at $\tau = 0$, by
\begin{eqnarray}
{\cal E}_{\scriptU}[h]^{'}\!(0) \,& = &\,
  \frac{\Re \, \Tr}{N_c\,d\, V} 
\,\mathlarger\sum_{b,\,\mu,\,\vec{x}}\, 
- i \, \left[ \, \gamma^{b}(\vecx) \, t^b \, U_{\mu}(h;\vecx)
          \, - \, U_{\mu}(h;\vecx) \,
          \gamma^{b}(\vec{x} + \hat{e}_{\mu}) \, t^b \,
            \right] \phantom{\mbox{oioi}} \nonumber \\[2mm]
   & = &\,
  \frac{2 \, \Re \, \Tr}{N_c\,d\, V} 
\,\mathlarger\sum_{b,\,\mu,\,\vec{x}}\,
     \frac{\gamma^{b}(\vecx) \, t^b}{2 \, i} \,
   \Big[ \, U_{\mu}(h;\vecx) \, - \, U_{\mu}(h;\vec{x} -
    \hat{e}_{\mu}) \, \Big] \label{eq:EderivPBCs2} \; ,
\end{eqnarray}
where $\vec{x}\in\Lambda_x$, the color index $b$ takes values $1,\ldots,N_c^2-1$
and $\mu=1,\ldots,d$.
At the same time, we define the gauge-fixed (lattice) gauge field $A_{\mu}(h;
\vecx)$ using the relation
\begin{eqnarray}
A_{\mu}(h;\vecx) \,& \equiv &\, \frac{1}{2 \,i}\, \left[ \,
    U_{\mu}(h;\vecx) - U_{\mu}^{\dagger}(h;\vecx)
    \, \right]_{\rm traceless} \label{eq:Atrac} \\[2mm]
    & = &\, \frac{1}{2 \,i}\, \left[ \,
   U_{\mu}(h;\vecx) - U_{\mu}^{\dagger}(h;\vecx) \,
        \right] \, - \,  \1 \, \frac{\Tr}{2 \,i\,N_c}\,
    \left[ \, U_{\mu}(h;\vecx) - U_{\mu}^{\dagger}(h;\vecx)
      \, \right] \label{eq:defAgen} \\[2mm]
    & = &\, \frac{1}{2 \,i}\, \left[ \,
      U_{\mu}(h;\vecx) - U_{\mu}^{\dagger}(h;\vecx) \,
        \right] \, - \,  \1 \, \frac{\Im\,\Tr}{N_c}\,
           \left[\, U_{\mu}(h;\vecx) \,\right] \; ,
        \label{eq:defAgenIm}
\end{eqnarray}
where $\Im$ selects the imaginary part of a complex number.
Also, we write
\begin{equation}
A_{\mu}(h;\vecx) \, \equiv \, \sum_b\,
    A_{\mu}^{b}(h;\vecx) \, t^b\,,
    \label{eq:Aofx}
\end{equation}
so that, recalling eq.\ (\ref{eq:Trlambda}), the color components 
$A_{\mu}^{b}(h;\vecx)$ are given by
\begin{equation}
A_{\mu}^{b}(h;\vecx) \, = \, \frac{1}{2} \, \Tr
              \left[\, A_{\mu}(h;\vecx)\, t^{b} \, \right]
\label{eq:A} \; .
\end{equation}
Then, since the generators $t^b$ are traceless, it is
evident that the term proportional to the identity matrix
$\1$ in eqs.\ (\ref{eq:defAgen})--(\ref{eq:defAgenIm}) does
not contribute to $A_{\mu}^{b}(h;\vecx)$, see eq.\ (\ref{eq:A}), i.e.\
\begin{equation}
A_{\mu}^{b}(h;\vecx) \, = \, \Tr \;
    \left\{\, t^{b} \, \left[ \, \frac{ U_{\mu}(h;\vecx)
    - U_{\mu}^{\dagger}(h;\vecx) }{4\,i} \,
    \right] \, \right\} \, = \,
    \Re \, \Tr \; \left[\, t^{b} \, \frac{ U_{\mu}(h;\vecx)}{2\,i}
    \, \right] \; . \label{eq:Adue}
\end{equation}
We may thus rewrite the first derivative of the minimizing
functional from eq.\ (\ref{eq:EderivPBCs2}) as
\begin{equation}
{\cal E}_{\scriptU}[h]^{'}\!(0) \, = \, \frac{2}{N_c\,d\, V} 
\,\mathlarger\sum_{b,\,\mu,\,\vec{x}}\, 
     \gamma^{b}(\vecx) \, \left[
    \, A_{\mu}^{b}(h;\vecx) \, - \, A_{\mu}^{b}(h;\vec{x}
   - \hat{e}_{\mu}) \, \right] \; ,
   \label{eq:EUprimeA}
\end{equation}
which provides a nice analogy with the continuum case, as shown 
next.\footnote{Equivalently, one could note, in eq.\ (\ref{eq:EderivPBCs2}), 
that
\begin{equation}
\Re \Tr \left[t^b U_{\mu}(h;\vecx) / i \right] \,=\,
 \Re \Tr \left[ t^b U_{\mu}(h;\vecx) / i \right]^{\dagger}
 \, = \, \Re \Tr \left[- t^b U_{\mu}^{\dagger}(h;\vecx)
      / i \right] \; .
\label{eq:ReTrlambdaU}
\end{equation}
This allows us to write
\begin{equation}
{\cal E}_{\scriptU}[h]^{'}\!(0) \, = \,
\mathlarger\sum_{b,\,\mu,\,\vec{x}} \,
\frac{\,2\, \gamma^{b}(\vecx)}{N_c\,d\, V} 
\;\Re \Tr \, \left\{\, t^b
\biggl[ 
\frac{U_{\mu}(h;\vecx) - U_{\mu}^{\dagger}(h;\vecx)}{4\,i} 
- \frac{
U_{\mu}(h;\vec{x} \!-\! \hat{e}_{\mu}) -
U_{\mu}^{\dagger}(h;\vec{x} \!-\! \hat{e}_{\mu})}{4 \, i}
\biggr]\,\right\}
 \,,
\end{equation}
which naturally suggests the definition (\ref{eq:Atrac}), see also
eq.\ (\ref{eq:Adue}), for the (gauge-transformed) gauge field.}

Of course, if $\{ U_{\mu}(h;\vecx) \}$ is a stationary point of
${\cal E}_{\scriptU}[h](\tau)$ at $\tau=0$, we must have
\begin{equation}
{\cal E}_{\scriptU}[h]^{'}\!(0)\, = \, 0
\label{eq:Eprime0orig}
\end{equation}
``along'' any direction $\sum_b \gamma^{b}(\vecx)
t^b$, i.e.\ for every set of $\gamma^{b}(\vecx)$ factors.
This implies that the lattice divergence
\begin{equation}
\left( \nabla\cdots A^b \right)\!(h;\vecx) \; \equiv \;
  \sum_{\mu = 1}^d \; \Bigl[A_{\mu}^{b}(h;\vecx) -
                  A_{\mu}^{b}(h;\vec{x} - \hat{e}_{\mu})\Bigr]
\label{eq:diverA}
\end{equation}
of the gauge-fixed gauge field $A_{\mu}(h;\vecx)$ is zero,
i.e.\
\begin{equation}
\left( \nabla \cdots A^{b} \right)\!(h;\vecx) \, = \,
  0 \qquad \qquad \forall \;\, \vec{x}, b \;\;\; ,
\label{eq:diverg0}
\end{equation}
and the gauge field $A_{\mu}(h;\vecx)$ is transverse.
The above eqs.\ (\ref{eq:diverA}) and (\ref{eq:diverg0})
give the lattice formulation of the usual Landau gauge-fixing condition 
in the continuum and, due to eq.\ (\ref{eq:Trlambda}), are clearly equivalent to
\begin{equation}
\left( \nabla \cdots A \right)\!(h;\vecx)\,=\,0
\quad \quad \forall \;\, \vec{x}
\label{eq:diverg00}
\end{equation}
with, see eqs.\ (\ref{eq:Aofx}) and (\ref{eq:diverA}),
\begin{equation}
\left( \nabla \cdots A \right)\!(h;\vecx) \,\equiv\,
  \sum_{\mu = 1}^d \; \Bigl[A_{\mu}(h;\vecx) -
          A_{\mu}(h;\vec{x} - \hat{e}_{\mu})\Bigr]
 \,=\, \sum_{b = 1}^{N_c^2-1} \, t^b \,
       \left( \nabla\cdots A^{b} \right)\!(h;\vecx) \; .
\label{eq:diverAnob}
\end{equation}
Let us stress that the gauge transformation $\{ h(\vecx) \}$ depends
on ${\cal N}_p \equiv V\,(N_c^2 - 1)$ free parameters $\gamma^{b}(\vecx)$ and 
the minimization process enforces the corresponding ${\cal N}_p$ 
constraints (\ref{eq:diverg0}).

Clearly, since the link variables $U_{\mu}(h;\vecx)$ satisfy PBCs,
the same is true for the gauge fields $A_{\mu}(h;\vecx)$, defined in
eqs.\ (\ref{eq:Atrac})--(\ref{eq:defAgenIm}).
Thus, it is convenient to consider the Fourier transform (see, e.g.\
\cite{Leinweber:1998uu})
\begin{equation}
{\widetilde A}_{\mu}^{b}(h;\veck)\,\equiv\,
    \sum_{\vec{x} \in \Lambda_x} \, A_{\mu}^{b}(h;\vecx)
    \, \exp{\left[ - \frac{2 \pi i}{N} \left(\vec{k}\cdots
       \vec{x} + \frac{k_{\mu}}{2}\right)\right]}\; ,
\label{eq:AFourier}
\end{equation}
where the wave-number vectors $\vec{k}$ have integer components $k_{\mu}$,
which are usually restricted to the so-called first Brillouin
zone\footnote{One could also take $k_{\mu} = -N/2, -N/2+1,
\ldots, N/2\!-\!1$ for even $N$ and $k_{\mu} = -(N\!-\!1)/2, -(N\!-\!3)/2,
\ldots, (N\!-\!1)/2$ for odd $N$ or, equivalently, 
$k_{\mu} = -\floor*{N/2}, -\floor*{N/2}+1, \ldots, \ceil*{(N/2)-1}$ for
general $N$, where $\floor*{x}$ is the largest integer less than or equal 
to $x$ and $\ceil*{x}$ is the smallest integer greater than or equal to $x$.
This convention, however, would make the formulae --- and the
corresponding numerical code --- more cumbersome (see also footnotes
\ref{foot:m-even} and \ref{foot:k-decomp}).
\label{foot:symmetrick}}
$k_{\mu} = 0, 1, \ldots, N\!-\!1$.
Let us notice that, according
to this definition, the contribution to the Fourier transform coming from
the link between $\vec{x}$ and $\vec{x}+\hat{e}_{\mu}$ is calculated at
its midpoint $\vec{x}+\hat{e}_{\mu}/2$.
For later convenience, let us also define the Fourier transform of the
gauge link
\begin{equation}
{\widetilde U}_{\mu}(h;\veck)\,\equiv\,
    \sum_{\vec{x} \in \Lambda_x} \, U_{\mu}(h;\vecx) \, 
\exp{\left[ - \frac{2 \pi i}{N} \left(\vec{k}\cdots\vec{x}\right)\right]}\;.
\label{eq:UhxFourier}
\end{equation}

Now, in order to write down the inverse Fourier transform, we recall
that, in one dimension (and with $k$ taking values $0, 1, \ldots,
N\!-\!1$), we find \cite{Smit,Gattringer:2010zz}
\begin{equation}
\sum_{x=1}^N \,e^{-\frac{2\pi i}{N}\,k\,x}\;=\;
\sum_{x=0}^{N-1} \,\left(e^{-\frac{2\pi i}{N}\,k}\right)^x
\;=\; \frac{1\,-\,\left[\, \exp{\left(-2\pi i\,k/N\right)}
\,\right]^N}{1\,-\,\exp{\left(-2\pi i\,k/N\right)}} \,=\, 0
\label{eq:dKronecker}
\end{equation}
for $k \neq 0$.
Thus, the above expression is equal to $\,N\,\delta(k,0)$, where
$\delta(\,,\,)$ stands for the Kronecker delta function.
Analogously, in the $d$-dimensional case, we have
\begin{equation}
\sum_{\vec{x} \in \Lambda_x} \,
     e^{- \frac{2 \pi i}{N}\,\vec{k} \cdot \vec{x}}
\;=\; \prod_{\nu = 1}^{d} \,\Biggl[\,
  \sum_{x_{\nu} = 1}^{N} \,
    e^{-\frac{2\pi i}{N}\,k_{\nu}x_{\nu}}
       \, \Biggr] 
\;=\;N^d\, \delta(\vec{k},\veczero)
     \,=\,V\,\delta(\vec{k},\veczero) \; ,
\label{eq:deltakappa}
\end{equation}
where $\delta(\vec{k},\veczero)$ is a shorthand for
$\prod_{\nu=1}^d \delta(k_{\nu},0)$.
Conversely, we have
\begin{equation}
\sum_{\vec{k} \in {\widetilde \Lambda}_x} \,
     e^{\frac{2 \pi i}{N}\,
     \vec{k} \cdot \vec{x}}\,=\,
       V\,\delta(\vec{x},\veczero) \; ,
\label{eq:deltax}
\end{equation}
where ${\widetilde \Lambda}_x$ stands for the first Brillouin
zone (for the $\Lambda_x$ lattice).
Hence, it is straightforward to verify that the inverse Fourier 
transform, corresponding to eq.\ (\ref{eq:AFourier}), is given by
\begin{equation}
A_{\mu}^{b}(h;\vecx)\,\equiv\, \frac{1}{V} \,
     \sum_{\vec{k} \in {\widetilde \Lambda}_x} \,
    {\widetilde A}_{\mu}^{b}(h;\veck) \,
       \exp{\left[ \frac{2 \pi i}{N} \left(\vec{k} \cdots
       \vec{x} + \frac{k_{\mu}}{2}\right)\right]} \; .
\label{eq:AantiFourier}
\end{equation}

\vskip 3mm
As mentioned above, the term $\,i \pi k_{\mu} / N\,$ in the
exponent of eq.\ (\ref{eq:AFourier}) is obtained by considering
the gauge field at the midpoint $\vec{x} + \hat{e}_{\mu}/2$ of
a lattice link.\footnote{Of course, it should be specified in all formulae
that the gauge field relative to the lattice point $\vec{x}$ is actually 
evaluated at $\vec{x} + \hat{e}_{\mu}/2$, e.g.\ by writing
$U_{\mu}(h;\vecx) \equiv \exp{\left[ i \,
A_{\mu}(h;\vec{x} + \hat{e}_{\mu}/2) \, \right]}$.
This is especially relevant when considering the Fourier transform, as in 
eq.\ (\ref{eq:AFourier}), and in the (lattice) weak-coupling 
expansion \cite{Rothe:1992nt}.
Here, however, in order to keep the notation simpler, we do
not indicate this explicitly.}
This term is essential in order to show that, in momentum space, eq.\
(\ref{eq:diverg0}) becomes
\begin{equation}
0 \, = \,  \frac{1}{V} \, \sum_{\vec{k} \in {\widetilde
   \Lambda}_x} \, \sum_{\mu = 1}^d \, {\widetilde A}_{\mu}^{
     b}(h;\veck) \, \exp{\left( \frac{2 \pi i}{N} \vec{k}
       \cdots \vec{x} \right)} \,
2 \, i \, \sin{\left(\frac{\pi \, k_{\mu}}{N}\right)} \; ,
\label{eq:diverg0k}
\end{equation}
yielding (for each $\veck$) the lattice transversality
condition
\begin{equation}
\sum_{\mu = 1}^d \, {\widetilde A}_{\mu}^{b}(h;\veck)
        \, p_{\mu}(\veck) \, = \, 0 \; ,
\label{eq:Atransv}
\end{equation}
where
\begin{equation}
p_{\mu}(\veck) \, \equiv \,
  2 \, \sin{\left(\frac{\pi \, k_{\mu}}{N}\right)}
\label{eq:pmu}
\end{equation}
are the components of the lattice momentum $\vec{p}\mkh(\veck)$
\cite{Smit,Gattringer:2010zz}.
Indeed, without the factor $\exp{(i \pi k_{\mu} / N)}$, we would
obtain the condition
\begin{equation}
\sum_{\mu = 1}^d \, {\widetilde A}_{\mu}^{b}(h;\veck)
   \, \left[ \, 1 \, - \,
\cos{\left(\frac{2 \, \pi \, k_{\mu}}{N}\right)} \, + \,
    i \, \sin{\left(\frac{2 \pi \, k_{\mu}}{N}\right)} \,
    \right] \, =  \, 0 \; , \label{eq:Atransv2}
\end{equation}
which looks very different from the Landau gauge condition 
in the continuum.

Actually, one can verify that eqs.\ (\ref{eq:Atransv}) and
(\ref{eq:Atransv2}) have the same (formal) continuum 
limit \cite{Leinweber:1998uu}, but with different discretization
errors.
To this end, we write
\begin{equation}
\frac{2 \, \pi \, k_{\mu}}{N} \, = \, a \,
\frac{2 \, \pi \, k_{\mu}}{a\,N} \, \equiv \, a \,
   {\hat p}_{\mu} \; ,
\end{equation}
where $a$ is the lattice spacing and ${\hat p}_{\mu}$ is now
a continuum momentum in physical units, and take the
limit $a \to 0$ with ${\hat p}_{\mu}$ kept fixed.
We find, in both cases, that the term multiplying
${\widetilde A}_{\mu}^{b}(h;\veck)$ is proportional to
${\hat p}_{\mu}$, yielding the desired transversality condition.
However, in the first case the discretization error is of
order $a^2$, while in the second it is of order $a$.
Moreover, eq.\ (\ref{eq:pmu}) provides a more natural definition
of the lattice-momentum components than the expression
in square brackets in eq.\ (\ref{eq:Atransv2}), since
\begin{equation}
p^2(\veck) \, = \, \sum_{\mu = 1}^d \, p_{\mu}^2(\veck) 
\,\equiv\, \sum_{\mu = 1}^d \, 4\,\sin^2{\left(\frac{\pi \, k_{\mu}}{N}\right)}
\label{eq:p2def}
\end{equation}
are the eigenvalues of (minus) the usual lattice Laplacian
\begin{equation}
- \Delta (\vec{x},\vecy) \, \equiv \,
  \sum_{\mu = 1}^d \, \left[ \,
     2 \, \delta(\vec{x},\vec{y}) \, - \,
    \delta(\vec{x}+\hat{e}_{\mu},\vec{y}) \, - \,
    \delta(\vec{x}-\hat{e}_{\mu},\vec{y}) \,
      \right] \; ,
\end{equation}
corresponding to the plane-wave eigenvectors $\exp{\left( - 2 \pi i
\vec{k} \cdots \vec{y} / N \right)}$.


\subsection{Numerical gauge fixing}
\label{sec:numPBCs}

In order to minimize ${\cal E}_{\scriptU}[h]$ numerically,
it is sufficient to implement an iterative algorithm that monotonically 
decreases the value of the minimizing functional.
Indeed, since ${\cal E}_{\scriptU}[h]$ is bounded from
below, an algorithm of this kind is expected to converge.
As the simplest approach, one can sweep through the lattice
$\Lambda_{x}$ and apply --- for each lattice site $\vec{x}$ ---
a convenient update
\begin{equation}
h(\vecx) \, \to \, h'(\vecx) = r(\vecx) \, h(\vecx) \; ,
\label{eq:htorh}
\end{equation}
where $r(\vecx)\in$ SU($N_c$),
while keeping all the other matrices $h(\vecx)$ fixed.
In other words, a single-site update at $\vecx$ corresponds to 
$\{ h(\vecx) \}\to\{ h'(\vecx) \}$, where the new set of gauge transformations is unaltered
except for applying $r(\vecx)$ to $h(\vecx)$ as above.
From eqs.\ (\ref{eq:minimizing}) and (\ref{eq:Uofh}) we see that
the corresponding change ${\cal E}_{\scriptU}[h']\,-\,{\cal E}_{\scriptU}[h]$
in the minimizing functional due to this update is given by 
\begin{eqnarray}
& & \frac{\Re \, \Tr}{N_c\,d\,V} \,
\sum_{\mu = 1}^d \, \left[ \,
   U_{\mu}(h;\vecx) \,+\, U_{\mu}(h;\vec{x} \!-\! \hat{e}_{\mu})^{\dagger} \,-\,
   r(\vecx)\,U_{\mu}(h;\vecx) \,-\, U_{\mu}(h;\vec{x} \!-\! \hat{e}_{\mu})\,
   r(\vecx)^{\dagger}\, \right] \nonumber \\
&\,=\,& \frac{\Re \, \Tr\,
\left[ \, w(\vecx) \, \right]}{N_c\,d\,V} \,-\, \frac{\Re \, \Tr\,
\left[ \, r(\vecx) \, w(\vecx) \, \right]}{N_c\,d\,V} \,,
   \label{EUofr}
\end{eqnarray}
with
\begin{equation}
 w(\vecx) \, \equiv \, \sum_{\mu = 1}^d \,
          \left[ \, U_{\mu}(h;\vecx) \, + \,
             U_{\mu}(h;\vec{x} \!-\! \hat{e}_{\mu})^{\dagger}
             \, \right] \; .
\label{eq:w}
\end{equation}
Then, for the change to be negative,
the single-site update must satisfy the inequality
\begin{equation}
-\!\Re \, \Tr \, \left[ \, r(\vecx)
  \, w(\vecx) \, \right] \, \leq \, -\Re \, \Tr \,
    \left[ \, w(\vecx) \, \right]\,.
\label{eq:rineq}
\end{equation}
Common possible choices\footnote{Note that the inequality
(\ref{eq:rineq}) is linear in the updating matrix $r(\vecx)$.
This makes the minimization problem within the chosen approach rather simple.}
for $r(\vecx)$ --- usually written as a linear combination of the
identity matrix $\1$ and of the matrix $w(\vecx)$ --- can be
found in refs.\ \cite{Suman:1993mg,Cucchieri:1995pn,
Cucchieri:1996jm,Cucchieri:2003fb,Leal:2022ojc}.
In particular, in the SU(2) case, the matrix $w(\vecx)$ is
proportional to an SU(2) matrix.
On the contrary, in the general SU($N_c$) case, it is simply
an $N_c \times N_c$ complex matrix and one needs to project
this matrix onto the gauge group (see, e.g., refs.\
\cite{Suman:1993mg} and \cite{Leal:2022ojc}).
Let us note that, from the point of view of the organization of
the numerical algorithm, one does not need to store both the gauge
transformation $\{ h(\vecx) \}$ and the link configuration $\{
U_{\mu}(\vecx) \}$.
Indeed, every time a single-site update (\ref{eq:htorh}) is
performed, one can modify the gauge configuration directly, by evaluating
the products\footnote{Of course, in a numerical simulation, one
should verify that these transformations of the link variables
do not spoil their unitarity due to accumulation of roundoff errors.}
\begin{equation}
U_{\mu}(h;\vecx) \,\to\, r(\vecx)\,U_{\mu}(h;\vecx) \;
   \qquad \mbox{and}\ \qquad
U_{\mu}(h;\vec{x} - \hat{e}_{\mu})\, \to \,
   U_{\mu}(h;\vec{x} - \hat{e}_{\mu})\,
   r(\vecx)^{\dagger} \; ,
   \label{eq:Uupdates}
\end{equation}
for each direction $\mu=1,\ldots,d$.
An iteration of the method corresponds to a full sweep of the lattice, 
applying the above single-site updates at each point $\vecx$.

\vskip 3mm
As a check of convergence of the (iterative) minimization algorithm 
after $t$ sweeps over the lattice, one can ``monitor'' the behavior of 
several different quantities\footnote{Note that, compared to refs.\
\cite{Cucchieri:1995pn,Cucchieri:1996jm,Cucchieri:2003fb}, here
we have slightly changed the definition of $\Sigma_Q$, in order
to have a quantity that is invariant under global gauge transformations.} 
\cite{Cucchieri:1995pn,Cucchieri:1996jm,Cucchieri:2003fb}, e.g.\
\vspace{2mm}
\begin{eqnarray}
\Delta {\cal E} \,& \equiv &\, {\cal E}_{\scriptU}[h;t]
      \, - \, {\cal E}_{\scriptU}[h;t-1]\,,
\label{eq:e1} \\[3mm]
(\nabla A)^2 \,& \equiv &\, \frac{1}{(N_c^2 - 1) \, V} \,
    \sum_b \, \sum_{\vec{x}\in\Lambda_x}
       \, \Big[ \left( \nabla \cdots A^{b} \right)\!(h;\vecx)
              \, \Big]^{2}\,,
\label{eq:e2} \\[2mm]
\Sigma_Q \,& \equiv &\, \frac{1}{N} \,
\sum_{b,\,\mu,\,x_{\mu}} \,
    \left[ \, Q_{\mu}^{b}(h;x_{\mu}) - {\widehat Q}_{\mu}^{b}(h)
       \, \right]^{2} \; / \;\;\;
   \sum_{b,\,\mu} \left[ {\widehat Q}_{\mu}^{b}(h) \right]^2
          \; ,
\label{eq:e6}
\end{eqnarray}
\protect\vskip 1mm
\noindent
where all quantities are evaluated using the
gauge-transformed configuration $\{ U_{\mu}(h;\vecx) \}$, the
color index $b$ takes values $1,\ldots,N_c^2 -1$ and, as always
throughout this work, $\mu=1,\ldots,d$ and $x_\mu=1,\ldots,N$.
In eq.\ (\ref{eq:e6}) above, we have defined
\vspace{2mm}
\begin{equation}
\label{eq:chargeQ}
Q_{\mu}^b(h;x_{\mu}) \, \equiv \,
\sum_{\substack{x_{\nu} \\[0.3mm] \nu \neq \mu}}
     A_{\mu}^{b}(h;\vecx) 
\end{equation}
and
\begin{equation}
 {\widehat Q}_{\mu}^{b}(h) \, \equiv \, \frac{1}{N} \,
     \sum_{x_{\mu}} \, Q_{\mu}^{b}(h;x_{\mu}) \,=\,
     \frac{1}{N} \,\sum_{\vecx} \, A_{\mu}^{b}(h;\vecx)\; .
\label{eq:chargeQhat}
\end{equation}
One can check that, if the Landau-gauge-fixing condition (\ref{eq:diverg0}) 
is satisfied, then $Q_{\mu}^b(h;x_{\mu})$ must be independent of $x_{\mu}$.
Indeed, from eqs.\ (\ref{eq:diverA}) and (\ref{eq:diverg0}) we
obtain\footnote{This proof is equivalent to the usual proof
that a continuity equation implies a conserved charge.}
\begin{eqnarray}
0 \,& = &\, \sum_{\substack{\,x_{\nu} \\[0.3mm] \nu \neq \mu}}
\left( \nabla\cdots A^{b} \right)\!(h;\vecx) \, = \,
\sum_{\substack{x_{\nu} \\[0.3mm] \nu \neq \mu}}
  \sum_{\sigma = 1}^d \left[\, A_{\sigma}^{b}(h;\vecx) -
      A_{\sigma}^{b}(h;\vec{x} - \hat{e}_{\sigma})\,\right]
    \nonumber \\[2mm]
    & = & \sum_{\substack{x_{\nu} \\[0.3mm] \nu \neq \mu}}
\!\left[A_{\mu}^{b}(h;\vecx) -
   A_{\mu}^{b}(h;\vec{x} - \hat{e}_{\mu})\right] 
+\! \sum_{\substack{x_{\nu} \\[0.3mm] \nu \neq \mu}}
        \sum_{\sigma\neq\mu}
      \left[A_{\sigma}^{b}(h;\vecx) -
      A_{\sigma}^{b}(h;\vec{x} - \hat{e}_{\sigma})\right] \; ,
                  \label{eq:qconstant0}
\end{eqnarray}
for any $\mu=1,\ldots, d\,$ and $x_\mu=1,\ldots,N$.
Here, the first term on the r.h.s.\ is simply given 
by $\,Q_{\mu}^b(h;x_{\mu}) - Q_{\mu}^b(h;x_{\mu} - 1) \,$, 
while the second one may be written as
\begin{equation}
    \sum_{\sigma\neq\mu} \,\left\{\,
          \sum_{\substack{x_{\nu} \\[0.3mm] \nu \neq \mu,\sigma}}
          \sum_{x_{\sigma}}
             \left[ \,
          A_{\sigma}^{b}(h;\vecx) -
          A_{\sigma}^{b}(h;\vec{x} - \hat{e}_{\sigma})
                  \, \right] \,\right\}  \; .
                  \label{eq:qconstant}
\end{equation}
Let us stress that, in the above formulae, the coordinate 
$x_{\mu}$ is fixed and all other coordinates are summed over.
In particular, in eq.\ (\ref{eq:qconstant}), we have singled out the 
sum over the coordinate $x_{\sigma}$.
This makes it evident that, with respect to this coordinate, one has
a {\em telescopic sum}, yielding (for each direction $\sigma \neq \mu$)
\begin{equation}
          \sum_{\substack{x_{\nu} \\[0.3mm] \nu \neq \mu,\sigma}}
             \Biggl[ \,\left.
          A_{\sigma}^{b}(h;\vecx)\,\right|_{x_{\sigma}=N} -
                \left.
          A_{\sigma}^{b}(h;\vecx)\,\right|_{x_{\sigma}=0}
                  \, \Biggr] \; .
                  \label{eq:qconstant-sigma}
\end{equation}
Then, when PBCs are imposed along the direction $\sigma$, the last
expression cancels out and we get
\begin{equation}
Q_{\mu}^b(h;x_{\mu}) \, = \, Q_{\mu}^b(h;x_{\mu} - 1)\,,
\end{equation}
i.e.\ the ``charges'' $Q_{\mu}^b(h;x_{\mu})$ are constant,
they do not depend on $x_{\mu}$, for any direction $\mu$.

Also, note that the quantity $(\nabla A)^2$ is invariant under
global gauge transformations $v \in$ SU($N_c$).
Indeed, from eqs.\ (\ref{eq:Uofh}) and (\ref{eq:defAgen}) we
have that, for $h(\vecx)\to
v\, h(\vecx)$, the gauge field $A_{\mu}(h;\vecx)$ changes as
\begin{equation}
A_{\mu}(h;\vecx) \, \to \, v \, A_{\mu}(h;\vecx)
   \, v^{\dagger} \label{eq:Aglobal}
\end{equation}
and the same form holds for the transformation of
$\left( \nabla \cdots A \right)\!(h;\vecx)$, see
eq.\ (\ref{eq:diverAnob}).
The above statement then follows if we write, see eq.\ (\ref{eq:Trlambda}),
\begin{equation}
(\nabla A)^2 \, \equiv \, \frac{\Tr}{2\,(N_c^2 - 1)\,V} \,
                     \sum_{\vec{x}\in\Lambda_x}
       \, \Big[ \left( \nabla \cdots A \right)\!(h;\vecx)
           \, \Big]^{2}
\label{eq:defdiv2}
\end{equation}
and use the cyclicity of the trace.
This result is expected if we interpret eq.\ (\ref{eq:EUprimeA}) as
a directional derivative of the minimizing functional
${\cal E}_{\scriptU}[h]$ along the ``direction'' specified by
the vector with components $\gamma^b(\vec{x})$, so that
$\left( \nabla \cdots A^b \right)\!(h;\vecx)$ are the
(color) components of its gradient.
Then,
since ${\cal E}_{\scriptU}[h]$ is invariant
under global gauge transformations,\footnote{This, of course,
implies that $\Delta {\cal E}$ is also invariant under global gauge
transformations.}
one should have that the magnitude of its gradient
--- which
quantifies the steepness of the minimizing function at a given point
in the link-configuration space --- is also invariant under such
global transformations, even though its components $\left( \nabla
\cdots A^b \right)\!(h;\vecx)$ are not.

Similarly, we can write $\Sigma_Q$ as
\begin{equation}
\Sigma_Q \,=\, \frac{1}{N} \,
\sum_{\mu=1}^d\,\sum_{x_{\mu}=1}^N \, \Tr \,
    \left[ \, Q_{\mu}(h;x_{\mu}) - {\widehat Q}_{\mu}(h)
       \, \right]^{2} \; / \;\;\;
   \sum_{\mu=1}^d \,\Tr\, \left[ {\widehat Q}_{\mu}(h) \right]^2\,,
\label{eq:SigQ}
\end{equation}
with
\begin{equation}
\label{eq:Qmux}
Q_{\mu}(h;x_{\mu})\,\equiv\,
  \sum_{b = 1}^{N_c^2-1} \, t^b \, Q_{\mu}^b(h;x_{\mu})
  \qquad \quad \mbox{and} \qquad \quad
{\widehat Q}_{\mu}(h)\,\equiv\,
  \sum_{b = 1}^{N_c^2-1} \, t^b \,
  {\widehat Q}_{\mu}^b(h) \; .
\end{equation}
Then, clearly we have invariance under a global gauge transformation
$v$, see eqs.\ (\ref{eq:chargeQ}) and (\ref{eq:chargeQhat}),
since $Q_{\mu}(h;x_{\mu})\to v\, Q_{\mu}(h;x_{\mu})\,
v^{\dagger}$ and ${\widehat Q}_{\mu}(h) \, \to \, v\,
{\widehat Q}_{\mu}(h) \, v^{\dagger}$.

We see, therefore, that all the three quantities proposed to monitor the
convergence of the algorithm, given in eqs.\ (\ref{eq:e1})--(\ref{eq:e6}),
are invariant under a global gauge transformation, just as the minimizing
functional in eq.\ (\ref{eq:minimizingV}).


\subsection{Gluon propagator}
\label{sec:gluonPBCs}

The lattice space-time gluon propagator is defined as\footnote{Here,
in order to simplify the notation, we do not make explicit the
dependence of the gluon propagator on the gauge transformation
$\{ h(\vecx) \}$.}
\begin{equation}
D_{\mu \nu}^{b c}(\vec{x}_1,\vec{x}_2) \, \equiv \,
   \Bigl< A^{b}_{\mu}(h;\vec{x}_1)\, A^{c}_{\nu}(h;\vec{x}_2) \Bigr> 
\; , \label{eq:Dmunubc}
\end{equation}
where $\langle \, \cdot \, \rangle$ stands for the path-integral
(Monte Carlo) average.
If we impose translational invariance, i.e.\ if we consider
the quantity $D_{\mu \nu}^{b c}(\vec{x}_1\!-\!\vec{x}_2)\equiv
D_{\mu \nu}^{b c}(\vec{x}_1,\vec{x}_2)$, 
corresponding to total momentum conservation, we can also write
\begin{equation}
D_{\mu \nu}^{b c}(\vec{x})\,=\,
\Bigl< A_{\mu}^{b}(h;\vecx)\,
       A_{\nu}^{c}(h;\veczero) \Bigr> \,=\,
\frac{1}{V}\, \sum_{\vec{x}_2 \in \Lambda_x}
\Bigl< A_{\mu}^{b}(h;\vec{x}+\vec{x}_2)\,A_{\nu}^{c}(h;\vec{x}_2) \Bigr>
\label{eq:Dtranslation}
\; .
\end{equation}
Then, the associated (double)
Fourier transform $D_{\mu \nu}^{b c}(\vec{k}_1,\vec{k}_2)$ is diagonal in 
momentum space, see eq.\ (\ref{eq:AFourier}), i.e.\ 
\begin{eqnarray}
D_{\mu \nu}^{b c}(\vec{k}_1,\vec{k}_2) \,& = &\,
    \sum_{\vec{x}_1,\vec{x}_2}
    D_{\mu \nu}^{b c}(\vec{x}_1\!-\!\vec{x}_2) \,
     \exp{\left\{ - \frac{2 \pi i}{N}
     \left[\vec{k}_1 \cdots \Bigl(\vec{x}_1 +
              \frac{{\hat e}_{\mu}}{2}\Bigr) +
           \vec{k}_2 \cdots \Bigl(\vec{x}_2 +
              \frac{{\hat e}_{\nu}}{2}\Bigr)
        \right]\right\}} \nonumber \\[2mm]
 \,& = &\, 
    \sum_{\vec{x},\,\vec{x}_2}
    D_{\mu \nu}^{b c}(\vecx) \,
     \exp{\left\{ - \frac{2 \pi i}{N}
     \left[\vec{k}_1 \cdots \Bigl(\vec{x} +
              \frac{{\hat e}_{\mu}}{2}\Bigr) +
       \left(\vec{k}_2+\vec{k}_1\right) \cdots \vec{x}_2
    + \vec{k}_2 \cdots \frac{{\hat e}_{\nu}}{2}\right]\right\}}
                         \nonumber \\[2mm]
 \,& = &\, V \,
\delta(\vec{k}_1+\vec{k}_2,\veczero) \,
    \sum_{\vec{x}} \, 
    D_{\mu \nu}^{b c}(\vecx) \,
     \exp{\left\{ - \frac{2 \pi i}{N}
     \left[\vec{k}_1 \cdots \Bigl(\vec{x} +
              \frac{{\hat e}_{\mu}}{2}\Bigr) +
    \vec{k}_2 \cdots \frac{{\hat e}_{\nu}}{2}\right]\right\}}
          \nonumber \\[2mm]
 \,& = &\, V \,
\delta(\vec{k}_1,-\vec{k}_2) \,
\sum_{\vec{x}} \, D_{\mu \nu}^{b c}(\vecx)
   \, \exp{\left[ - \frac{2 \pi i}{N} \,\vec{k}_1 \cdots \left(\vec{x}
          + \frac{{\hat e}_{\mu}}{2}
          - \frac{{\hat e}_{\nu}}{2} \right)\right]} \; ,
\end{eqnarray}
where we defined $\vec{x} \equiv \vec{x}_1 - \vec{x}_2\,$ (with
$\vec{x},\vec{x}_1,\vec{x}_2 \in\Lambda_x$) and we used 
eq.\ (\ref{eq:deltakappa}).
Thus, after setting $\vec{k} \equiv \vec{k}_1 = - \vec{k}_2$, we can
write
\begin{equation}
D_{\mu \nu}^{b c}(\vec{k},-\veck) \, = \, V \,
\sum_{\vec{x}} \, D_{\mu \nu}^{b c}(\vecx)
   \, \exp{\left[ - \frac{2 \pi i}{N} \, \left(
   \vec{k} \cdots \vec{x} + \frac{k_{\mu} - k_{\nu}}{2}
          \right)\right]} \,\equiv\,
V \,  D_{\mu \nu}^{b c}(\veck)
\label{eq:Dofk}
\end{equation}
and
\begin{equation}
D_{\mu \nu}^{b c}(\vecx) \, = \, 
   \Bigl< \,A^{b}_{\mu}(h;\vecx)\, A^{c}_{\nu}(h;\veczero)
    \,\Bigr> \,=\,\frac{1}{V} \,
\sum_{\vec{k} \in {\widetilde \Lambda}_x} \,
         D_{\mu \nu}^{b c}(\veck) \,
\exp{\left[ \frac{2 \pi i}{N} \left(\vec{k} \cdots \vec{x}
                   + \frac{k_{\mu} - k_{\nu}}{2}\right)\right]} \; ,
\label{eq:Dofx}
\end{equation}
as can be seen
by substituting the rightmost expression above into eq.\ ({\ref{eq:Dofk})
and using eq.\ (\ref{eq:deltakappa}).
This defines, in a natural way,
the inverse Fourier transform for the gluon propagator.
Note that it is also in
agreement with the corresponding definition given in the case of the 
gauge field in eq.\ (\ref{eq:AantiFourier}), since it is equivalent to
\begin{eqnarray}
D_{\mu \nu}^{b c}(\veck,-\veck) \,&=&\, 
    \sum_{\vec{x},\vec{x}_2} \:
    \biggl<\,  A_{\mu}^{b}(h;\vec{x}+\vec{x}_2)\,A_{\mu}^{b}(h;\vec{x}_2)\,
         \biggr> \,
     \exp{\!\left\{\! - \frac{2 \pi i}{N} \left[ \vec{k} \cdots
     (\vec{x} +\vec{x}_2 -\vec{x}_2) + \frac{k_{\mu} -k_{\nu}}{2}\,
     \right]\right\}}
\nonumber \\[2mm]
&=&\, \Bigl< \,{\widetilde A}^{b}_{\mu}(h;\veck)\, 
            {\widetilde A}^{c}_{\nu}(h;-\veck) \,\Bigr> \,,
\end{eqnarray}
where we substituted (\ref{eq:Dtranslation}) into eq.\ (\ref{eq:Dofk}),
applied the translation $\vec{x}+\vec{x}_2 \to \vec{x}$ (with
$\vec{x}_2$ fixed) before summing over $\vec{x}\in \Lambda_x$, and used
(\ref{eq:AFourier}).

At the same time, due to global color invariance and to the
transversality condition (\ref{eq:Atransv}), the Landau-gauge propagator
must be given by (see, e.g., ref.\ \cite{Zwanziger:1991gz})
\begin{equation}
D_{\mu \nu}^{bc}(\vecx) \, = \,
    \frac{\delta^{bc}}{V} \, \Biggl\{
 \, D(\veczero) \, \delta_{\mu \nu} \, + \,
          \sum_{\substack{\vec{k} \in {\widetilde \Lambda}_x\\
                 \vec{k} \neq \vec{0}}} \, D(\veck)
    \, \exp{\left( \frac{2 \pi i}{N}
    \,\vec{k} \cdots \vec{x}
    \right)} \, P_{\mu \nu}(\veck) \,
    \exp{\left[ \frac{\pi i \left(
              k_{\mu} - k_{\nu}\right)}{N} \right]} \,
   \Biggr\} \; ,
\label{eq:Ddecomp}
\end{equation}
where $\vec{0}$ is the wave-number vector with all null components,
$\delta_{\mu \nu}$ stands for the Kronecker delta function of the lattice directions and
\begin{equation}
P_{\mu \nu}(\veck) \, \equiv \, \left[ \,
  \delta_{\mu \nu} \, - \, \frac{p_{\mu}(\veck) \,
    p_{\nu}(\veck)}{p^2(\veck)} \, \right]
\end{equation}
is the usual transverse projector, see eq.\ (\ref{eq:pmu}).
In particular, note that
\begin{equation}
D_{\mu \mu}^{bb}(\vecx) \, = \,
    \frac{D(\veczero)}{V} \, + \,
          \sum_{\substack{\vec{k} \in {\widetilde \Lambda}_x\\
                 \vec{k} \neq \vec{0}}} \, \frac{D(\veck)}{V}
    \, \exp{\left( \frac{2 \pi i}{N} \, \vec{k} \cdots \vec{x}
    \right)} \, P_{\mu \mu}(\veck) \; ,
\label{eq:Ddiag}
\end{equation}
where the repeated indices do not imply summation.
Then, the scalar function $D(\veczero)$ can be
evaluated, for example, using eqs.\ (\ref{eq:Ddiag}) and
(\ref{eq:deltakappa}), yielding\footnote{The
formulae reported here are those usually employed in lattice
numerical simulations.
However, it is evident that, in the evaluation of these scalar
functions, one could also make use of the off-diagonal Lorentz
components of $D_{\mu \nu}^{bb}(\vecx)$.
The evaluation of these (off-diagonal) components can be useful for
analyzing the breaking of rotational symmetry on the lattice
\cite{Leinweber:1998uu}.}
\begin{eqnarray}
D(\veczero) \,& \equiv &\, \frac{1}{d\,(N_c^2-1)}
   \sum_{b,\,\mu} \, \sum_{\vec{x}} 
              D_{\mu \mu}^{bb}(\vecx)
\,=\, \frac{1}{\cal N}
   \sum_{b, \mu}
   \sum_{\vec{x}, \,\vec{x}_2} \,
     \Bigl<  A_{\mu}^{b}(h;\vec{x}\!+\!\vec{x}_2)\,
             A_{\mu}^{b}(h;\vec{x}_2) \Bigr> \hspace{3mm}
\nonumber \\[2mm]
 & = &\, \frac{1}{\cal N}\,
   \sum_{b,\,\mu} \, \Bigl<\, \Bigl[\,\sum_{\vec{x}} \,
              A_{\mu}^{b}(h;\vecx)\,\Bigr]^{2} \Bigr> \; ,
\label{eq:D0def}
\end{eqnarray}
where we also used eq.\ (\ref{eq:Dtranslation}) and, in the last step,
we applied again the translation $\vec{x}+\vec{x}_2 \to \vec{x}$ (with
$\vec{x}_2$ fixed).
As always, in the sums we have $\mu=1,\ldots,d\,$, the color
index $b$ takes values $1,\ldots,N_c^2-1$ and
$\vec{x},\vec{x}_2 \in\Lambda_x$.
We also defined the normalization factor ${\cal N}\equiv d(N_c^2-1)V$.
Similarly, we have
\begin{eqnarray}
\!\!\!\!\! D(\veck) \,& \equiv &\, \frac{1}{(d-1)\,(N_c^2-1)} \,
   \sum_{b, \mu} \, \sum_{\vec{x}} \, D_{\mu \mu}^{bb}(\vecx)\,
    \exp{\!\left(\! - \frac{2 \pi i}{N} 
\:\vec{k} \cdots \vec{x}\,
\right)} \nonumber \\[2mm]
& = &\, \frac{1}{{\cal N}'}\, \sum_{b,\mu} \,
    \sum_{\vec{x},\vec{x}_2} \:
    \biggl<\,  A_{\mu}^{b}(h;\vec{x}+\vec{x}_2)\,A_{\mu}^{b}(h;\vec{x}_2)\,
         \biggr> \,
     \exp{\!\left[ - \frac{2 \pi i}{N} \; \vec{k} \cdots
                      (\vec{x} +\vec{x}_2 -\vec{x}_2) \right]}
       \nonumber \\[2mm]
& = &\, \frac{1}{{\cal N}'}\,
   \sum_{b,\mu}\, \biggl< \,
\sum_{\vec{x}}  A_{\mu}^{b}(h;\vecx)\,
     \exp{\!\left(\! - \frac{2 \pi i}{N} \: \vec{k} \cdots \vec{x}\right)} 
\; \sum_{\vec{x}_2} A_{\mu}^{b}(h;\vec{x}_2)\,
     \exp{\!\left( \frac{2 \pi i}{N}\: \vec{k} \cdots \vec{x}_2 \right)} 
\biggr> \label{eq:Dkexp} \\[2mm]
& = &\, \frac{1}{{\cal N}'}
   \sum_{b,\mu} \Biggl< \:\!
     \left[\,\sum_{\vec{x}} A_{\mu}^{b}(h;\vecx) 
\cos{\!\left( \frac{2 \pi}{N} \vec{k} \cdots \vec{x} \right)}
\right]^{2}
\!\!+ \left[\,\sum_{\vec{x}} A_{\mu}^{b}(h;\vecx)
 \sin{\!\left( \frac{2 \pi}{N} \vec{k} \cdots \vec{x} \right)}
   \right]^{2} \:\! \Biggr> \; , \hspace{8mm} \label{eq:Dkdef}
\end{eqnarray}
where we used one more time the translation $\vec{x}+\vec{x}_2 \to
\vec{x}$ and we have defined ${\cal N}'\equiv (d-1)(N_c^2-1)V$.

Let us remark that the above expressions, obtained in the lattice
formulation, are essentially the same as in the continuum, with only a
few subtleties.
In particular, in the continuum, the scalar
quantities $D(\veczero)$ and $D(\veck)$ depend only on the magnitude
$k$ of the wave-number vector $\vec{k}$ (or of the corresponding
momentum $p \propto k$) and are usually denoted by $D(0)$ and
$D(k)$.
This notation is also very often employed in lattice studies.
Here, however, we prefer to keep explicitly the dependence of the
gluon propagator on the components of $\vec{k}$ for two (related) reasons.
Firstly, due to the breaking of the rotational symmetry \cite{deSoto:2022scb},
it is no
longer true that the lattice results for the gluon propagator are
just a function of $k$.
Secondly, when representing $D(\veck)$ as a function of $p^2(\veck)$,
see eq.\ (\ref{eq:p2def}), it is necessary to consider all the
components of $\vec{k}$ --- and not simply its magnitude $k$ --- since
$p^2$ is not proportional to $k^2$.
Let us also recall \cite{Cucchieri:1997dx} that the factor $d\!-\!1$
in the denominator of the expression for $D(\veck)$ comes from
\begin{equation}
\sum_{\mu=1}^d\,P_{\mu \mu}(\veck)\,=\,d\!-\!1
\end{equation}
and tells us that, for each value of $b$, there are only
$d\!-\!1$ linearly independent components ${\widetilde A}_{
\mu}^{b}(h;\veck)$, due to the Landau-gauge-fixing condition,
see eq.\ (\ref{eq:Atransv}).
At the same time, the factor $d$ in the denominator of the
expression for $D(\veczero)$ reflects the fact that the
same equation does not impose any constraint on the gauge field for
$\vec{k}=\vec{0}$.
Also note that eq.\ (\ref{eq:Dkdef})
is invariant\footnote{The same invariance applies to the magnitude
of the lattice momenta $p^2(\veck)$, see eqs.\ (\ref{eq:pmu}) and
(\ref{eq:p2def}).}
under the reflection $\vec{k} \to -\vec{k}$ or, more
generally, under the reflection $\vec{k} \to -\vec{k}+N
\hat{e}_{\mu}$.

The gluon-propagator functions $D(\veczero)$ and $D(\veck)$ can also be 
written in terms of the momentum-space gauge field ${\widetilde A}_{\mu}^{
b}(h;\veck)$, see eq.\ (\ref{eq:AFourier}),
yielding
\begin{equation}
D(\veczero) \, = \, \frac{1}{\cal N}
   \sum_{b,\,\mu} \,\Bigl<\,
  {{\widetilde A}_{\mu}^{b}(h;\veczero)}^2\,
  \Bigr>
\,=\,\frac{1}{2\,{\cal N}}\, \sum_{\mu} \,\Tr\, \Bigl<\,
  {{\widetilde A}_{\mu}(h;\veczero)}^2 \,\Bigr>
  \label{eq:D0defTr} 
\end{equation}
and, see eq.\ (\ref{eq:Dkexp}),
\begin{eqnarray}
D(\veck) \!& = \!& \frac{1}{{\cal N}'} \!\sum_{b,\,\mu,\,\vec{x},\vec{x}_2}
 \! \biggl<\! A_{\mu}^{b}(h;\vec{x})
 \exp{\!\left[\!-\frac{2 \pi i}{N} \!\left(\!\vec{k} \cdots \vec{x} 
  \!+\! \frac{k_{\mu}}{2} \!\right)\!\right]}
     A_{\mu}^{b}(h;\vec{x}_2)
  \exp{\!\left[\frac{2 \pi i}{N} \!\left(\!\vec{k} \cdots \vec{x}_2 \!+\!
    \frac{k_{\mu}}{2}\!\right)\!\right]} \biggr> \nonumber \\[2mm]
&=&\, \frac{1}{{\cal N}'} \,\sum_{b,\,\mu}\, \Bigl<
     {\widetilde A}_{\mu}^{b}(h;\veck)\,
     {\widetilde A}_{\mu}^{b}(h;-\veck)
            \Bigr>
\,=\, \frac{1}{2\,{\cal N}'}\,
   \sum_{\mu} \,\Tr\,\Bigl<
     {\widetilde A}_{\mu}(h;\veck)\,
     {\widetilde A}_{\mu}(h;-\veck)
            \Bigr> \; ,
\label{eq:DkdefbisTr}
\end{eqnarray}
where we used (\ref{eq:Trlambda}) and the definition
\begin{equation}
{\widetilde A}_{\mu}(h;\veck)\,\equiv\,
\sum_{b=1}^{N_c^2-1}t^b\,
{\widetilde A}_{\mu}^{b}(h;\veck) \; ,
\end{equation}
in analogy with eq.\ (\ref{eq:Aofx}).
At the same time, eq.\ (\ref{eq:AFourier}) implies that
\begin{equation}
{\widetilde A}_{\mu}(h;\veck)\,=\,
    \sum_{b=1}^{N_c^2-1}t^b\,
  \sum_{\vec{x} \in \Lambda_x} \,
    A_{\mu}^{b}(h;\vecx)
    \, \exp{\left[ - \frac{2 \pi i}{N} \left(\vec{k} \cdots
       \vec{x} + \frac{k_{\mu}}{2}\right)\right]}\,.
\label{eq:AFourier2}
\end{equation}
Then, given that the generators $t^b$ of the SU($N_c$) group 
have been chosen to be Hermitian and the components
$A_{\mu}^{b}(h;\vecx)$ are real, see comment below eq.\
(\ref{eq:hofgamma}) [or eq.\ (\ref{eq:Adue})], we have
\begin{equation}
       \left[\, {\widetilde A}_{\mu}(h;\veck)\,
       \right]^{\dagger} \,=\,
{\widetilde A}_{\mu}(h;-\veck) \; .
\label{eq:Atilde-k}
\end{equation}
Thus, we can also write eq.\ (\ref{eq:DkdefbisTr}) as
\begin{equation}
D(\veck) \, = \,
 \frac{1}{2\,{\cal N}'}\,
   \sum_{\mu=1}^d \, \Tr \; \Bigl< \,
     {\widetilde A}_{\mu}(h;\veck)\,
     \left[\, {\widetilde A}_{\mu}(h;\veck)\,
       \right]^{\dagger}\,\Bigr> \; .
\end{equation}
Finally, when considering a global gauge transformation $v$,
${\widetilde A}_{\mu}(h;\veck)$ transforms, see eqs.\
(\ref{eq:Aofx}), (\ref{eq:Aglobal}) and (\ref{eq:AFourier2}), as
\begin{equation}
{\widetilde A}_{\mu}(h;\veck)\,\to\, v \,
{\widetilde A}_{\mu}(h;\veck)\,v^{\dagger} \; ,
\end{equation}
so that the scalar functions $D(\veczero)$ and $D(\veck)$ are
invariant under such (global) gauge transformations.
In other words,
the Landau-gauge gluon propagator has the same invariance of
the minimal-Landau-gauge condition and of the quantities
(\ref{eq:e1})--(\ref{eq:e6}), shown in the last section.


\section{Minimal Landau gauge on the extended lattice}
\label{sec:bloch-PBCs}

Here we define the extended-lattice version of the gauge-fixing problem 
presented in the previous section, highlighting the similarities with
Bloch's theorem and discussing the corresponding result for the minimal
Landau gauge in Yang-Mills theory.
More specifically, after describing the setup, we
review, in section \ref{sec:Bloch-theorem-solid}, the statement of
the theorem in solid-state physics, summarizing its demonstration.
Then, in section \ref{sec:Bloch-theorem},
we outline the analogous result for the gauge-fixing case, while
in section \ref{sec:proof} we present its proof.
Our notation for Cartan sub-algebras and other mathematical details
that are relevant in the gauge-theory case are given in the appendix.

\vskip 3mm
Following refs.\ \cite{Cucchieri:2016qyc,Zwanziger:1993dh} we
consider a thermalized link configuration $\{U_{\mu}(
\vecx)\}$, for the SU($N_c$) gauge group in $d$ dimensions,
defined on a lattice $\Lambda_x$ with volume $V=N^d$ and PBCs.
Then, we extend this configuration by replicating it $m$ times
along each direction, yielding a configuration on the extended
lattice $\Lambda_z$, with lattice volume $m^d V$.
We parametrize the sites of $\Lambda_z$ by
\begin{equation}
\vec{z} \, \equiv \, \vec{x} \, + \, N\,\vec{y}\; ,
\label{eq:zcoord}
\end{equation}
where $\vec{x}\in \Lambda_x$ and $\vec{y}$ belongs to the {\em
index lattice}\footnote{Note that in \cite{Cucchieri:2016qyc} we 
referred to $\Lambda_y$ as the ``replica'' lattice.}
$\{ \Lambda_y$: $y_{\mu} = 0, 1, \ldots, m-1 \}$,
so that the components $z_{\mu}$ take values $1, 2, \ldots, m N$.
We also denote by ${\Lambda_{x}}^{\!\!(\vecy)}$ each of the $m^d$ (identical)
replicas of the original lattice $\Lambda_x$, specified by the $\vec{y}$ index 
coordinates.
By construction, $\{ U_{\mu}(\vecz) \}$ is invariant under
translations by $N$ in any direction.

Then, as was done in the previous section for the original lattice 
$\Lambda_x$, we impose the minimal-Landau-gauge condition on $\Lambda_z$, 
i.e.\ we minimize the functional
\begin{eqnarray}
{\cal E}_{\scriptU}[g] \,& \equiv &\, \frac{\Re \,
   \Tr}{N_c\,d\,m^d\,V}
 \,\sum_{\mu = 1}^d \, \sum_{\vec{z}\in\Lambda_z} \, \left[ \,
 \1 \, - \, U_{\mu}(g;\vecz) \, \right] \,,
 \label{eq:minimizingLz} \\[2mm]
U_{\mu}(g;\vecz) \,& \equiv &\, g(\vecz)
\, U_{\mu}(\vecz) \, g(\vec{z} + \hat{e}_{\mu})^{\dagger}
 \;  \label{eq:Uofg}
\end{eqnarray}
with respect to the gauge transformation $\{ g(\vecz) \}$, while keeping 
the link configuration $\{ U_{\mu}( \vecz) \}$ fixed.
Here, $g(\vecz)$ are SU($N_c$) matrices subject to PBCs on
the extended lattice $\Lambda_z$, i.e.\
\begin{equation}
g(\vec{z} + m N \hat{e}_{\mu}) \, = \, g(\vecz) \; .
\label{eq:gPBCs}
\end{equation}

The resulting gauge-fixed field configuration is, of course, transverse 
on $\Lambda_z$, and it is also invariant under a translation by 
$m N \hat{e}_{\mu}$.
Indeed, as mentioned above, by construction of 
$\Lambda_z$ we have
$U_{\nu}(\vec{z}+N\hat{e}_{\mu}) = U_{\nu}(\vecz) =
U_{\nu}(\vec{z} + m N \hat{e}_{\mu})$ for $\mu, \nu = 1, \ldots, d$.
Then, from eqs.\ (\ref{eq:Uofg}) and (\ref{eq:gPBCs}) we get
\begin{equation}
U_{\nu}(g; \vec{z} + m N \hat{e}_{\mu})
 \, = \, U_{\nu}(g; \vecz) \; .
\label{eq:UPBCs}
\end{equation}
We thus have invariance under a translation by $m N \hat{e}_{\mu}$
 --- i.e.\ PBCs on $\Lambda_z$ --- for the transformed gauge field.
On the other hand, the original invariance under a translation by 
$N \hat{e}_{\mu}$ is
lost after the gauge-fixing process, since the gauge 
transformation $\{ g(\vecz) \}$ does not have it.


\subsection{Bloch's theorem for a crystalline solid}
\label{sec:Bloch-theorem-solid}

As explained in ref.\ \cite{Cucchieri:2016qyc}, the extended-lattice 
problem defined above on $\Lambda_{z}$ is very similar to the setup 
usually considered in the proof of Bloch's theorem \cite{AM} for an 
(ideal) crystalline solid in $d$ dimensions.
Indeed, the index lattice $\Lambda_y$ corresponds to a
finite cubic Bravais lattice, with $m$ unit cells in each
direction, equipped with PBCs.
Equivalently, this Bravais lattice is a
simple cubic lattice, with cells indexed by vectors $\vecy\in\Lambda_y$.
At the same time, the original lattice $\Lambda_x$ may be viewed as
a primitive cell of the Bravais lattice.
Let us recall that, in state-solid physics, the primitive cell is defined as the $d$-dimensional volume 
spanned by the (orthogonal) primitive vectors $l \hat{e}_{\mu}$, where $l$ 
is the length of the cell, i.e.\ a vector $\vec{r}$ restricted to the primitive cell 
is written as $\,l \sum_{\mu=1}^{d}\, r_{\mu} \hat{e}_{\mu}$, with 
$r_{\mu} \in [0,1)$.
Finally, the thermalized lattice configuration $\{ U_{\mu}(
\vecz) \}$, invariant under translation by $N\,\vec{y} = N \sum_{
\mu=1}^d y_{\mu} \hat{e}_{\mu}$ with $\vec{y} \in \Lambda_y$,
corresponds (for example) to a periodic electrostatic potential
$U(\vecr)$, invariant under translations by any vector $\vec{R} =
l \,\sum_{\mu = 1}^d R_{\mu} {\hat e}_{\mu}$ of the Bravais lattice,
where the integer components $R_{\mu}$ take values $0,1,\ldots,m-1$.

Bloch's theorem states that the solution of the Schrödinger equation
for this problem, i.e.\ the wave function $\psi(\vecr)$ for an electron in 
such a periodic potential, can be expressed as a combination of so-called
Bloch states --- or {\em Bloch waves} --- given by a plane wave (over the 
whole lattice) modulated by a function, which is obtained as a 
(periodic) solution to the restricted unit-cell problem.
More precisely, let us denote by $\psi(\vecr)$ any function defined on
the considered crystalline cubic lattice and by $L = l m$ the physical
size of the lattice.
Then, the use of PBCs, i.e.\ the condition 
$\psi(\vecr)=\psi(\vec{r} + L \hat{e}_{\mu})$ for any direction $\mu$,
implies that $\psi(\vecr)$ can be (Fourier) expanded in
plane waves $\exp{(2 \pi i \, \veck \cdots \vec{r} / L)}$ with
\begin{equation}
\exp{\left(2 \pi i \, \frac{\veck \cdots L \hat{e}_{\mu}}{L}\right)}\,=\,
\exp{\left(2 \pi i\, k_{\mu}\right)}\,=\,1 \; .
\label{eq:kappaL}
\end{equation}
This tells us that the components of $\veck$ are integer
numbers (i.e.\ $k_{\mu} \in {\cal Z})$ and that, when they
are restricted to the {\em first Brillouin zone}, we have\footnote{Here,
in order to simplify the notation, we consider an even value for
the integer $m$.
For $m$ odd, the integers $k_{\mu}$ take values in the interval
$[-(m-1)/2,(m-1)/2]$.
(See also footnote \ref{foot:symmetrick}.)
\label{foot:m-even}}
$k_{\mu} \in [-m/2, m/2)$, yielding discrete Fourier momenta
$\widetilde{k}_{\mu} \equiv 2 \pi k_{\mu}/ (lm) \in [-\pi/l,\pi/l)$.
Then, with this restriction, the allowed plane waves have components
$k_{\mu} + m K_{\mu}$, with $K_{\mu} \in {\cal Z}$, i.e.\ they
can be written as $\exp{[2 \pi i \, (\veck+m\,\vecK) \cdots
\vec{r} / L\,]}$.
Here, the vector $\,m\, \vec{K} / L = \sum_{\mu=1}^{d}\, K_{\mu} \hat{e}_{
\mu}/l$ corresponds to the so-called reciprocal lattice, i.e.\ it is such
that
\begin{equation}
\exp{\left(2 \pi i\; \frac{m\, \vec{K}}{L} \cdots \vecR \,\right)}\,=\,
\exp{\left( 2 \pi i \,\sum_{\mu=1}^{d}\,K_{\mu} \,R_{\mu} 
\right)}\,=\, 1
\label{eq:reciprocal}
\end{equation}
for any translation vector $\vec{R}$ of the Bravais lattice, yielding
\begin{equation}
\exp{\left[ 2 \pi i \left(\frac{\veck+m\,\vecK}{L}\right) \cdots
\vecR\right]}\,=\, 
\exp{\left(2 \pi i \,\frac{\veck}{L} \cdots \vecR\,\right)}\,=\,
\exp{\left( 2 \pi i \,\sum_{\mu=1}^{d}\,\frac{k_{\mu} \,R_{\mu}}{m} 
\right)} \; ,
\label{eq:waveR}
\end{equation}
with $\veck$ in the first Brillouin zone.

With this setup, one can prove Bloch's theorem (see, e.g., the first
proof in ref.\ \cite{AM}) by using the properties of the translation
operator
\begin{equation}
{\cal T}(\vecR) \, \psi(\vecr) \, = \,
  \psi(\vec{r} +\vecR) \; .
\label{eq:Tdef}
\end{equation}
In particular, we need to recall the relation
\begin{equation}
  {\cal T}(\vecR) \, {\cal T}(\vecRpp)
       \, = \, {\cal T}(\vecRpp) \, {\cal T}(\vecR) \, = \,
      {\cal T}(\vec{R} + \vecRpp) \; ,
\label{eq:TT}
\end{equation}
valid for all vectors $\vec{R}$ and $\vecRp$ on the Bravais lattice.
Hence, the translation operators form an Abelian group, with the
trivial identity element ${\cal T}(\veczero)$ and the inverse
element ${\cal T}^{\mk -1}(\vecR) = {\cal T}(-\vecR)$.
At the same time,
it is evident that
any plane wave $\exp{[2 \pi i \, (\veck+m\,\vecK) \cdots
\vec{r} / L]}$ --- with fixed $\veck$ (restricted to
the first Brillouin zone) and $\vecK$ as above --- is an
eigenfunction of ${\cal T}(\vecR)$ with eigenvalue $\exp{(2 \pi
i \veck \cdots \vecR/L)}$, see eq.\ (\ref{eq:waveR}).
Thus, in the most general case, we have the eigenvectors
\begin{equation}
{\cal T}(\vecR) \, \psi_{\veck}(\vecr)
  \, = \, \psi_{\veck}(\vec{r} +\vecR) \, = \,
    \exp{\left( 2 \pi i \,\frac{\veck}{L} \cdots \vecR\right)} \,
        \psi_{\veck}(\vecr)
\label{eq:Teigenvalue}
\end{equation}
with
\begin{equation}
  \psi_{\veck}(\vecr)\,=\,\sum_{\vec{K}}\,
    c_{\veck}(\vecK)\,
 \exp{\left[ 2 \pi i \left(\frac{\veck+m\,\vecK}{L}\right) \cdots
           \vec{r} \,\right]} \; ,
\end{equation}
where $\veck$ is fixed and taken in the first Brillouin zone, while
$\vecK$ refers to vectors of the reciprocal lattice.
The last result is usually written as
\begin{equation}
  \psi_{\veck}(\vecr)\,=\,
 \exp{\!\left(2 \pi i\, \frac{\veck}{L} \cdots \vec{r}\right)} \,
\sum_{\vecK} c_{\veck}(\vecK)
 \exp{\!\left( 2 \pi i \frac{m\,\vecK}{L} \!\cdot \vec{r} \right)} 
  \;\equiv\;
  \exp{\!\left( 2 \pi i \,\frac{\veck}{L} \cdots \vec{r} \right)}\,
  u_{\veck}(\vecr) \; ,
  \label{eq:psi-uk}
\end{equation}
where the function $u_{\veck}(\vecr)$ trivially satisfies,
see eq.\ (\ref{eq:reciprocal}), the condition 
\begin{equation}
u_{\veck}(\vec{r}+\vecR)\,=\,
u_{\veck}(\vecr) \;.
\label{eq:uBCs}
\end{equation}
Hence, $u_{\veck}(\vecr)$ is effectively specified by vectors $\vecr$ in
the primitive cell and may be obtained from a restricted version of the
original problem.

The proof of Bloch's theorem goes as follows.
The Hamiltonian ${\cal H}$ for the crystalline solid is,
by hypothesis, invariant under a translation by $\vec{R}$,
i.e.\ ${\cal H}$ commutes with ${\cal T}(\vecR)$.
Then, one can choose the eigenstates $\psi_{\veck}(\vecr)$
of ${\cal T}(\vecR)$ to also be eigenstates of ${\cal H}$, i.e.\
\begin{equation}
{\cal H}\,\psi_{\veck}(\vecr)\,=\,
\lambda_{\veck} \,\psi_{\veck}(\vecr) \; .
\label{eq:Hpsi}
\end{equation}
Equivalently, by using eq.\ (\ref{eq:psi-uk}), one can define
\cite{AM}
\begin{equation}
{\cal H}\,\psi_{\veck}(\vecr)
\,=\, \lambda_{\veck} \, 
\exp{\!\left( 2 \pi i \,\frac{\veck}{L} \cdots \vec{r} \right)}\,
u_{\veck}(\vecr) \; 
\,\equiv\,
\exp{\!\left( 2 \pi i \,\frac{\veck}{L} \cdots \vec{r} \right)}\;
{\cal H}_{\veck}\,u_{\veck}(\vecr)
\end{equation}
and consider, instead of the original problem (\ref{eq:Hpsi}) on the
Bravais lattice and with the Hamiltonian ${\cal H}$, the new problem
\begin{equation}
{\cal H}_{\veck}\,u_{\veck}(\vecr)\,=\,
\lambda_{\veck} \, u_{\veck}(\vecr) \; ,
\label{eq:Hpsikappa}
\end{equation}
which is restricted to a single primitive cell and subject to
the BCs (\ref{eq:uBCs}).
In the general case, one expects the last eigenvalue problem to
have infinite solutions (indexed by $n$), i.e.\ we can write
\begin{equation}
{\cal H}_{\veck}\,u_{\veck,n}(\vecr)\,=\,
\lambda_{\veck,n} \, u_{\veck,n}(\vecr)
\; . \label{eq:Hpsikappa-n}
\end{equation}
Clearly, ${\cal H}_{\veck}$ depends\footnote{In 
particular, the explicit form of ${\cal H}_{\veck}$
corresponds to a ``shifted'' kinetic term (by the momentum $\veck$) plus
the periodic potential $U(\vecr)$, defined for the primitive cell \cite{AM}.}
on the (discretized) components $\widetilde{k}_{\mu} \equiv 2 \pi k_{\mu}/ (lm)
\in [-\pi/l,\pi/l)$.
Hence, when one considers the infinite-volume limit $m \to +
\infty$, the new Hamiltonian depends on the (now continuous)
parameters $\widetilde{k}_{\mu}$ and one expects the energy
levels $\lambda_{\veck,n}$ to be also a continuous
function of these parameters.
Then, for each $n$, these values constitute a so-called
energy band, leading to the description of the solid in terms of a
band structure.


\subsection{Bloch's theorem for the gauge-fixing problem}
\label{sec:Bloch-theorem}

The above setup applies --- in a rather straightforward manner --- also
to the gauge link configuration on the extended lattice $\Lambda_z$.
The main difference is that, here, the primitive cell, i.e.\
the original lattice $\Lambda_x$, is also discretized, since it is given
by the vectors $\vec{x}
= a \sum_{\mu =1}^{d}\, x_{\mu} \hat{e}_{\mu}$, where $a$ is the 
{\em lattice spacing} and the components $x_{\mu}$ take integer values 
in $[1,N]$.
Thus, in the above formulae for the crystalline solid, we just have
to substitute the magnitude $l$ with $N a$ (and, therefore, $L$ with $m N a$).
Then, after setting the lattice spacing equal to 1, as usually done in 
lattice gauge theory, we find that the vectors of the Bravais lattice 
become $\vec{R} = N \sum_{\mu=1}^{d}\, R_{\mu} \hat{e}_{\mu}$, with
$R_\mu=0,1,\ldots,m-1$.
Finally, by combining the original lattice $\Lambda_x$ with the index
lattice $\Lambda_y$, we recover our notation for $\Lambda_z$,
identifying the components $R_{\mu}$ with $y_{\mu}$ and $\vec{r}$
with $\vec{z}=\vec{x}+N\vec{y}$.
In particular, we find that the generic plane waves $\,\exp{[2 \pi i \, 
\veckp \!\cdot\! \vecz/ (m N)]}$ are written in terms of wave-number
vectors with components $k_{\mu}^{\prime} = k_{\mu} + m\, K_{\mu}$,
as above.
However, as stressed before (see footnote \ref{foot:symmetrick}),
instead of the symmetric interval around $0$ usually taken for
the first Brillouin zone, here we consider integers $k_{\mu}$
in the interval $[0,m-1]$ and
$K_{\mu}$ in $[0,N\!-\!1]$.

In analogy with the Bloch theorem described in the previous section,
one can prove (see appendix F of ref.\ \cite{Zwanziger:1993dh}
and section \ref{sec:proof} below) that the gauge transformation
$g(\vecz)$ that minimizes
the functional ${\cal E}_{\scriptU}[g]$, see eqs.\
(\ref{eq:minimizingLz}) and (\ref{eq:Uofg}), defined for the extended
lattice through $\vec{z}=\vec{x}+N\vec{y}$, can be written as
\begin{equation}
g(\vecz) \, =\, \exp{\left( i \sum_{\mu = 1}^d \,
                  \frac{\Theta_{\mu} \, z_{\mu}}{N}
                    \right)} \, h(\vecx) \; ,
\label{eq:bloch-g}
\end{equation}
where $h(\vecx)=h(\vecx+N\vec{y})$ has the periodicity of the original lattice 
$\Lambda_x$ and the matrices $\Theta_{\mu}$
belong to a Cartan sub-algebra of the $su(N_c)$ Lie algebra,
i.e.\ they commute.
In appendix \ref{sec:Cartan} we discuss the main
properties of these matrices, which can be written as
\begin{equation}
\Theta_{\mu} \, = \, \sum_{b=1}^{N_c-1} \,
    \theta^b_{\mu}\, t^b_{\scriptC}\,,
\label{eq:tautheta}
\end{equation}
where $\theta^b_{\mu}$ ($\mu=1,\ldots,d$) are real parameters and
the matrices $t^b_{\scriptC}$ are the generators of the
Cartan sub-algebra of $su(N_c)$, which has dimension $N_c\!-\!1$.

\vskip 3mm
As a result of eq.\ (\ref{eq:bloch-g}) above and the cyclicity of
the trace, the minimizing functional ${\cal E}_{\scriptU}[g]$
in eq.\ (\ref{eq:minimizingLz}) becomes
\begin{equation}
{\cal E}_{\scriptU}[g] \, = \, \frac{\Re \,
    \Tr}{N_c\,d\,m^d\,V} \,
\sum_{\mu = 1}^d \, \sum_{\vec{z}\in\Lambda_z} \, \left[ \, \1
    \, - \, U_{\mu}(h;\vecx) \,
            e^{- i \frac{\Theta_{\mu}}{N}} \, \right] \; ,
            \label{eq:minfunctUhTheta}
\end{equation}
which is independent of $\vec{y}$.
Thus, we can write
\begin{equation}
{\cal E}_{\scriptU}[g] \, \equiv \,
{\cal E}_{\scriptU,\scriptT}[h] \,=\,
\, \frac{\Re \,
    \Tr}{N_c\,d\,V} \,
\sum_{\mu = 1}^d \, \sum_{\vec{x}\in\Lambda_x} \, \left[ \, \1
    \, - \, U_{\mu}(h;\vecx) \,
            e^{- i \frac{\Theta_{\mu}}{N}} \, \right] \;
            \label{eq:minfunctUhThetax}
\end{equation}
and define
\begin{equation}
{\cal E}_{\scriptU,\scriptT}[h]
\,\equiv\,\, \frac{\Re \, \Tr}{N_c\,d} \,
 \sum_{\mu = 1}^d \, \left[ \, \1 \, - \, Z_{\mu}(h) \,
      \frac{e^{- i \frac{\Theta_{\mu}}{N}}}{V}
    \, \right] \; ,
\label{eq:minimizing2}
\end{equation}
where
\begin{equation}
  Z_{\mu}(h) \, \equiv \,
    \sum_{\vec{x}\in\Lambda_x}
        \, U_{\mu}(h;\vecx)
\label{eq:Z}
\end{equation}
is the zero mode of the (gauge-transformed) link variable
$U_{\mu}(h;\vecx)$ in a given direction, and it
is evident that the numerical minimization can now be carried out
on the original lattice $\Lambda_x$.
At the same time, imposing PBCs on $\Lambda_z$ in eq.\ (\ref{eq:bloch-g}),
we see that the expression (with no summation over the index $\mu$)
\begin{equation}
  \exp{\left( i \,\frac{\Theta_{\mu}\, z_{\mu}}{N} \right)}  \; ,
\end{equation}
evaluated for $z_{\mu}= m N$, should be equal to
\begin{equation}
  \exp{\left( i m\Theta_{\mu} \right)} \, = \,
  \left[ \, \exp{\left( i \Theta_{\mu} \right)} \,\right]^m
             \, = \, \1 \; .
\label{eq:pbctheta}
\end{equation}
Thus, the matrices $\Theta_{\mu}$ have eigenvalues of the type
$2 \pi n_{\mu} / m$, where $n_{\mu}$ is an integer.
Equivalently,
the matrices $\exp{( i \Theta_{\mu} )}$ have eigenvalues $\exp{(
2 \pi i n_{\mu} / m )}$.

By comparing eq.\ (\ref{eq:bloch-g}) with eq.\ (\ref{eq:psi-uk}),
and also eq.\ (\ref{eq:pbctheta}) with eq.\ (\ref{eq:kappaL}), its is evident 
that the matrices $\Theta_{\mu}$ play the role of the momentum
$\veck$ in the crystalline-solid problem.
It is also interesting to observe that, from the numerical point of
view, the minimizing functional (\ref{eq:minimizing2})--(\ref{eq:Z})
can be interpreted as the usual minimizing functional
(\ref{eq:minimizing}) on the lattice $\Lambda_x$, using a periodic
gauge transformation $h(\vecx)$, together with an ``extended''
(i.e.\ nonperiodic) gauge transformation $\exp{( i \sum_{\nu = 1}^d
\, \Theta_{\nu} \, x_{\nu}/ N )}$.
The functional ${\cal E}_{\scriptU, \scriptT}[h]$, however,
still depends (implicitly) on the size $m$ of the index lattice
$\Lambda_y$ through eq.\ (\ref{eq:pbctheta}).
One should also note that the substitution of the  original
minimizing function ${\cal E}_{\scriptU}[g]$ --- which considers
the gauge transformation $g(\vecz)$ on the extended lattice
$\Lambda_z$ --- with the modified minimizing function ${\cal E}_{
\scriptU, \scriptT}[h]$ --- which is restricted to the original lattice 
$\Lambda_x$ and depends on the $\Theta_{\mu}$ matrices, see again 
eqs.\ (\ref{eq:minimizing2}) and (\ref{eq:Z}) --- is completely analogous
to the substitution of the eigenvalue problem (\ref{eq:Hpsi}) with the
problem (\ref{eq:Hpsikappa}).
The main difference is that,
while the vector $\veck$
is fixed in the Hamiltonian ${\cal H}_{\veck}$, the matrices
$\Theta_{\mu}$ are chosen by the minimization algorithm (see
section \ref{sec:numerical-PBCs} below).
On the other hand, one could also consider --- in analogy with the
usual condensed-matter approach --- a given (fixed) set of matrices
$\Theta_{\mu}$ and look (for example) at the different Gribov
copies corresponding to different solutions $\{ h(\vecx)\}$ of the
small-lattice problem (\ref{eq:minfunctUhThetax}) defined by the
$\Theta_{\mu}$'s.

We should note here that we are using the same notation as
in section \ref{sec:mlg-PBCs} for the solution $\{ h(\vecx)\}$, meaning a 
periodic gauge transformation --- i.e.\ effectively restricted to the small
lattice $\Lambda_x$ --- that solves the optimization problem defined by the 
minimizing functional on $\Lambda_x$.
However, one must remember that, in the extended-lattice problem, 
the corresponding functional does not depend only on 
$\{U_\mu(\vecx)\}$ and $\{h(\vecx)\}$, but also on $\{\Theta_{\mu}\}$. 
In fact, as it is evident from eq.\ (\ref{eq:bloch-g}),
here the gauge transformation $h(\vecx)$ is {\em not} just the 
restriction of $g(\vecz)$ to the small lattice $\Lambda_x$,
but it is the solution
to the modified small-lattice problem (\ref{eq:minfunctUhThetax}).
Hence, if we want to relate the two objects, we might
say that the transformation $\{ h(\vecx)\}$ in section \ref{sec:mlg-PBCs} 
is the minimum of ${\cal E}_{\scriptU, \scriptT}[h]$ with all
matrices $\Theta_{\mu}$ trivially given by $\1$.
This distinction will be made clearer in the next few sections.


\subsection{Proof of equation (\ref{eq:bloch-g})}
\label{sec:proof}

Expression (\ref{eq:Tdef}) can of course be applied also to the
lattice setup considered in section \ref{sec:Bloch-theorem}.
For example, the translation operator ${\cal T}(N \hat{e}_{\mu})$
acts on $U_{\nu}(\vecz)$ and $g(\vecz)$ by shifting them to the
site $\vec{z} + N \hat{e}_{\mu}$, i.e.\
\begin{eqnarray}
{\cal T}(N \hat{e}_{\mu}) \, U_{\nu}(\vecz) \,& = &\,
                     U_{\nu}(\vec{z} + N \hat{e}_{\mu})\,, 
\label{eq:translU} \\[2mm]
{\cal T}(N \hat{e}_{\mu}) \, g(\vecz) \,& = &\,
                     g(\vec{z} + N \hat{e}_{\mu}) \; .
\end{eqnarray}
Moreover, the use of PBCs, see eqs.\
(\ref{eq:gPBCs})--(\ref{eq:UPBCs}), implies that
\begin{equation}
{\cal T}(m N \hat{e}_{\mu}) \, = \, 
\left[\ {\cal T}(N \hat{e}_{\mu})\ \right]^{\,m}
\ = \, \1 \; ,
\label{eq:TmN1}
\end{equation}
where $\1$ is the identity operator.
Also, with our setup,
the effect of ${\cal T}(N \hat{e}_{\mu})$ in eq.\ (\ref{eq:translU}) 
is simply that of the identity.

In order to prove eq.\ (\ref{eq:bloch-g}), a key point is
that the minimizing problem for the extended lattice,
defined by the functional in eq.\ (\ref{eq:minimizingLz}), is invariant 
if we consider a shift of the lattice sites $\vec{z}$ by $N$ in any
direction $\mu$, since this amounts to a simple redefinition
of the origin for the extended lattice $\Lambda_z$. 
This implies that, if $g(\vecz)$ is a solution of the minimizing
problem satisfying the BCs (\ref{eq:gPBCs}), then $g^{\mk\prime}(
\vecz) = g(\vec{z} + N {\hat e}_{\mu})$ is (trivially) a solution
too, satisfying the same BCs.
Moreover, these two solutions select the same local minimum within the
first Gribov region.
At the same time, as already stressed above, due to cyclicity of the
trace, ${\cal E}_{\scriptU}[g]$ is invariant under global
gauge transformations $v$ and
the same is true for the quantities introduced in
eqs.\ (\ref{eq:e1})--(\ref{eq:e6}),
when applied to the extended lattice $\Lambda_z$.
Note that this corresponds to left multiplication\footnote{In this sense, 
right multiplication by $v$ does {\em not} produce an equivalent solution, 
since $\{ g(\vecz)\}$ is not necessarily a solution to the gauge-fixing 
problem defined by applying a global gauge transformation $v$ to the original
link configuration, i.e., in general $\{ g(\vecz)\}$ does not minimize
the functional ${\cal E}$ when the link configuration is
$v\,U_{\mu}(\vecz) \,v^{\dagger}$.} of the solution to
the gauge-fixing problem by a fixed group element, 
mapping $\{ g(\vecz)\}$ onto $\{ v\, g(\vecz)\}$.
Thus, the gauge transformation $\{ g(\vecz)\}$ --- i.e.\ a given minimum
solution --- is always determined modulo a global (left) transformation, and
(with our setup) remains a solution under translations by $N$ in any direction.

The above observation needs some comments.
In particular, we recall that, in ref.\ \cite{Zwanziger:1993dh}, the
proof of eq.\ (\ref{eq:bloch-g}) is presented only for the absolute
minima (of the minimizing functional) that belong to the interior
of the so-called fundamental modular region.
Indeed, as shown in the appendix A of the same reference, these
minima are unique, i.e.\ non degenerate, implying that the gauge
transformation $\{ g(\vecz) \}$ connecting the (unfixed)
thermalized configuration $\{ U_{\mu}(\vecz) \}$ to the
(gauge-fixed) absolute minimum $\{U_{\mu}(g;\vecz) \}$ is unique,
modulo a global gauge transformation.
However, as stressed at the end of the {\em Bloch waves} section
of ref.\ \cite{Cucchieri:2016qyc}, even in the case of local
minima one can make the (reasonable) hypothesis that a specific
realization of one of these minima
also corresponds to a specific and unique transformation $\{ g(\vecz)
\}$ (up to a global transformation) when considering a given
configuration $\{ U_{ \mu}(\vecz) \}$.
Indeed, this has been verified numerically (see, e.g., ref.\
\cite{Marinari:1991zv}) for small lattice volumes and for the local
minima of the minimizing functional (\ref{eq:minimizing}).
We thus assume, as in ref.\ \cite{Cucchieri:2016qyc}, that
local minima of ${\cal E}_{\scriptU}[g]$ also define unique gauge 
transformations.
In other words, here
we are considering truly degenerate local minima, i.e.\ connected
by a nontrivial gauge transformation, as different minima.
Also, we assume that --- at least for numerical
simulations on finite lattice volumes --- these
degenerate minima will not have identical values of the quantities
characterizing the minimum solution, such as ${\cal E},
\Delta {\cal E}, (\nabla A)^2$ and
$\Sigma_Q$, described\footnote{Clearly, these
quantities are unaffected by a shift of the origin.
Also, as discussed above, they are invariant under global gauge
transformations.
On the other hand, we are not considering here the
possibility that nontrivially different solutions might have all identical 
numerical values for these quantities, when
performing a numerical simulation. \label{foot:nondeg}}
in section \ref{sec:mlg-PBCs} (see also
section \ref{sec:num-min-convergence} below).
As a matter of fact, at the numerical level, the only degeneracy
that can likely occur is the trivial one, i.e.\ when the
corresponding link configurations are related by a global gauge
transformation.

Based on the above discussion, we proceed to prove eq.\
(\ref{eq:bloch-g}) by writing
\begin{equation}
{\cal T}(N \hat{e}_{\mu}) \, g(\vecz) \, = \,
\left[\ {\cal T}(\hat{e}_{\mu})\ \right]^N \, g(\vecz) \, = \,
 g(\vec{z} + N \hat{e}_{\mu}) \, = \, g^{\mk\prime}(
 \vecz) \,=\, s_{\mu} \, g(\vecz) \; ,
\label{eq:vmu}
\end{equation}
where $s_{\mu}$ is a $\vec{z}$-independent SU($N_c$) matrix.
This is the main hypothesis considered in refs.\
\cite{Zwanziger:1993dh,Cucchieri:2016qyc} and it is supported by 
our arguments above, i.e.\ that a shift of $\{ g(\vecz)\}$ 
by $N$ along a given direction $\mu$ produces an equivalent solution, and 
can therefore be parametrized as left multiplication by a fixed element 
$s_\mu$ of the group.
Then, due to eq.\ (\ref{eq:TT}), we have that
the $s_{\mu}$'s are commuting SU($N_c$) matrices, i.e.\
they can be written as $\exp{( i \Theta_{\mu} )}$,
with $\Theta_{\mu}$ given in eq.\ (\ref{eq:tautheta}).
Also, due to the PBCs for $\Lambda_z$, we need to
impose the condition (\ref{eq:TmN1}).
Hence, the relations
\begin{equation}
{\cal T}(m N \hat{e}_{\mu}) \, g(\vecz) \, = \,
\left[ \, {\cal T}(N \hat{e}_{\mu}) \, \right]^m
  \, g(\vecz) \, = \, s_{\mu}^m \, g(\vecz)
\label{eq:vmuNm}
\end{equation}
and
\begin{equation}
{\cal T}(m N \hat{e}_{\mu}) \, g(\vecz) \, = \,
g(\vec{z} + m N \hat{e}_{\mu}) \, = \, g(\vecz)
\label{eq:vmuNm2}
\end{equation}
yield
\begin{equation}
s_{\mu}^m \, = \, \1 \; .
\label{eq:labdam1}
\end{equation}
We stress that the action of the translation operator
${\cal T}(N \hat{e}_{\mu})$ in eq.\ (\ref{eq:vmu}),
i.e.\ the matrix $s_{\mu}=\exp{( i \Theta_{\mu} )}$, depends on the solution $\{ g(\vecz)\}$
to which it is applied, i.e.\ the parametrization of the matrices
$\Theta_{\mu}$ is determined by the considered solution of the gauge-fixing
problem, see also the comment below eq.\ (\ref{eq:gNmu}).

The above eq.\ (\ref{eq:vmu}) is the matrix analogue of the eigenvalue
equation (\ref{eq:Teigenvalue}) [for $\vec{R} = N \hat{e}_{\mu}$
and $l \to N$, so that $L \to m N$].
Indeed, instead of the wavefunction $\psi_{\veck}(\vecr)$,
eq.\ (\ref{eq:vmu}) applies to a solution $\{ g(\vecz) \}$
of the minimizing problem ${\cal E}_{\scriptU}[g]$, corresponding to
a specific local minimum.
Also, on the r.h.s.\ of the equation, the 
matrix $s_{\mu}$ appears\footnote{Based on this analogy, it is natural 
that the matrices $s_{\mu}$ be characteristic of the considered
solution $\{ g(\vecz) \}$.}
instead of the phase $\exp{( 2 \pi i\, k_{\mu} / m)}$, i.e.\ the 
corresponding eigenvalue in eq.\ (\ref{eq:Teigenvalue}).
Moreover, the action of the
translation operators ${\cal T}(N \hat{e}_{\mu})$ in eq.\
(\ref{eq:vmu}) can likewise be expressed in terms of phase factors,
if we write the gauge transformation $g(\vecz)$ as
\begin{equation}
g(\vecz)\,=\,
\sum_{i,j=1}^{N_c}\, g^{{\textstyle\mathstrut}ij}(\vecz)\;
{\bf W}^{ij} \; ,
\end{equation}
where the matrices ${\bf W}^{ij} = w_i w_j^{\dagger}$ are
defined in section \ref{sec:lambda-basis} of the appendix and 
$g^{{\textstyle\mathstrut}ij}(\vecz)$ denotes the coefficient 
of ${\bf W}^{ij}$ in the expansion of $g(\vecz)$.
Then, we immediately find\footnote{Here, we used
the definition $g^{{\textstyle\mathstrut}ij}(\vecz) =
w_i^{\dagger}\,g(\vecz)\, w_j$, as in eq.\ (\ref{eq:gencoeff}), and the
property (\ref{eq:Theta-wi}) of the matrices $\Theta_\mu$.
See also eqs.\ (\ref{eq:thetawj}) and (\ref{eq:thetawidagger})
below.
\label{foot:ghj}}
\begin{equation}
 {\cal T}(N \hat{e}_{\mu}) \, g(\vecz) \, = \,
       s_{\mu} \, g(\vecz)
  \,=\, \exp{(i \Theta_{\mu})}\, g(\vecz) \,=\,
   \sum_{i,j=1}^{N_c}\, e^{2\pi i n_{\mu}^{i}/m}\,
    g^{{\textstyle\mathstrut}ij}(\vecz)\; {\bf W}^{ij} \;, 
   \label{eq:TNg-components}
\end{equation}
with integer $n_{\mu}^{i}$,
so that each coefficient $g^{{\textstyle\mathstrut}ij}(\vecz)\,{\bf W}^{ij}$
gains a phase factor $\,\exp{(2\pi i \,n_{\mu}^{i}/m)}$.
These factors are the usual eigenvalues $\tau_{\mu}$ of the
translation operator ${\cal T}(N \hat{e}_{\mu})$ that satisfy the
relation $(\tau_{\mu})^{m} = 1$, implying that they can be written
as $\tau_{\mu}=\exp{\left( 2 \pi i \, k_{\mu}^{\mk \prime}/m \right)}$
with $k_{\mu}^{\mk \prime} \in {\cal Z}$.
In particular, in the first Brillouin zone, we have $\tau_{\mu}=
\exp{\left( 2 \pi i \,k_{\mu}/m \right)}$ with $k_{\mu} =k_{\mu}^{
\mk\prime}\pmod{m}$.

The above result
\begin{equation}
g(\vecz+N\hat{e}_{\mu})\,=\,\exp(i\Theta_{\mu})\,g(\vecz)
\label{eq:gNmu}
\end{equation}
is already equivalent to one of the usual formulations of the
Bloch theorem (see, e.g., eq.\ (8.6) in ref.\ \cite{AM}).
Indeed, by paraphrasing the statement in ref.\ \cite{Ziman}, we can say 
that 
\begin{quote}
\em
For any solution
$g(\vecz)$ of the minimizing problem ${\cal E}_{\scriptU}[g]$
there exists a set of commuting matrices $\Theta_{\mu}$ such
that the translation by a vector $N\hat{e}_{\mu}$ is equivalent
to multiplying the solution by the factor $\exp(i\Theta_{\mu})$.
\end{quote}

This provides a way to construct the solution $g(\vecz)$ --- at
a point $\vecz$ of the extended lattice $\Lambda_z$ --- as
the successive application of $\exp(i\Theta_{\mu})$ to 
$g(\vecx)$, which is the same solution but restricted to the primitive cell
${\Lambda_{x}}$.
Hence, by taking into account the displacement, from point $\vecx$,
along each direction $\mu$ --- given by the indices $y_\mu$ ---
we can write
\begin{equation}
g(\vecz) \, = \, g(\vec{x} + N\vecy) \, = \,
 \exp{\left( i \sum_{\nu = 1}^d \,
      \Theta_{\nu} \, y_{\nu} \right)} \, g(\vecx)\; .
\label{eq:TRg}
\end{equation}
We stress that the above expression tells us that the extended-lattice 
solution $g(\vecz)$ is
obtained by successive ``block-rotations'' of the primitive-cell portion of
the solution $g(\vecx)$:
each time we move to a neighboring cell along the direction
$\mu$, the solution picks up a factor $\exp(i\Theta_{\mu})$.
As a consequence, by substituting eq.\ (\ref{eq:TRg}) into the
expressions (\ref{eq:minimizingLz})--(\ref{eq:Uofg}) and in
analogy with the discussion presented
in Section \ref{sec:Bloch-theorem} above, the minimization problem is 
broken down (due to cyclicity of the trace) into $m^d$ copies of
the minimization problem\footnote{At the same time, the gauge-fixed
link configuration $\{ U_{\mu}(g;\vecz) \}$ can also be visualized
as made up of $m^d$ domains, related by block-rotations (see
discussion in sections \ref{sec:minimizing-revisited} and
\ref{sec:numerical-PBCs} below). \label{foot:domains}}
\begin{equation}
  \frac{\Re \, \Tr}{N_c\,d\,V}
 \,\sum_{\mu = 1}^d \, \sum_{\vec{x}\in\Lambda_x} \, \left[ \,
 \1 \, - \, g(\vecx)
\, U_{\mu}(\vecx) \, g(\vec{x} + \hat{e}_{\mu})^{\dagger}
 \, \right] \; .
\end{equation}
For each of these copies, it corresponds to the minimization of the
original functional for the lattice $\Lambda_x$, i.e.\ the expression in
eq.\ (\ref{eq:minimizing}) [with $g(\vecx)$ instead of $h(\vecx)$], but 
with the 
boundary condition, see eq.\ (\ref{eq:gNmu}) with $\vecz=\vecx$,
\begin{equation}
g(\vecx+N\hat{e}_{\mu})\,=\,\exp(i\Theta_{\mu})\,g(\vecx)\,.
\label{eq:gperiod}
\end{equation}
Thus, the function $g(\vecx)$ is {\em not} a solution to the usual
gauge-fixing problem restricted to the primitive cell $\Lambda_x$ --- which 
would correspond to a periodic function under translations by $N$
in all directions --- but is closely related to it by the above rotations.

We now note that the BCs (\ref{eq:gperiod}) may be incorporated automatically if we write,
in analogy with the usual proof of the Bloch theorem \cite{AM,Ziman},
\begin{equation}
g(\vecx) \, = \, \exp{\left( i \sum_{\nu = 1}^d \,
      \frac{\Theta_{\nu} \, x_{\nu}}{N} \right)} \, h(\vecx)\; ,
\label{eq:gofh}
\end{equation}
where $h(\vecx)$ is a solution to the gauge-fixing problem 
restricted to $\Lambda_x$, redefined\footnote{This is discussed in
detail in the next section.} in terms of a modified gauge-transformed
link configuration
$\{ U_{ \mu}(h;\vecx) \exp(-i\Theta_\mu/N) \}$,
see eq.\ (\ref{eq:minfunctUhThetax}).
In this way, the condition (\ref{eq:gperiod}) is clearly satisfied.
Moreover, it is straightforward to verify that the function $h(\vecx)$
is periodic on $\Lambda_x$.
Indeed, by inverting (\ref{eq:gofh}), i.e., by writing
\begin{equation}
h(\vecx) \, = \, \exp{\left( - i
  \sum_{\nu = 1}^d \, \frac{\Theta_{\nu} \, x_{\nu}}{N}
              \right)} \, g(\vecx)\,,
\label{eq:defhx}
\end{equation}
we have that
\begin{eqnarray}
h(\vecx + N \hat{e}_{\mu}) \, & = & \,
        \exp{\left( - i \sum_{\nu = 1}^d \,
            \frac{\Theta_{\nu} \, x_{\nu}}{N}
              \right)} \exp{( - i \Theta_{\mu} )} \,
     g(\vecx + N \hat{e}_{\mu}) \nonumber \\[2mm]
     & = &\, 
        \exp{\left( - i \sum_{\nu = 1}^d \,
            \frac{\Theta_{\nu} \, x_{\nu}}{N}
              \right)} \,
     g(\vecx) \,=\, h(\vecx) \; ,
\end{eqnarray}
where we used (\ref{eq:gperiod}) and the fact that the matrices
$\Theta_{\nu}$ commute with each other.
Therefore, the above eq.\ (\ref{eq:gofh}) provides the desired solution
to the modified minimization problem on $\Lambda_x$, written
in terms of the periodic function $h(\vecx)$, up to
choice of parameters for the $\Theta_\mu$ matrices,
which are also fixed by the minimization problem.\footnote{But
this is precisely what enlarges the
set of solutions and allows a more efficient way to deal with the
extended-lattice problem!
As said at the end of the previous section,
an approach closer to the one usually employed in condensed
matter theory would require to consider a given (fixed) set of
matrices $\Theta_{\mu}$ and use the minimization procedure
only to determine $h(\vecx)$.\label{foot:but}}

This completes our proof of eq.\ (\ref{eq:bloch-g}), which may also be
written as
\begin{equation}
g(\vecz) \, = \, \exp{\left( i
  \sum_{\nu = 1}^d \, \frac{\Theta_{\nu} \, z_{\nu}}{N}
              \right)} \, h(\vecz) \; ,
\label{eq:gofhz}
\end{equation}
where the function $h(\vecz)$ is defined on the extended lattice but
has periodicity under translations by $N$ in all directions, i.e.\ it
is a ``clone'' of the primitive-cell solution $h(\vecx)$ above.
Hence, as done for the original Bloch theorem, we can write the solution
$g(\vecz)$
as a product of a ``plane wave'' by a (periodic) solution of a modified 
version of the primitive-cell problem.


\section{The minimizing problem revisited}
\label{sec:minimizing-revisited}

Using the analogue of Bloch's theorem, i.e.\ eq.\ (\ref{eq:bloch-g}),
the gauge-transformed link variable (\ref{eq:Uofg}) is given by
\begin{equation}
U_{\mu}(g;\vecz) \, = \, 
\exp{\left( i \sum_{\nu = 1}^d \,
                  \frac{\Theta_{\nu} \, z_{\nu}}{N}
             \right)} \, U_{\mu}(h;\vecx) \,
\exp{\left( - i \, \frac{\Theta_{\mu}}{N} \right)} \,
\exp{\left( - i \sum_{\nu = 1}^d \,
                  \frac{\Theta_{\nu} \, z_{\nu}}{N} 
              \right)} \; ,
\label{eq:UPBCs1}
\end{equation}
with $h(\vecx)$ discussed in the previous two sections and recalling the
general expression for a gauge-transformed link, eq.\ (\ref{eq:Uofh}).
Since $h(\vecx)$ satisfies PBCs with respect to the original
lattice $\Lambda_x$, it is clear that $\{ U_{\mu}(h; \vecx) \}$
is also a periodic, gauge-transformed link configuration on $\Lambda_x$.
Thus, the effect of the index lattice is completely encoded in
the exponential factors and in the matrices $\Theta_{\mu}$.
Let us stress that, even though we use the same notation\footnote{See
also the comment in the last paragraph of section \ref{sec:Bloch-theorem}.}
considered in Sec.\ \ref{sec:mlg-PBCs}, in the present case
$\{ U_{\mu}(h;\vecx) \}$ is not transverse on $\Lambda_x$.
Indeed, transversality\footnote{Here we mean the property
(\ref{eq:diverA})--(\ref{eq:diverg0}), i.e.\ the fact that
the Landau-gauge condition --- applied to the lattice gauge fields 
defined by the gauge-link configuration and now written for the 
(gauge-fixed) links on $\Lambda_z$ --- is satisfied.
One of the goals of this section is to understand what this
implies for the gauge field when restricted to the original
lattice $\Lambda_x$. \label{transv}}
applies to $\{ U_{\mu}(g;\vecz) \}$,
taken for the extended lattice $\Lambda_z$.
By considering the relation (\ref{eq:zcoord}) we can, however,
rewrite the above result in a different way, i.e.\
\begin{equation}
U_{\mu}(g;\vecz) \, = \,
U_{\mu}(g;\vec{x},\vecy) \, = \,
\exp{\left( i \sum_{\nu = 1}^d \, \Theta_{\nu} \, y_{\nu}
              \right)} \, U_{\mu}(l;\vecx) \,
\exp{\left( - i \sum_{\nu = 1}^d \, \Theta_{\nu} \, y_{\nu}
              \right)} \; ,
\label{eq:Uofg2}
\end{equation}
where the $\vec{y}$ coordinates characterize the replicated lattice
$\Lambda_x^{(\vecy)}$ and we have defined a ``local'' version of the
transformed gauge link
\begin{eqnarray}
U_{\mu}(l;\vecx) \,& = &\, l(\vecx) \, U_{\mu}(\vecx) \,
l(\vec{x}+{\hat e}_{\mu})^{\dagger} \nonumber \\[2mm]
& \equiv &\, \exp\!{\left( i \sum_{\nu = 1}^d \,
                  \frac{\Theta_{\nu} \, x_{\nu}}{N}
             \right)} \Biggl[\, U_{\mu}(h;\vecx) \,
\exp\!{\left( - i \, \frac{\Theta_{\mu}}{N} \right)} \,
\Biggr]\,\exp\!{\left( - i \sum_{\nu = 1}^d \,
                  \frac{\Theta_{\nu} \, x_{\nu}}{N}
             \right)} \;\;\;\;\;
\label{eq:Uofl}
\end{eqnarray}
where the gauge transformation restricted to $\Lambda_x$, 
see eq.\ (\ref{eq:bloch-g}), is given as
\begin{eqnarray}
\label{eq:gofx0}
      l(\vecx) \,&=&\, \exp{\left( i
        \sum_{\nu = 1}^d \, \frac{\Theta_{\nu} \, x_{\nu}}{N}
                    \right)} \, h(\vecx) \; .
\end{eqnarray}
Similarly, we can write\footnote{Note that in eqs.\ (\ref{eq:Uofl})
and (\ref{eq:Uofl-}) the external factors, i.e.\ $\exp{\left( \pm i
\sum_{\nu = 1}^d \, \Theta_{\nu} \, x_{\nu}/ N \right)}$, are the
same.
The implied expressions for $U_{\mu}(h;\vecx)$ and $U_{\mu}(h; \vec{x}
- \hat{e}_{\mu})$ are clearly compatible with each other and in principle
there is no need to define them separately.
This is done for later
convenience, since these expressions are used
for the (gauge-fixed) gauge field entering the transversality condition.
See, in section \ref{sec:transversality-Lambdaz}, eqs.\
(\ref{eq:Adivlb}), (\ref{eq:Adivlnobmod}) or (\ref{eq:Adivlnob}),
and footnote \ref{foot:Amug}. \label{foot:Ulgeneral}}
\begin{equation}
U_{\mu}(l;\vec{x} - \hat{e}_{\mu}) \, \equiv \,
\exp\!{\left( i \sum_{\nu = 1}^d \,
  \frac{\Theta_{\nu} \, x_{\nu}}{N} \right)}
\left[\,
\exp\!{\left( - i \, \frac{\Theta_{\mu}}{N} \right)} \,
  U_{\mu}(h;\vec{x}-\hat{e}_{\mu}) \, \right]\,
\exp\!{\left( - i \sum_{\nu = 1}^d \,
                  \frac{\Theta_{\nu} \, x_{\nu}}{N}
             \right)} \; .
\label{eq:Uofl-}
\end{equation}

Let us point out that the quantity $l(\vecx)$ is actually a redefinition of
$g(\vecx)$ in (\ref{eq:gofh}), which is however never extended to $\Lambda_z$.
This is done to single out the $\Lambda_x$
portion of the solution $g(\vecz)$ and will be important from now on
in our analysis.
In particular, we make use of the fact that both
$l(\vecx)$ and $h(\vecx)$ ``exist'' only on $\Lambda_x$, and are therefore
simply replicated identically to other cells ${\Lambda_{x}}^{\!\!(\vecy)}$.
We stress, however, that the properties of these two small-lattice
gauge transformations differ: indeed, while $l(\vecx)$ is the nonperiodic
solution of the minimization problem defined by the original functional 
${\cal E}_{\scriptU}[l]$ on $\Lambda_x$, see eq.\ (\ref{eq:EUg-Ul}) below,
$h(\vecx)$ is the periodic 
solution of the modified minimization problem (\ref{eq:minfunctUhThetax}),
which depends on the $\Theta_{\mu}$'s.
Thus, $\{ U_{\mu}(l;\vecx) \}$ is transverse on 
$\Lambda_x$ and $\{ U_{\mu}(h;\vecx) \}$ is not, as already mentioned above.

The definition of $l(\vecx)$ implies that, see eq.\ (\ref{eq:gperiod}),
\begin{eqnarray}
   l(\vec{x}+N \hat{e}_{\mu}) \, & = &\,
          \exp{( i \Theta_{\mu} )} \,
     \exp{\left( i \sum_{\nu = 1}^d \,
           \frac{\Theta_{\nu} \, x_{\nu}}{N}
          \right)} \,  h(\vec{x}+N \hat{e}_{\mu})
          \nonumber \\[2mm]
          & = & \, \exp{( i \Theta_{\mu} )} \,
     \exp{\left( i \sum_{\nu = 1}^d \,
   \frac{\Theta_{\nu} \, x_{\nu}}{N} \right)} \,
   h(\vecx) \,=\, \exp{( i \Theta_{\mu} )} \, l(\vecx) \; ,
\label{eq:lofx-nonperiodic}
\end{eqnarray}
yielding
\begin{eqnarray}
U_{\mu}(l;\vec{x} + N \hat{e}_{\nu}) \,&=&\,
 l(\vec{x} + N \hat{e}_{\nu}) \, U_{\mu}(\vec{x} + N \hat{e}_{\nu}) \,
l(\vec{x}+N \hat{e}_{\nu}+{\hat e}_{\mu})^{\dagger} \nonumber \\[2mm]
   &=&\,\exp{( i \Theta_{\nu} )} \, l(\vecx) \,
              U_{\mu}(\vecx) \, l(\vec{x} + {\hat e}_{\mu})^{\dagger}
   \exp{( - i \Theta_{\nu} )} \nonumber \\[2mm]
   &=&\, \exp{( i \Theta_{\nu} )} \, U_{\mu}(l;\vecx) \, 
   \exp{( - i \Theta_{\nu} )} \; ,
\label{eq:UlBCs}
\end{eqnarray}
which is reminiscent of the so-called twisted BCs
\cite{Gonzalez-Arroyo:1997ugn} with constant transition 
matrices\footnote{However, in that case, one needs to satisfy the relation
  $ \Omega_{\mu} \, \Omega_{\nu}\,=\, z_{\mu \nu}\,
   \Omega_{\nu} \, \Omega_{\mu}$,
where the constants $z_{\mu \nu}$ are elements of the center of
the group.
Then, since the $\Theta_{\nu}$ are commuting matrices, we have
(in our case) the trivial condition $z_{\mu \nu} = 1$ for any
$\mu$ and $\nu$, i.e.\ no twist.}
$\Omega_{\nu}=\exp{( i \Theta_{\nu} )}$.
One should also note that, if we expand the link variable
$U_{\mu}(l;\vecx)$ in terms of the ${\bf W}^{ij}$ matrices,
as done in the previous section for the $g(\vecz)$ matrices,
we can rewrite eq.\ (\ref{eq:UlBCs}) as\footnote{The proof 
follows the same steps explained in footnote \ref{foot:ghj}.}
\begin{equation}
\!\!\! U_{\mu}(l;\vec{x} + N \hat{e}_{\nu}) =
\sum_{i,j=1}^{N_c}\, U^{ij}_{\mu}(l;\vec{x} + N \hat{e}_{\nu})\; {\bf W}^{ij}
= \sum_{i,j=1}^{N_c}
   e^{2\pi i (n_{\nu}^{i}-n_{\nu}^{j})/m}\,
   \,U^{ij}_{\mu}(l;\vecx)\;{\bf W}^{ij} \;,
\end{equation}
where $n_{\nu}^{i}$, $n_{\nu}^{j}$ are integers.
Hence, the coefficients of $U_{\mu}(l;\vecx)$ satisfy toroidal
BCs (see appendix A.3 in ref.\ \cite{Gattringer:2010zz})
\begin{equation}
U^{ij}_{\mu}(l;\vec{x} + N \hat{e}_{\nu}) \,=\,
   e^{2\pi i (n_{\nu}^{i}-n_{\nu}^{j})/m}\,
   U^{ij}_{\mu}(l;\vecx) \; ,
\label{eq:Ul-toroidal}
\end{equation}
which, depending on the values of $n_{\nu}^{i}$ and
$n_{\nu}^{j}$, include periodic as well as anti-periodic BCs, given 
respectively by $e^{2\pi i (n_{\nu}^{i} -n_{\nu}^{j})/m}=1$ and
$e^{2\pi i (n_{\nu}^{i} -n_{\nu}^{j})/m}=-1$.

The above eq.\ (\ref{eq:Uofg2}) implies that gauge-fixed
configurations in different replicated lattices $\Lambda_x^{(\vecy)}$
differ only by the exponential factors $\exp{(\pm i \sum_{\nu =
1}^d \, \Theta_{\nu} \, y_{\nu} )}$, which correspond to a global
gauge transformation within each $\Lambda_x^{(\vecy)}$.
Moreover, we have that $\{ U_{\mu}(l;\vecx) \}$ is transverse on
each replicated lattice $\Lambda_x^{(\vecy)}$.
Indeed, by noting that
\begin{equation}
\Tr \, \Bigl[ \, U_{\mu}(l;\vecx) \, \Bigr] \, = \,
\Tr \, \Biggl[ \, U_{\mu}(h;\vecx) \,
\exp{\left( - i \,\frac{\Theta_{\mu}}{N}\right)} \, \Biggr] \; ,
\label{eq:TrUl}
\end{equation}
we can rewrite eq.\
(\ref{eq:minfunctUhThetax}) as
\begin{equation}
{\cal E}_{\scriptU}[g] \, = \,
{\cal E}_{\scriptU,\scriptT}[h] \, = \,
{\cal E}_{\scriptU}[l] \,\equiv\,
\frac{\Re \, \Tr}{N_c\,d\,V} \,
 \sum_{\mu = 1}^d \, \sum_{\vec{x}\in\Lambda_x} \,
    \Bigl[ \, \1 \, - \, U_{\mu}(l;\vecx)
    \, \Bigr]
    \label{eq:EUg-Ul}
\end{equation}
and, therefore, $\{ U_{\mu}(l;\vecx) \}$ is transverse\footnote{We
will address the transversality condition in detail 
in the next section. See also footnote \ref{transv}.}
when the functional ${\cal E}_{\scriptU}[l]$ is minimized.

We can summarize these results by saying that, with the
consideration of the extended lattice $\Lambda_z$, we trade
the periodic transverse link configuration $\{ U_{\mu}(h;\vecx)\}$ on 
the original lattice $\Lambda_x$ --- in the small-lattice problem --- with 
the nonperiodic, but still transverse, link configuration 
$\{ U_{\mu}(l;\vecx)\}$, also defined on $\Lambda_x$.\footnote{Or, 
equivalently, with the periodic and not transverse configuration
$\{ U_{\mu}(h;\vecx)\}$ obtained from the modified minimization problem
(\ref{eq:minfunctUhThetax}), determined by the $\Theta_\mu$'s.}
Moreover, this transverse link configuration is replicated on each
$\Lambda_x^{(\vecy)}$, indexed by the $y_{\mu}$ coordinates,
and then globally rotated using the gauge transformation 
$\,\exp{( i \sum_{\nu = 1}^d \, \Theta_{\nu} \, y_{\nu} )}$, 
see eq.\ (\ref{eq:Uofg2}), in such
a way that PBCs are satisfied on $\Lambda_z$.
One could visualize this lattice setup by making an analogy with
some of the works by M.C. Escher, such as those called {\em Metamorphosis
I, II} and {\em III} (see, for example, \cite{Escher}), in which one
starts from a simple geometrical form, e.g.\ a square, and replicates
it several times on a plane, by adding a small rotation
(and a distortion) at each step.
As already stressed in footnote \ref{foot:domains}, the description
of the gauge-fixed configuration --- in terms of $\{ U_{\mu}(l;\vecx)\}$
and of global rotations $\exp{( i \sum_{\nu = 1}^d \, \Theta_{\nu} \,
y_{\nu} )}$ --- naturally singles out domains, which can be
characterized (for example) in terms of color magnetization, as
done in section \ref{sec:numerical-PBCs}.

The above observations have important consequences also for the type of
Gribov copies that one can obtain when using the extended lattice
$\Lambda_z$.
Indeed, they are essentially given by the Gribov copies that can
be found on the original lattice $\Lambda_x$ where, however,
the transverse link configuration $\{ U_{\mu}(l;\vecx) \}$ is now
nonperiodic.
As a consequence,
the set of local minima generated by the usual small-lattice gauge-fixing 
procedure, i.e.\ by the gauge transformation $\{ h(\vecx) \}$ as in
section \ref{sec:mlg-PBCs}, are (in principle) not related to the local 
minima generated by the new gauge-fixing approach, i.e.\ by the gauge 
transformation $\{ l(\vecx) \}$.
In fact, one should recall that $\{h(\vecx)\}$ in the extended problem
is also (implicitly) determined by the $\Theta_{\nu}$ matrices, and 
vice versa, through the minimization process.
Moreover, due to the extra freedom allowed by the Bloch waves (see
footnote \ref{foot:but}), we expect
\begin{equation}
{\cal E}_{\scriptU}[l] \, = \,
{\cal E}_{\scriptU,\scriptT}[h] \,\leq\,
{\cal E}_{\scriptU}[h] \;
\end{equation}
for a fixed (thermalized) gauge-link configuration $\{ U_{\mu}(\vecx) \}$.
At the same time, not much can be said about a comparison of different
Gribov copies due to the $\{ l(\vecx) \}$ gauge transformation
and those obtained by gauge fixing a configuration that is directly
thermalized on the extended lattice $\Lambda_z$, i.e.\
which has (at any step) an invariance under translation by
$m N \hat{e}_{\mu}$ only.


\subsection{The transversality condition}
\label{sec:transversality-Lambdaz}

We turn now to the constraints imposed by the minimization of the
functional ${\cal E}_{\scriptU}[l]$.
Our goal is to obtain expressions for observables constructed from the 
transformed gauge links $U_\mu(l;\vecx)$, both to characterize the
transversality condition, i.e.\ to obtain the gauge-fixing criteria
from the minimizing functional ${\cal E}_{\scriptU}[l]$, and to define
the quantities that will be measured in our simulations.
However, since we want to explore the similarities between the minimization
problem on the extended lattice and the original problem on the small 
lattice $\Lambda_x$ (as addressed in section \ref{sec:mlg-PBCs}), we 
also express our results in terms of the periodic transformation
$\{ h(\vecx) \}$, stressing that it now refers to the modified 
minimization condition depending on the matrices $\Theta_\mu$.
To this end, we note that these matrices
(detailed in appendix \ref{sec:Cartan}) are conveniently parametrized in 
terms of an $SU(N_c)$ matrix $v$ and a set of integers $\{n_{\mu}^{j}\}$ characterizing the
plane waves.

We first recall that, see eqs.\
(\ref{eq:minimizing2})--(\ref{eq:Z}),
\begin{eqnarray}
\!\!\!\!\!\!\!\!\!\!\!\!
{\cal E}_{\scriptU}[g] \,  =  \,
{\cal E}_{\scriptU}[l] \, & = & \,
{\cal E}_{\scriptU,\scriptT}[h] \nonumber \\[2mm]
   &=& \, \frac{\Re \, \Tr}{N_c\,d} \,
 \sum_{\mu = 1}^d \, \left\{ \, \1 \, - \,
    \left[ \, \sum_{\vec{x}\in\Lambda_x}
        h(\vecx) \, U_{\mu}(\vecx)\,
        h^{\dagger}(\vec{x} + \hat{e}_{\mu}) \,
     \right] \, \frac{e^{-i \frac{\Theta_{\mu}}{N}}}{V}
    \, \right\}
    \label{eq:minimizing-Theta2}
\end{eqnarray}
and that, when the matrices $\Theta_{\mu}$ are written in the basis
$\{ {\bf W}^{ij} = w_i\,w_j^{\dagger} = v^{\dagger} \, {\bf M}^{ij}\, v\}$ introduced 
in section \ref{sec:lambda-basis}, we have, see eq.\ (\ref{eq:Thetadiag-exp}),
\begin{equation}
  e^{-i \frac{\Theta_{\mu}}{N}} \,=\, v^{\dagger}\,
             T_\mu(m N;\{n_{\mu}^{j}\}) \, v \; ,
\label{eq:vDjj}
\end{equation}
where the diagonal matrix $T_\mu(m N;\{n_{\mu}^{j}\})$ has elements
\begin{equation}
{T_\mu}^{jj}\,\equiv\,
\exp{\left(\!- 2\pi i\; \frac{n_{\mu}^{j}}{m\,N}\!\right)} \, .
\label{eq:Djj}
\end{equation}
Then, from the above equations it is evident that, when analyzing
the consequences of the gauge-fixing condition, we have to treat
differently the gauge transformations $h(\vecx)$ and $v$, which
depend on real parameters,\footnote{The matrix elements of
$h(\vecx)$ and $v$ are complex when considering the SU($N_c)$
gauge group.
Here, we will consider separately the real and imaginary parts of
$h_{ij}(\vecx)$ and $v_{ij}$.}
and the transformation $T(m N;\{n_{\mu}^{j}\})$,
which is defined in terms of the integer parameters $n_{\mu}^{j}$.
In particular, the minimizing functional (\ref{eq:minimizing-Theta2})
is quadratic with respect to the matrix elements $h_{ij}(\vecx)$ 
(see also appendix C.3 in ref.\ \cite{Grabenstein:1994jk})
and $v_{ij}$, and has to satisfy the (also quadratic) constraints
$h(\vecx)\, h^{\dagger}(\vecx) = v\,v^{\dagger} = \1$.
At the same time, ${\cal E}_{\scriptU,\scriptT}[h]$ depends
nonlinearly on the (integer) parameters $n_{\mu}^{j}$, which
are subject to the linear constraint (\ref{eq:tracewith-n}).
Thus, the minimizing problem we are interested in is a mixed-integer
nonlinear optimization problem, which can be formulated as
\cite{Floudas}
\begin{equation}
   \min_{x,\,n} \, f(x,n)
\label{eq:fmin}
\end{equation}
with
\begin{equation}
    f: \left[\,{\cal R}^{d_r} \times {\cal Z}^{d_i} \right] , \;\;\;
    x\,\in\,\Omega_r \,\subset\, {\cal R}^{d_r}, \;\;
    \mbox{and} \;\;
    n\,\in\,\Omega_i \,\subset\, {\cal Z}^{d_i} \; ,
    \label{eq:constraints}
\end{equation}
where the subsets $\Omega_r$ and $\Omega_i$ (respectively of dimensions
$d_r$ and $d_i$) are determined by the
constraints imposed on the real variables $x$ and on the integer
variables $n$.
It is important to stress that, in these cases, the determination
of the global minimum is, in general, an NP-hard problem.
 
In order to obtain an explicit expression for the stationarity
condition imposed by the minimization of ${\cal E}_{\scriptU,
\scriptT}[h]$, let us first examine the case
in which the matrices $\Theta_{\mu}$ are fixed.
For this, we can repeat the analysis carried out in section \ref{sec:mlg-PBCs}
and consider the gauge transformation
$ h(\vecx) \to R(\tau; \vecx) \, h(\vecx)$ with the
one-parameter subgroup (\ref{eq:hofgamma}).
Hence,
we obtain,
see eq.\ (\ref{eq:minimizing-Theta2}),
\begin{eqnarray}
\hskip -4mm
{\cal E}_{\scriptU,\scriptT}[h]^{'}\!(0) & = &
\frac{\Re \, \Tr}{N_c\,d\, V} \, 
\mathlarger\sum_{b,\,\mu,\,\vec{x}}
\!-i\, \Bigl[\gamma^{b}(\vecx) \, t^b \,
        U_{\mu}(h;\vecx)\,e^{- i \frac{\Theta_{\mu}}{N}}
       -  e^{- i \frac{\Theta_{\mu}}{N}} U_{\mu}(h;\vecx)
     \gamma^{b}(\vec{x} + \hat{e}_{\mu}) \, t^b\Bigr] \hskip 5mm 
    \nonumber \\[2mm]
&=& \frac{2 \, \Re \, \Tr}{N_c\,d\, V} \, 
\mathlarger\sum_{b,\,\mu,\,\vec{x}}\,
     \frac{\gamma^{b}(\vecx) \, t^b}{2 \, i} \,
   \Bigr[ \, U_{\mu}(h;\vecx) \, e^{- i \frac{\Theta_{\mu}}{N}}
   \, - \, e^{- i \frac{\Theta_{\mu}}{N}} \,
   U_{\mu}(h;\vec{x} - \hat{e}_{\mu}) \, \Bigr] \; ,
\label{eq:Eprime0}
\end{eqnarray}
which should be compared
to eq.\ (\ref{eq:EderivPBCs2}).
Here, as usual,
$\vec{x}\in\Lambda_x$, the color index $b$ takes values $1,\ldots,N_c^2-1$
and $\mu=1,\ldots,d$.
The above expression is also equal to
\begin{eqnarray}
\hskip -3.5mm
\frac{2 \, \Re \, \Tr}{N_c\,d\, V}
\mathlarger\sum_{b,\,\mu,\,\vec{x}}
     \frac{\gamma^{b}(\vecx)}{2 \, i} \,
    e^{i \sum_{\nu}\!\! \frac{\Theta_{\nu}  x_{\nu}}{N}}
   \Bigr[ \, t^b\, U_{\mu}(h;\vecx) \, e^{- i \frac{\Theta_{\mu}}{N}}
-e^{- i \frac{\Theta_{\mu}}{N}} \,
   U_{\mu}(h;\vec{x} - \hat{e}_{\mu}) t^b\, \Bigr]
   e^{- i \sum_{\nu}\!\! \frac{\Theta_{\nu}  x_{\nu}}{N}}\,,\quad
\label{eq:ElprimeUl}
\end{eqnarray}
since the external factors $\exp{(\pm i \sum_{\nu = 1}^d \,
\Theta_{\nu} \, x_{\nu} / N )}$ are simplified by using the
cyclicity of the trace.
Then, the first derivative of the minimizing functional --- with respect to
$\{h(\vecx)\}$ and considering fixed $\Theta_\mu$'s --- can be written
in terms of the link variables $U_{\mu}(l;\vecx)$, see eq.\ (\ref{eq:Uofl}), as
\begin{eqnarray}
\hskip -2mm
{\cal E}_{\scriptU,\scriptT}[h]^{'}\!(0) & = &
\frac{2 \, \Re \, \Tr}{N_c\,d\, V}
\mathlarger\sum_{b,\,\mu,\,\vec{x}}\,
     \frac{\gamma^{b}(\vecx)}{2 \, i} \,
   \Bigr[ \, \tilde{t}^{\,b}(\vecx)\, U_{\mu}(l;\vecx) \, - \, 
   U_{\mu}(l;\vec{x} - \hat{e}_{\mu}) \, \tilde{t}^{\,b}(\vecx)\, \Bigr]
\nonumber \\[2mm]
& = &
\frac{2 \, \Re \, \Tr}{N_c\,d\, V}
\mathlarger\sum_{b,\,\mu,\,\vec{x}}\,
\frac{\gamma^{b}(\vecx)\,\tilde{t}^{\,b}(\vecx)}{2 \, i} \,
   \Bigr[ \, U_{\mu}(l;\vecx) \, - \,
           U_{\mu}(l;\vec{x} - \hat{e}_{\mu}) \, \Bigr]\,,
\end{eqnarray}
where we have defined the new set of Hermitian and traceless
generators\footnote{This is a
similarity transformation which preserves the orthogonality
relation (\ref{eq:Trlambda}) and the structure constants $f^{abc}$
of the $su(N_c)$ Lie algebra.
Moreover, it does not change the Cartan generators $\{ t^b_{\scriptC} \}$
(see appendix \ref{sec:Cartan}), which trivially commute with the
$\Theta_{\mu}$ matrices. \label{foot:ttilde}}
\begin{equation}
\tilde{t}^{\,b}(\vecx) \, \equiv \,
\exp{\left( i \sum_{\nu = 1}^d \, \frac{\Theta_{\nu} \, x_{\nu}}{N}
                  \right)}\, t^{b}\,
\exp{\left( - i \sum_{\nu = 1}^d \, \frac{\Theta_{\nu} \, x_{\nu}}{N}
\right)} \,.
\label{eq:ttilde}
\end{equation}
Now, we impose the stationarity condition
${\cal E}_{\scriptU,\scriptT}[h]^{'}\!(0)=0$,
which must hold
for any set of parameters $\gamma^{b}(\vecx)$.
Clearly, this
means that, for each lattice site $\vecx$ and color index $b$, 
we have the condition
\begin{equation}
\Re \, \Tr \, \sum_{\mu = 1}^d\, \frac{\tilde{t}^{\,b}(\vecx)}{2 \, i} \,
   \Bigr[ \, U_{\mu}(l;\vecx) \, - \,
           U_{\mu}(l;\vec{x} - \hat{e}_{\mu}) \, \Bigr]\,=\,0\,.
\label{eq:mincond}
\end{equation}
In analogy with eqs.\ (\ref{eq:Atrac}) and (\ref{eq:diverAnob}), let us define
\begin{eqnarray}
\hskip -2mm
 A_{\mu}(l;\vecx) \,&\equiv&\, \frac{1}{2 \,i}\, \left[ \,
  U_{\mu}(l;\vecx) \, -\, U_{\mu}^{\dagger}(l;\vecx) \,
     \right]_{\rm traceless}\;
\label{eq:Altraceless}
         \\[2mm]
    & = &\, \frac{1}{2 \,i}\, \left[ \,
   U_{\mu}(l;\vecx) - U_{\mu}^{\dagger}(l;\vecx) \,
        \right] \, - \,  \1 \, \frac{\Tr}{2 \,i\,N_c}\,
    \left[ \, U_{\mu}(l;\vecx) - U_{\mu}^{\dagger}(l;\vecx)
      \, \right]
\label{eq:AlUU}
\end{eqnarray}
and
\begin{equation}
\left( \nabla\cdots A \right)\!(l;\vecx) \; \equiv \;
  \sum_{\mu = 1}^d \; \Bigl[A_{\mu}(l;\vecx) -
       A_{\mu}(l;\vec{x} - \hat{e}_{\mu})\Bigr]\,.
\label{eq:defdivl}
\end{equation}
We can now write the minimization condition (\ref{eq:mincond})
above
in terms of color components of the gauge-field gradient, using the 
site-dependent generators in eq.\ (\ref{eq:ttilde}), as
\begin{equation}
\left( \nabla\cdots A^{b} \right)\!(l;\vecx) \, = \,
\Tr \left[\,\tilde{t}^{\,b}(\vecx)\,
\frac{\left( \nabla\cdots A \right)\!(l;\vecx)}{2}\,\right]
 \, = \,
  0 \quad \quad \forall \;\, \vec{x}, b\;\;\;\; ,
\label{eq:Adivlb}
\end{equation}
by noting, see eq.\ (\ref{eq:AlUU}), that
\begin{equation}
\Tr \!\left[\, \tilde{t}^{\,b}(\vecx)\,
\frac{A_{\mu}(l;\vecx)}{2}\,\right] 
\,=\, 
\Tr \left\{\, \tilde{t}^{\,b}(\vecx)
\left[\frac{U_{\mu}(l;\vecx) \,-\, U_{\mu}^{\dagger}(l;\vecx)}{4\,i}
\right]\right\}
=
\Re\,\Tr \!\left[\, \tilde{t}^{\,b}(\vecx)\,
\frac{U_{\mu}(l;\vecx)}{2\,i}\,\right] \;\quad\;
\label{eq:Albnew}
\end{equation}
and
\begin{equation}
\Tr \!\left[\, \tilde{t}^{\,b}(\vecx)\,
\frac{A_{\mu}(l;\vecx - \hat{e}_{\mu})}{2}\,\right] 
\,=\,
\Re \, \Tr \!\left[\, \tilde{t}^{\,b}(\vecx)\,
\frac{U_{\mu}(l;\vecx - \hat{e}_{\mu})}{2\,i}\,\right] \; .
\label{eq:Amulxb-new}
\end{equation}
Hence, the ${\cal N}_p=V\,(N_c^2 - 1)$ constraints needed to characterize the
stationary point of
${\cal E}_{\scriptU,\scriptT}[h](\tau)$,
with respect to the gauge transformation 
$\{ h(\vecx) \}$ --- obtained in eq.\ (\ref{eq:mincond}) and rewritten in
eq.\ (\ref{eq:Adivlb}) --- may be interpreted as a transversality
condition for the color components of the gauge-transformed gauge 
field $A_{\mu}(l;\vecx)$, as will be defined below.
Actually, as already mentioned, to implement these conditions in practice, 
it is 
convenient\footnote{See also the beginning of section \ref{sec:numerical-PBCs},
where it is stressed that, in the numerical code, it is more natural to save
the values of $U_{\mu}(h;\vecx)$ and $\Theta_{\mu}$, instead of the values of
$U_{\mu}(l;\vecx)$.}
to write the above expressions in terms of $U_{\mu}(h;\vecx)$ and $\Theta_{\mu}$.
We then get, from eq.\ (\ref{eq:Albnew}),
\begin{eqnarray}
\Tr \!\left[\, \tilde{t}^{\,b}(\vecx)\,
\frac{A_{\mu}(l;\vecx)}{2}\,\right]
&=&\,
\Tr \left\{ \frac{t^{b}}{4\,i} \!\left[  U_{\mu}(h;\vecx)
       \, e^{- i \frac{\Theta_{\mu}}{N}}-\, e^{i\frac{\Theta_{\mu}}{N}} \,
    U_{\mu}^{\dagger}(h;\vecx) \right] \right\} \nonumber \\[2mm]
&=&\,
\Re\Tr \left[t^b\,\frac{U_{\mu}(h;\vecx)\,
e^{- i \frac{\Theta_{\mu}}{N}}}{2 i} \right] \; ,
\label{eq:Albsimple}
\label{eq:Axmodified}
\end{eqnarray}
using eq.\ (\ref{eq:Uofl}) and the definition (\ref{eq:ttilde}).
In like manner, see eqs.\ (\ref{eq:Uofl-}) and (\ref{eq:Amulxb-new}), we have
\begin{eqnarray}
\label{eq:Ax-modified0}
\hskip -6mm
\Tr \!\left[\, \tilde{t}^{\,b}(\vecx)\,
\frac{A_{\mu}(l;\vecx - \hat{e}_{\mu})}{2}\,\right]
    &=& \Tr \, \Biggl\{\frac{t^{b}}{4\,i}
            \left[ \, e^{- i \frac{\Theta_{\mu}}{N}}\,
       U_{\mu}(h;\vec{x} - \hat{e}_{\mu}) 
       - U_{\mu}^{\dagger}(h;\vec{x} - \hat{e}_{\mu})\,
       e^{i \frac{\Theta_{\mu}}{N}} \,
    \right] \, \Biggr\} \nonumber \\[2mm]
     &=& \Re \, \Tr \left[\, t^{b}\,
    \frac{e^{- i \frac{\Theta_{\mu}}{N}}\,
       U_{\mu}(h;\vec{x} - \hat{e}_{\mu})}{2\,i}
      \, \right]
      \; . 
\label{eq:Ax-modified}
\end{eqnarray}
Notice that, contrary to
eqs.\ (\ref{eq:Albnew}) and (\ref{eq:Amulxb-new}), 
the expressions on the r.h.s.\ of 
eqs.\ (\ref{eq:Albsimple}) and (\ref{eq:Ax-modified0})
are written in terms of the original (site-independent) generators 
$\{ t^b \}$ and involve only $U_{\mu}(h;\vecx)$ and $\Theta_{\mu}$.
They are the natural
choice to be employed in a numerical simulation.
Of course, the above connection between the expressions in terms of
$\{ \tilde{t}^{\,b}(\vecx) \}$ and of $\{ t^{\,b} \}$ can also be seen directly
after rewriting eq.\ (\ref{eq:Adivlb}) as
\begin{equation}
0\,=\,\Tr \left[\,\tilde{t}^{\,b}(\vecx)\,
\frac{\left( \nabla\cdots A \right)\!(l;\vecx)}{2}\,\right]
 \, = \,
\Tr \left[\,t^b\;
  \frac{e^{-i \sum_{\nu}\!\!\!\frac{\Theta_{\nu} \, x_{\nu}}{N}}
        \!\left( \nabla\cdots A \right)\!(l;\vecx)\;
      e^{i \sum_{\nu}\!\!\!\frac{\Theta_{\nu} \, x_{\nu}}{N}}
}{2}\right]\;,
\end{equation}
where the r.h.s.\ is in agreement with eq.\ (\ref{eq:Eprime0}),
see also eqs.\ (\ref{eq:Uofl}) and (\ref{eq:Uofl-}).

Using the above results and in analogy with section \ref{sec:mlg-PBCs},
we can define the color components of the gauge-transformed
gauge field as
\begin{equation}
A_{\mu}^{b}(l;\vecx) \,\equiv
\Tr \!\left[\, \tilde{t}^{\,b}(\vecx)\,
\frac{A_{\mu}(l;\vecx)}{2}\,\right] 
\,=\, 
\Re\,\Tr \!\left[\, \tilde{t}^{\,b}(\vecx)\,
\frac{U_{\mu}(l;\vecx)}{2\,i}\,\right]
\label{eq:Alb}
\,=\,
\Re\Tr \left[t^b\,\frac{U_{\mu}(h;\vecx)\,
e^{- i \frac{\Theta_{\mu}}{N}}}{2 i} \right]
\end{equation}
and
\begin{eqnarray}
A_{\mu}^{b}(l;\vecx - \hat{e}_{\mu}) & \equiv &
\Tr \!\left[\, \tilde{t}^{\,b}(\vecx)\,
\frac{A_{\mu}(l;\vecx - \hat{e}_{\mu})}{2}\,\right] 
\,=\,
\Re \, \Tr \!\left[\, \tilde{t}^{\,b}(\vecx)\,
\frac{U_{\mu}(l;\vecx - \hat{e}_{\mu})}{2\,i}\,\right]
\nonumber \\[2mm]
&=&\,
\Re \, \Tr \left[\, t^{b}\,
    \frac{e^{- i \frac{\Theta_{\mu}}{N}}\,
       U_{\mu}(h;\vec{x} - \hat{e}_{\mu})}{2\,i}
      \, \right] \; ,
\label{eq:Amulxb-}
\end{eqnarray}
which imply the relations
\begin{equation}
A_{\mu}(l;\vecx) \,=\, \sum_{b=1}^{N_c^2-1}\,
A_{\mu}^{b}(l;\vecx) \,\tilde{t}^{\,b}(\vecx)
\quad\quad{\rm and}\quad\quad
A_{\mu}(l;\vecx - \hat{e}_{\mu}) \,=\, \sum_{b=1}^{N_c^2-1}\,
A_{\mu}^{b}(l;\vecx - \hat{e}_{\mu}) \,\tilde{t}^{\,b}(\vecx) \; ,
\label{eq:Amulvecx}
\end{equation}
since $\{ \tilde{t}^{\,b}(\vecx) \}$ is a basis of the $su(N_c)$
Lie algebra.
Then, eq.\ (\ref{eq:Adivlb}) can be written as
\begin{equation}
\sum_{\mu = 1}^d \; \Bigl[A_{\mu}^{b}(l;\vecx) -
       A_{\mu}^{b}(l;\vec{x} - \hat{e}_{\mu})\Bigr]
 \, = \,
  0 \quad \quad \forall \;\, \vec{x}, b
\label{eq:Adivlnobmod}
\end{equation}
and it is also equivalent to the transversality condition
\begin{equation}
\left( \nabla\cdots A \right)\!(l;\vecx) \; = \;
  \sum_{\mu = 1}^d \; \Bigl[A_{\mu}(l;\vecx) -
       A_{\mu}(l;\vec{x} - \hat{e}_{\mu})\Bigr]
 \, = \,
  0 \quad \quad \forall \;\, \vec{x}\;\;\;\; .
\label{eq:Adivlnob}
\end{equation}

One should stress that the above expressions are
valid only ``locally'', i.e.\ when evaluating the lattice
divergence of the gauge field at site $\vecx$, and that
they have to be modified accordingly when moving to
the next site, e.g.\ when evaluating $\left( \nabla\cdots
A^{b} \right)\!(l;\vecx + \hat{e}_{\mu})$.
In particular, we
consider a new set of
generators $\{ \tilde{t}^{\,b}(\vecx) \}$ for each site
$\vecx$, where the divergence is evaluated, and these 
generators are used both to define $A_{\mu}^{b}(l;\vecx)$
and $A_{\mu}^{b}(l;\vecx - \hat{e}_{\mu})$, in terms of the
matrices $U_{\mu}(l;\vecx)$ and $U_{\mu}(l;\vecx - \hat{e}_{\mu})$.
Indeed, the lattice
divergence is just a simple (backward) discretization of
the usual continuum divergence and, when written explicitly
for the color components of the gauge field, it should be
based on the same generators at points $\vecx$ and $\vec{x} - \hat{e}_{\mu}$,
namely $\{ \tilde{t}^{\,b}(\vecx) \}$.
This is the origin of the different expressions obtained
for the gauge field at site $\vecx$ and at site $\vecx - \hat{e}_{\mu}$
--- respectively eqs.\ (\ref{eq:Albnew}) and (\ref{eq:Amulxb-new}), 
or eqs.\ (\ref{eq:Albsimple}) and (\ref{eq:Ax-modified0}) ---
considering that the generators $\tilde{t}^{\,b}(\vecx)$ are defined
as a function of $\vecx$, and that the generators $t^b$ do not generally 
commute with the matrices $\Theta_{\mu}$.
At the same time, note that the combination $U_{\mu}(h;\vecx)\,
e^{- i \frac{\Theta_{\mu}}{N}}$ or, equivalently,
$e^{- i \frac{\Theta_{\mu}}{N}} \, U_{\mu}(h;\vecx)$,
also appears in the
minimizing functional (\ref{eq:minfunctUhThetax}), which enforces
the transversality
condition on the lattice $\Lambda_x$, but
applied to this modified link configuration, see comment below
eq.\ (\ref{eq:gofh}).

Of course, as done in section \ref{sec:linkNP}, a more natural approach
would be to consider an expansion in the basis $\{ {\bf W}^{ij} \}$,
which is constructed using the common eigenvectors of the matrices $\Theta_\mu$.
Then the matrices $\,\Theta_\mu$ are diagonal, see eq.\ (\ref{eq:Thetadiag}), and we get a unique
definition of the gauge-field components at $\vecx$ and $\vecx - \hat{e}_{\mu}$.
Here, however, we work with the color components, in order
to obtain expressions that can be easily compared with those presented in
section \ref{sec:mlg-PBCs}.
Indeed, all the expressions above clearly reduce to the ones in section
\ref{sec:mlg-PBCs} in the trivial case $\,\Theta_\mu=\1$ for all $\mu$.

\vskip 3mm
As for the minimization with respect to the matrices $\Theta_{\mu}$,
it does not introduce any other constraint, even though --- when
varying the parameters $v_{ij}$ and $n_{\mu}^{j}$, see eqs.\
(\ref{eq:vDjj}) and (\ref{eq:Djj}) --- we need to verify the
inequalities imposed by the considered definition of local minimum,
see eqs.\ (\ref{eq:fmin})--(\ref{eq:constraints}).
This becomes evident if we consider the stationarity condition for
the whole (extended) lattice $\Lambda_z$, i.e.\
\begin{eqnarray}
  0\,=\,\left( \nabla \cdots A \right)\!(g;\vecz) \,&=&\,
  \left( \nabla \cdots A \right)\!(g;\vec{x} + \vec{y} N)
  \nonumber \\[2mm]
  &=&\,
   \sum_{\mu=1}^{d}\, A_{\mu}(g;\vec{x} + \vec{y} N)\,-\,
   A_{\mu}(g;\vec{x} + \vec{y} N - \hat{e}_{\mu}) \; ,
\label{eq:divAz}
\end{eqnarray}
which enforces the ${\cal N}_{p,m}=V m^d (N_c^2-1)$ constraints
expected\footnote{Clearly, the value of ${\cal N}_{p,m}$ is independent of
the way in which we write the gauge transformation $\{ g(\vecz) \}$,
i.e.\ as a Bloch function or as a general transformation, as long as
$g(\vecz) \in$ SU($N_c$).}
from the minimization of ${\cal E}_{\scriptU}[g]$.
At the same time, we know that
\begin{eqnarray}
\!\!\!\!\!\!\!\!\!\!
\nonumber
  A_{\mu}(g;\vecz) \, & = & \,
  A_{\mu}(g;\vec{x} + \vec{y} N) \\[2mm]
\label{eq:Agzetadef}
 & \equiv & \,
 \frac{1}{2 \,i}\, \left[ \,
    U_{\mu}(g;\vec{x} + \vec{y} N) -
    U_{\mu}^{\dagger}(g;\vec{x} + \vec{y} N)
    \, \right]_{\rm traceless} \\[2mm]
\label{eq:Agzeta}
 &=&\, \exp{\left( i \sum_{\nu = 1}^d \, \Theta_{\nu} \, y_{\nu} \right)} 
\, \left[ \frac{U_{\mu}(l;\vecx) - U_{\mu}^{\dagger}(l;\vecx)}{2\,i}
    \right]_{\rm traceless}\!\!\!
\exp{\left( - i \sum_{\nu = 1}^d \, \Theta_{\nu} \, y_{\nu}
              \right)} \\[2mm]
 &=&\, \exp{\left( i \sum_{\nu = 1}^d \, \Theta_{\nu} \, y_{\nu} \right)} 
\, A_{\mu}(l;\vecx) \,
\exp{\left( - i \sum_{\nu = 1}^d \, \Theta_{\nu} \, y_{\nu}
              \right)} \label{eq:Aglx}
\;\;\;\;\;
\label{eq:Agzeta2}
\end{eqnarray}
where we used eqs.\ (\ref{eq:Uofg2}), (\ref{eq:Uofl}) and (\ref{eq:Altraceless}),
and similarly\footnote{Clearly, similar expressions
hold for $A_{\mu}(g;\vec{x} + \vec{y} N - \hat{e}_{\mu})$, which can
be written in terms of $U_{\mu}(l;\vecx - \hat{e}_{\mu})$ or of
$A_{\mu}(l;\vec{x} - \hat{e}_{\mu})$,
see eqs.\ (\ref{eq:Uofl-}) and (\ref{eq:Altraceless}).
\label{foot:Amug}}
for $A_{\mu}(g;\vecz- \hat{e}_{\mu}) = A_{\mu}(g;\vec{x} + \vec{y} N - \hat{e}_{\mu})$.
Hence, we find that
\begin{equation}
  \left( \nabla \cdots A \right)\!(g;\vecz) \,=\,
  \exp{\left( i \sum_{\nu = 1}^d \, \Theta_{\nu} \, y_{\nu}
    \right)}\,\, \left( \nabla \cdots A \right)\!(l;\vecx) \,\,
    \exp{\left( - i \sum_{\nu = 1}^d \, \Theta_{\nu} \, y_{\nu}
    \right)}
\label{eq:Ag-AhT}
\end{equation}
and it is evident that eq.\ (\ref{eq:divAz}) does not add any
information to
eq.\ (\ref{eq:Adivlnob}).

In summary, the transversality condition for the lattice gauge field
$A_\mu(g;\vecz)$ defined in (\ref{eq:Agzetadef}) is imposed by requiring
the small-lattice field $A_\mu(l;\vecx)$, defined in (\ref{eq:Altraceless}),
to be transverse.
This can be verified by using the expressions in eqs.\
(\ref{eq:Adivlb}), (\ref{eq:Adivlnobmod}) or (\ref{eq:Adivlnob}).


\subsection{The limit \texorpdfstring{$m \to +\infty$}{mtoinfinity}}
\label{sec:minfty}

We consider now the limit of $m$ going to infinity, i.e.\
when the eigenvalues $\exp{( 2 \pi i \bar{n}_{\mu} / m )}$ of the
matrices $\exp{( i \Theta_{\mu})}$ --- with $\bar{n}_{\mu}=n_{\mu}
\pmod{m} \in [0, m-1]$ --- can be written as $\exp{( 2 \pi i
\epsilon_{\mu} )}$ with the real (continuous) parameters $\epsilon_{\mu}
\equiv \bar{n}_{\mu} / m$
taking values in the interval $[0,1)$.
Then, as noticed in ref.\ \cite{Zwanziger:1993dh}, the
minimization process imposes also the stationarity condition
with respect to variation of the $\Theta_{\mu}$ matrices.
In this case, it is convenient to consider eq.\
(\ref{eq:minimizing-Theta2}) with the matrices $\Theta_{\mu}$
written in terms of the Cartan generators $\{ t^b_{\scriptC}
\}$, as in eq.\ (\ref{eq:tautheta}).
Next, we can consider small variations of the parameters $\theta^b_{
\mu}$, i.e.\ write the matrices
\begin{equation}
\Theta_{\mu}^{\mk \prime} \, = \, \sum_{b=1}^{N_c-1} \,
    t^b_{\scriptC} \, \left(\, \theta^b_{\mu}\,+\,
       \tau\,\eta^b_{\mu}\,\right) \,=\,
\Theta_{\mu} \, + \,\tau\, \sum_{b=1}^{N_c-1} \,
    t^b_{\scriptC} \,\eta^b_{\mu}\,,
\end{equation}
where $\eta^b_{\mu}$ are general parameters and $\tau$ is small, so that
\begin{equation}
e^{-i \frac{\Theta_{\mu}^{\mk \prime}}{N}} \,\approx\,
e^{-i \frac{\Theta_{\mu}}{N}} \,
\left(\,\1\,-\,i\,\frac{\tau}{N}\, \sum_{b=1}^{N_c-1} \,
    t^b_{\scriptC} \, \eta^b_{\mu}\, \right) \; .
\end{equation}
Hence, by imposing a null first variation of the minimizing functional with
respect to $\tau$, as above, 
we must have, see eqs.\ (\ref{eq:minimizing2}) and (\ref{eq:Z}),
\begin{equation}
   0 \, = \, \frac{\Re \, \Tr}{N_c\,d}
 \sum_{\mu = 1}^d \left[ i \,
    Z_{\mu}(h) \, \frac{e^{-i \frac{\Theta_{\mu}}{N}}}{V\,N}
      \sum_{b=1}^{N_c-1}
    t^b_{\scriptC} \, \eta^b_{\mu}
    \right]
= \frac{\Re \, \Tr}{N_c\,d\,N\,V}
    \sum_{\mu,\,b} \,
    t^b_{\scriptC} \, \eta^b_{\mu}\,
     \left[ \, i \,
    Z_{\mu}(h) \, e^{-i \frac{\Theta_{\mu}}{N}} \,
     \right]
\end{equation}
and we find
\begin{equation}
   0 \,=\, \Re \, \Tr \, \left\{\,
    t^b_{\scriptC} \left[ \, i \,
    Z_{\mu}(h) \, e^{-i \frac{\Theta_{\mu}}{N}}
     \right]\right\} =\, \frac{1}{2} \Tr \left\{\,i\,
    t^b_{\scriptC} \left[ \,
    Z_{\mu}(h) \, e^{-i \frac{\Theta_{\mu}}{N}}-\,
    e^{i \frac{\Theta_{\mu}}{N}}\, Z^{\dagger}_{\mu}(h)
        \, \right]\right\} \;
\end{equation}
for all $\mu$ and $b$, since the equality must hold for any set of parameters
$\{ \eta^b_{\mu} \}$. 
Finally, using eq.\ (\ref{eq:Z}) we obtain
\begin{equation}
   0 \,=\, \Tr \, \left\{\,
    \frac{t^b_{\scriptC}}{2i} \sum_{\vec{x}\in\Lambda_x}\,
      \left[ \,
    U_{\mu}(h;\vecx) \, e^{-i \frac{\Theta_{\mu}}{N}}\,- \,
    e^{i \frac{\Theta_{\mu}}{N}}\, U^{\dagger}_{\mu}(h;\vecx)
        \, \right]\right\} \; ,
\label{eq:zeromode0}
\end{equation}
which can be written as,\footnote{We stress that, even though we are using here the same 
index $b$ to denote the color components with respect to the generators 
$t^b$ of the Lie algebra, the constraint in 
eq.\ (\ref{eq:zeromode0}) is written in terms of color components
with respect to the Cartan generators $t^b_{\scriptC}$.
The same holds for
the color components of
$Q_{\mu}(l)$ and $A_{\mu}(l;\vecx)$ in the equations below.}
see eq.\ (\ref{eq:Axmodified}),
\begin{equation}
   Q_{\mu}^b(l)\,\equiv\, \sum_{\vec{x}\in\Lambda_x}\,
          A_{\mu}^{b}(l;\vecx)
          \,=\, 0 \; .
\label{eq:zeromodeb}
\end{equation}
At the same time,
we can define, see eq.\ (\ref{eq:Altraceless}),
\begin{equation}
Q_{\mu}(l)\,\equiv\,
     \sum_{\vec{x}\in\Lambda_x}
        A_{\mu}(l;\vecx)
\label{eq:QmulofA}
\end{equation}
so that
\begin{equation}
Q_{\mu}^b(l)\,=\,
\frac{\Tr}{2} \, \left[\,
    t^b_{\scriptC}\; Q_{\mu}(l)
        \, \right] \; ,
\end{equation}
where we used eq.\ (\ref{eq:Alb}) and the definition (\ref{eq:ttilde}).

The above gauge-fixing condition tells us that the color
components of the gauge field
$A_{\mu}(l;\vecx)$,
corresponding to the generators $t^b_{\scriptC}$ of the Cartan 
sub-algebra,
have zero constant mode in the infinite-volume limit
$m \to + \infty$, yielding
\begin{equation}
 \sum_{b=1}^{N_c-1}\,
    Q_{\mu}^b(l) \, t^b_{\scriptC} \, = \, 0 \; .
\label{eq:Qzero}
\end{equation}
Then, using the result obtained at the end of section
\ref{sec:lambda-basis} of the appendix, see eqs.\
(\ref{eq:M_Cgen})--(\ref{eq:all}), which relates the coefficients
$m^i$ --- in the expansion of a matrix $M_{\scriptC}$ of the Cartan
sub-algebra, such as the (null) expression in eq.\ (\ref{eq:Qzero}),
relative to the generators $t^i_{\scriptC}$ --- with its coefficients
$a^{jj}$ in the basis of the matrices ${\bf W}^{jj}$, we can also write,
see eq.\ (\ref{eq:zeromodeb}),
\begin{equation}
Q_{\mu}^{jj}(l)\,=\,\sum_{i=1}^{N_c-1}\,Q_{\mu}^i(j)
  \,\left[\,R^{ij}\,
    \xi^j \,-\,R^{i(j-1)}\,\xi^{j-1}\right] 
\,=\, 0
\label{eq:aQzero}
\end{equation}
for the coefficients $Q_{\mu}^{jj}(l)$ of $Q_{\mu}(l)$,
which implies
\begin{equation}
\sum_{j=1}^{N_c}\, Q_{\mu}^{jj}(l)\, {\bf W}^{jj} \, =\, 0\; .
\end{equation}
We will comment again on this outcome in section
\ref{sec:gluon-extended}.
For the moment we only stress that the condition
(\ref{eq:zeromodeb}) --- or (\ref{eq:Qzero}) ---
is weaker than the one presented
in ref.\ \cite{Zwanziger:1993dh}, which, however, has
been obtained considering the absolute minimum
of the minimizing functional ${\cal E}_{\scriptU,\scriptT}[h]$.
Per contra, here we prefer to focus on a minimizing condition
that can be verified in a numerical simulation,
given that --- in the general case --- we have access only to
local minima of ${\cal E}_{\scriptU,\scriptT}[h]$.


\subsection{Convergence of the numerical minimization}
\label{sec:num-min-convergence}

The numerical convergence of a gauge-fixing algorithm can be checked,
also when using the extended lattice $\Lambda_z$, by considering the
three quantities defined in eqs.\ (\ref{eq:e1})--(\ref{eq:e6}).
Moreover, as for the minimizing functional ${\cal E}_{\scriptU}[l]=
{\cal E}_{\scriptU,\scriptT}[h]$, they can be evaluated on the
original lattice $\Lambda_x$, (essentially) without the need to
consider the whole extended lattice $\Lambda_z$.
For the quantity $\Delta {\cal E}$, this has already been proven in eq.\
(\ref{eq:minimizing2}).
In the case of $(\nabla A)^2$ we can write, as in eq.\ (\ref{eq:defdiv2}),
\begin{eqnarray}
(\nabla A)^2 \,& \equiv &\, \frac{\Tr}{2\,(N_c^2 - 1)\,m^d\,V} \,
    \sum_{\vec{z}\in\Lambda_z}
       \, \Big[ \left( \nabla \cdots A \right)\!(g;\vecz)
                     \Big]^{2} \nonumber \\[2mm]
& = &\, \frac{1}{m^d} \, \sum_{\vec{y}\in\Lambda_y} \,
               \Biggl\{ \,
              \frac{\Tr}{2\,(N_c^2 - 1)\,V} \,
    \sum_{\vec{x}\in\Lambda_x}
       \, \Big[ \left( \nabla \cdots A \right)\!(g;\vec{x} +
                           \vec{y} N) \Big]^{2} \,\Biggr\}
\label{eq:A2Lambdaz}
\end{eqnarray}
and use the expression for $\left( \nabla \cdots A \right)\!(g;\vec{x}
+ \vec{y} N)$ reported in the previous section, see eq.\ (\ref{eq:Ag-AhT}).
Then, due to the trace, it is clear that the exponential factors
$\exp{( \pm i \sum_{\nu = 1}^d \Theta_{\nu} y_{\nu})}$
cancel
for each site $\vecx$.
In particular --- after evaluating the trace --- there is no dependence 
on the $\vec{y}$ coordinates in eq.\ (\ref{eq:A2Lambdaz}) 
and we have
\begin{equation}
(\nabla A)^2 \, = \,
              \frac{\Tr}{2\,(N_c^2 - 1)\,V} \,
    \sum_{\vec{x}\in\Lambda_x}
       \, \Big[ \left( \nabla \cdots A \right)\!(
           l;\vecx) \Big]^{2}\,=\,
\frac{1}{(N_c^2 - 1)\,V} \,
    \sum_{\vec{x}\in\Lambda_x} \!\sum_{b=1}^{N_c^2 - 1}
       \, \Big[ \left( \nabla \cdots A^{b} \right)\!(
           l;\vecx) \Big]^{2} \; ,
\label{eq:nablaAx}
\end{equation}
with $\left( \nabla \cdots A^{b} \right)\!(l;\vecx)$
defined in
section \ref{sec:transversality-Lambdaz}.
The above result is, of course, expected, since the
gauge-fixed gauge configuration $\{ A_{\mu}(l;\vecx) \}$
is transverse
on each replicated lattice,
for any lattice site $\vecx$.
 
Finally, see eqs.\ (\ref{eq:e6}) and (\ref{eq:SigQ}),
for the quantity 
\begin{eqnarray}
\Sigma_Q \, &=& \, \frac{1}{m N} \,
\sum_{\mu = 1}^d \, \sum_{b\, = 1}^{N_c^2 -1} \,
\sum_{z_{\mu} = 1}^{m N} \,
    \left[ \, Q_{\mu}^{b}(g;z_{\mu}) - {\widehat Q}_{\mu}^{b}(g)  \,
      \right]^{2} \;/\;\;\;
   \sum_{\mu = 1}^d \, \sum_{b\, = 1}^{N_c^2 -1} \,
      \left[ {\widehat Q}_{\mu}^{b}(g) \right]^2
\nonumber \\[2mm]
 &=& \, \frac{1}{m N} \, \sum_{\mu = 1}^d \, \,
\sum_{z_{\mu} = 1}^{m N} \, \Tr\,
    \left[ \, Q_{\mu}(g;z_{\mu}) - {\widehat Q}_{\mu}(g) \,
      \right]^{2} \;/\;\;\;
   \sum_{\mu = 1}^d \, \Tr \,
      \left[ {\widehat Q}_{\mu}(g) \right]^2 \; ,
\end{eqnarray}
we define
\begin{equation}
Q_{\mu}^b(g;z_{\mu}) \,\equiv\, 
\sum_{\substack{z_{\nu}\\ \nu \neq \mu}}
     A^{b}_{\mu}(g;\vecz)
     \quad\quad\mbox{and}
\quad\quad {\widehat Q}_{\mu}^{b}(g) \,\equiv\,\frac{1}{m N}\,
\sum_{z_{\mu} = 1}^{m N} \,
     Q_{\mu}^{b}(g;z_{\mu}) \; ,
\end{equation}
in analogy with section \ref{sec:numPBCs}.
On the other hand, similarly to eq.\ (\ref{eq:Adue}),
we can write
\begin{equation}
Q_{\mu}^b(g;z_{\mu}) \,=\, 
    \Re \, \Tr\,
 \sum_{\substack{z_{\nu} \\ \nu \neq \mu}}
\,\frac{U_{\mu}(g;\vecz) \, t^{b} }{2\,i}\,
\label{eq:Qbzmu}
\end{equation}
so that we can use the expression
\begin{equation}
Q_{\mu}^b(g;z_{\mu}) \, = \, 
   \Re \, \Tr \;
\left[\, \frac{Q_{\mu}(g;z_{\mu}) \, t^{b}}{2\,i} \, \right]
\label{eq:Qbz}
\end{equation}
with, see eq.\ (\ref{eq:Uofg2}),
\begin{equation}
     Q_{\mu}(g;z_\mu=x_{\mu}\!+\!Ny_{\mu}) =
     \sum_{\substack{z_{\nu} \\ \nu \neq \mu}} U_{\mu}(g;\vecz)
         = \!\!\!\sum_{\substack{y_{\nu} = 1, m\\ \nu \neq \mu}}\!\!\!
   \exp{\!\left(\! i \sum_{\rho = 1}^d  \Theta_{\rho}
             \, y_{\rho} \!\right)\!} \, Q_{\mu}(l;x_{\mu})
   \exp{\!\left(\!\!- i \sum_{\rho = 1}^d  \Theta_{\rho}
             \, y_{\rho} \!\right)}
\label{eq:Qz}
\end{equation}
and
\begin{equation}
Q_{\mu}(l;x_{\mu}) \, \equiv \,
\sum_{\substack{x_{\nu} = 1, N\\ \nu \neq \mu}} U_{\mu}(l;\vecx) \; .
    \label{eq:Qtildex}
\end{equation}
Then, it is evident from the above equations that, in the evaluation
of $\Sigma_Q$, we do not need a full loop over the extended lattice 
$\Lambda_z$, but it suffices to consider a loop over $\Lambda_x$ (see
the last equation), followed by a loop over $\Lambda_y$, see the
r.h.s.\ of eq.\ (\ref{eq:Qz}).
Thus, the computational cost is still of order $V$ (if $m^d \ltapprox
V$).
Let us stress that the quantities $Q_{\mu}(l;x_{\mu})$ are not
constant on $\Lambda_x$, since the transverse gauge-fixed link
configuration $\{ U_{\mu}(l;\vecx)\}$ is nonperiodic and,
therefore, when repeating the steps in eq.\ (\ref{eq:qconstant0}),
the second term is different from zero, see also eqs.\
(\ref{eq:qconstant}) and (\ref{eq:qconstant-sigma}).
As a consequence, we cannot expect to write $\Sigma_Q$ by averaging
only over the fluctuations $\left[ Q_{\mu}^{b}(l;x_{\mu}) -
{\widehat Q}_{\mu}^{b}(l) \right]^{2}$, where
\begin{equation}
Q_{\mu}^b(l;x_{\mu}) \,\equiv\, 
\sum_{\substack{x_{\nu}\\ \nu \neq \mu}}
     A^{b}_{\mu}(l;\vecx)
     \quad\quad\mbox{and}
\quad\quad {\widehat Q}_{\mu}^{b}(l) \,\equiv\,\frac{1}{N}\,
\sum_{x_{\mu} = 1}^{N} \,
     Q_{\mu}^{b}(l;x_{\mu}) \; .
\end{equation}
On the contrary, the quantities $Q_{\mu}^{b}(g;z_{\mu})$ in 
eq.\ (\ref{eq:Qbz}) are independent
of $z_{\mu}$, since $U_{\mu}(g;\vecz)$ is periodic
in $\Lambda_z$ and the gauge field $A_{\mu}(g;\vecz)$
is transverse.
Therefore, for the evaluation of $\Sigma_Q$ we need to consider the
global rotations $\exp{\left( i \sum_{\rho = 1}^d \Theta_{\rho}
y_{\rho} \right)}$, on each replicated lattice $\Lambda_x^{(\vecy)}$,
see eq.\ (\ref{eq:Qz}), and we cannot avoid the double sum,
i.e.\ the sum over the $y_{\nu}$ coordinates in eq.\ (\ref{eq:Qz})
and the sum over the $x_{\nu}$ coordinates in eq.\ (\ref{eq:Qtildex}).


\section{Link variables in momentum space and the gluon propagator}
\label{sec:linkNP}

The formulae discussed in section \ref{sec:gluonPBCs} for the
gluon propagator in momentum space $D(\veck)$ --- when
the usual lattice $\Lambda_x$ is considered --- clearly apply also
to the case of the extended lattice $\Lambda_z$, simply by
exchanging the sum over $\vec{x} \in \Lambda_x$ with the sum over
$\vec{z} \in \Lambda_z$ and, correspondingly, the sum over $\vec{k}
\in {\widetilde \Lambda}_x$ with the sum over $\veckp
\in {\widetilde \Lambda}_z$, i.e.\ the wave-number vectors
have now components $k_{\mu}^{\mk\prime} = 0, 1, \ldots, m N\!-\!1$
(when restricted to the first Brillouin zone).
However, in order to understand the impact of the extended lattice
on the evaluation of the gluon propagator (see section
\ref{sec:gluon-extended} below), it is useful to first evaluate the
Fourier transform
\begin{equation}
{\widetilde U}_{\mu}(g;\veckpp) \, \equiv \,
  \sum_{\vec{z}\in\Lambda_z} \, U_{\mu}(g;\vecz) \,
   \exp{\left[ - \frac{2 \pi i}{m\,N} \left(\veckp
   \cdots \vec{z} \right)\right]}
\label{eq:UFourier}
\end{equation}
of $U_{\mu}(g;\vecz)$, for $\mu=1,\ldots,d$.
Notice that this definition is based on the extended 
lattice, differing from the small-lattice definition (\ref{eq:UhxFourier})
in the range of the sum and in the exponential factor.
Also,
it is natural to consider 
the coefficients\footnote{Equivalently,
we can say that we write the matrix $U_{\mu}(g;\vecz)$ as a linear
combination of the matrices ${\bf W}^{ij}=w_i\,w_j^{\dagger} = v^{\dagger}
\,{\bf M}^{ij} \,v$, introduced in section \ref{sec:lambda-basis}.
This yields
\begin{equation}
U_{\mu}(g;\vecz) \, = \,
  \sum_{i,j=1}^{N_c}\,
          U^{ij}_{\mu}(g;\vecz) \, {\bf W}^{ij} \,=\,
  v^{\dagger}\, \Biggl\{ \,\sum_{i,j=1}^{N_c}\,
          U^{ij}_{\mu}(g;\vecz)\, {\bf M}^{ij} \,\Biggr\} \, v \; .
\label{eq:UvMv}
\end{equation}
\label{foot:SU-M}}
\begin{equation}
U^{ij}_{\mu}(g;\vecz) \, \equiv \, w_i^{\dagger} \,
   U_{\mu}(g;\vecz) \, w_j \qquad\qquad
   (\mbox{with} \;\;\; i,j = 1, 2, \ldots, N_c)
\end{equation}
in the basis of the common eigenvectors $w_j$ of the Cartan generators
and of the matrices $\Theta_{\mu}$, see eqs.\ (\ref{eq:Theta-wi}) and
(\ref{eq:Amatrix})--(\ref{eq:gencoeff}).
More exactly, we use
\begin{equation}
\exp{\left( - i \, \Theta_{\mu}  \right)} \, w_j \,=\,
   \exp{\biggl[ - \frac{2 \pi i}{m} \, n_{\mu}^{j} \biggr]}
    \, w_j 
    \label{eq:thetawj}
\end{equation}
as well as
\begin{equation}
w_i^{\dagger}\, \exp{\left( i \, \Theta_{\mu}  \right)} \,=\,
   w_i^{\dagger}\,
\exp{\biggl[ \frac{2 \pi i}{m} \, n_{\mu}^{i} \biggr]} \; ,
    \label{eq:thetawidagger}
\end{equation}
and find, see also eqs.\ (\ref{eq:zcoord}) and (\ref{eq:Uofg2}),
\begin{eqnarray}
U^{ij}_{\mu}(g;\vecz)
   & = & \, w_i^{\dagger} \,
\exp{\left( i \sum_{\nu = 1}^d \, \Theta_{\nu} \, y_{\nu}
              \right)} \, U_{\mu}(l;\vecx) \,
\exp{\left( - i \sum_{\nu = 1}^d \, \Theta_{\nu} \, y_{\nu}
              \right)} \, w_j \nonumber \\[2mm]
   & = & \, \exp{\biggl[ - \frac{2 \pi i}{m} \, \sum_{\nu = 1}^d \,
      \left( \, n_{\nu}^{j} - n_{\nu}^{i}
    \, \right) \, y_{\nu} \biggr]} \,
    w_i^{\dagger} \, U_{\mu}(l;\vecx) \, w_j \nonumber \\[2mm]
   & \equiv & \,  \exp{\biggl[ - \frac{2 \pi i}{m} \, \sum_{\nu = 1}^d \,
      \left( \, n_{\nu}^{j} - n_{\nu}^{i}
    \, \right) \, y_{\nu} \biggr]} \,
U^{ij}_{\mu}(l;\vecx)
\label{eq:Ugzij}
\end{eqnarray}
where, recalling eq.\ (\ref{eq:Uofl}) and that the $\Theta_\mu$'s commute
with each other,
\begin{eqnarray}
U^{ij}_{\mu}(l;\vecx) \,& = &\,
    w_i^{\dagger} \, \exp{\left( i \sum_{\nu = 1}^d \,
    \frac{\Theta_{\nu} \, x_{\nu}}{N} \right)} \, U_{\mu}(h;\vecx) \,
\exp{\left( - i \, \frac{\Theta_{\mu}}{N} \,
            - i \sum_{\nu = 1}^d \,
                  \frac{\Theta_{\nu} \, x_{\nu}}{N}
             \right)} \, w_j \mbox{\phantom{oioi}} \nonumber \\[2mm]
    & = & \, \exp{\Biggl\{ - \frac{2 \pi i}{m\,N} \, \Biggl[\,
       \sum_{\nu = 1}^d \, \left( \, n_{\nu}^{j} - n_{\nu}^{i}
    \, \right) \, x_{\nu} \, + n_{\mu}^{j}\,\Biggr]
          \, \Biggr\}} \,
    w_i^{\dagger} \, U_{\mu}(h;\vecx) \, w_j \nonumber \\[2mm]
    & = & \, \exp{\Biggl\{ - \frac{2 \pi i}{m\,N} \, \Biggl[\,
       \sum_{\nu = 1}^d \, \left( \, n_{\nu}^{j} - n_{\nu}^{i}
    \, \right) \, x_{\nu} \, + n_{\mu}^{j}\,\Biggr]
          \, \Biggr\}} \,
    U^{ij}_{\mu}(h;\vecx) \; .
\label{eq:Ulzij}
\end{eqnarray}
Then, we obtain
\begin{eqnarray}
\!\!\!\!
\label{eq:UFourier1} 
{\widetilde U}^{ij}_{\mu}(g;\veckpp) \,& \equiv &\, w_i^{\dagger} \,
   {\widetilde U}_{\mu}(g;\veckpp)
   \, w_j \\[2mm] 
  & = &\, \sum_{\vec{z}\in\Lambda_z} \, 
  U^{ij}_{\mu}(g;\vecz)\, \exp{\left[ - \frac{2 \pi i}{m\,N}
  \left(\veckp \cdots \vec{z} \right)\right]}
\label{eq:UFourier2} \\[2mm]
& = &\, 
{\widetilde U}^{ij}_{\mu}\!\left(l;\,\frac{\veckpp}{m}\right) \, 
\sum_{\vec{y}\in\Lambda_y} \,
\exp{\biggl[ \, - \frac{2 \pi i}{m} \, \sum_{\nu = 1}^d \,
 \left( \, k_{\nu}^{\mk \prime} + n_{\nu}^{j} - n_{\nu}^{i}
    \, \right) y_{\nu} \, \biggr]}
\label{eq:kvaluesS} \\[2mm]
 & = &\, 
{\widetilde U}^{ij}_{\mu}\!\left(l;\,\frac{\veckpp}{m}\right)\, 
\prod_{\nu = 1}^{d} \, \Biggl\{\,
   \sum_{y_{\nu} = 0}^{m-1} \, \exp{\biggl[ \, - \frac{2 \pi i}{m} \,
  \left( \, k_{\nu}^{\mk \prime} + n_{\nu}^{j} - n_{\nu}^{i}
     \, \right) y_{\nu} \, \biggr]} \, \Biggr\}
       \; , \mbox{\phantom{oioi}}
\label{eq:kvalues}
\end{eqnarray}
where we used eqs.\ (\ref{eq:UFourier}) and (\ref{eq:Ugzij}).
We also introduced the coefficients of the
Fourier transform ${\widetilde U}_{\mu}(l;\veckpp/m)$ of the matrix
$U_{\mu}(l;\vecx)$ on the $\Lambda_x$ lattice, see eq.\ 
(\ref{eq:UhxFourier}), given by
\begin{eqnarray}
{\widetilde U}^{ij}_{\mu}\!\left(l;\,\frac{\veckpp}{m}\right) & = & 
\sum_{\vec{x}\in\Lambda_x} \, U^{ij}_{\mu}(l;\vecx) \,
   \exp{\left[ - \frac{2 \pi i}{m\,N} \left(\veckp
   \cdots \vec{x} \right)\right]} \nonumber \\[2mm]
&=& \sum_{\vec{x}\in\Lambda_x} \, U^{ij}_{\mu}(h;\vecx) \,
\exp{\Biggl\{ - \frac{2 \pi i}{m\,N} \, \Biggl[\, \sum_{\nu = 1}^d \,
   \left( \, k_{\nu}^{\mk \prime} + n_{\nu}^{j} - n_{\nu}^{i}
    \, \right) \, x_{\nu} \, + n_{\mu}^{j} \, \Biggr]
          \, \Biggr\}} \; ,\quad
\label{eq:kvaluesx}
\end{eqnarray}
where we make use of the expression (\ref{eq:Ulzij}).
Thus, from eqs.\ (\ref{eq:kvalues}) and (\ref{eq:deltakappa}), we find that
${\widetilde U}^{ij}_{\mu}(g;\veckpp)$ is zero unless the quantity 
$k_{\nu}^{\mk \prime} + n_{\nu}^{j} - n_{\nu}^{i}$ is a multiple
of $m$, for every direction $\nu$, and in this case we have
\begin{equation}
{\widetilde U}^{ij}_{\mu}(g;\veckpp) \; = \; m^d \,\,
{\widetilde U}^{ij}_{\mu}\!\left(l;\,\frac{\veckpp}{m}\right)\; .
\label{eq:Utildel}
\end{equation}

In order to better understand the above result, we note that
the integers $k_{\nu}^{\mk \prime}$
can be written as 
\begin{equation}
k_{\nu}^{\mk \prime} \,=\, k_{\nu} + K_{\nu} \,m\,,
\end{equation}
where $k_{\nu} \in [0, m-1]$ and $K_{\nu} \in {\cal Z}$.
This is the decomposition we choose for representing the wave numbers of
the Fourier momenta on the extended lattice for the gauge-fixing problem, 
as explained at the beginning of section \ref{sec:Bloch-theorem}, and it 
is completely analogous to the one introduced for the crystalline-solid 
problem in section \ref{sec:Bloch-theorem-solid}. In particular,
for $\veckpp$ in the first Brillouin zone, corresponding to
$k_{\nu}^{\mk \prime} \in [0,mN\!-\!1]$, we have $K_{\nu}$
in the interval $[0, N\!-\!1]$.\footnote{One should also note that, 
if instead of the
nonsymmetric interval $[0, mN\!-\!1]$ one contemplates the symmetric
interval $k_{\nu}^{\mk \prime} \in [-(mN/2),(mN/2)-1]$ for $m N$
even (see footnote \ref{foot:symmetrick} for the general
case), this decomposition applies with $k_{\nu} \in [-(m/2),
(m/2)-1]$ and $K_{\nu} \in [-(N\!-\!1)/2,(N\!-\!1)/2]$, at least for $m$
even and $N$ odd, and with slightly different formulae for $m$
odd and/or $N$ even.
Thus, the use of the nonsymmetric interval (around the origin)
makes our notation much simpler and straightforward.
\label{foot:k-decomp}}
Indeed, this implies that the vector with components $2 \pi
k_{\nu}^{\mk \prime}/ (m N)$ becomes the sum of two terms, 
$2 \pi k_{\nu} / (m N)$ and $2 \pi K_{\nu}/N$, with the latter
one --- corresponding to $2 \pi \vec{K} /N$ (with $K_{\nu}=0,1,\ldots,
N\!-\!1$) --- belonging to the reciprocal lattice, since $\exp{(2 \pi i
\,\vec{K} \cdots \vecR / N)}=1$ for any translation vector $\vec{R}
=N \vec{y} = N \sum_{\mu=1}^{d} y_{\mu} \hat{e}_{\mu}$.
At the same time, the former one --- i.e.\ $2 \pi \vec{k} /(m N)$ 
(with $k_{\nu}=0,1,\ldots,m-1$) --- is generated
by the translation operator ${\mathcal T}$.
In fact, as already noted in section \ref{sec:proof}, the
coefficients $g^{{\textstyle\mathstrut}ij}(\vecz)$ of $g(\vecz)$
in the ${\bf W}^{ij}$ basis get
multiplied by the phase $\exp{\left(2 \pi i n^{i}_{\mu}/m\right)}$
under a translation by
$\vec{R}=N\hat{e}_{\mu}$, see eq.\ (\ref{eq:TNg-components}),
in agreement with the above observation
if we identify $k_{\mu}$ with $n^{i}_{\mu}$.

The same observation applies to the integers\footnote{Here,
we suppose that the integers $n_{\nu}^{j}$ and $n_{\nu}^{i}$
have been fixed, either by the numerical minimization of ${\cal
E}_{\scriptU,\scriptT}[h]$ or set a priori (as in the
case of fixed matrices $\Theta_{\mu}$).}
$n_{\nu}^{j}$ and $n_{\nu}^{i}$, so that we can write 
\begin{equation}
n_{\nu}^{j} \,\equiv\, \bar{n}_{\nu}^j+ m\,\widetilde{n}_{\nu}^j\,,
\label{eq:ndecomp}
\end{equation}
with $\bar{n}_{\mu}^j \in [0, m-1]$ and 
$\widetilde{n}_{\nu}^j \in {\cal Z}$ (and similarly for $n_{\nu}^{i}$).
This implies that the quantity
\begin{equation}
 \chi_\nu\;\equiv\; k_{\nu}\,+\,\bar{n}_{\nu}^j\,-\, \bar{n}_{\nu}^i
\;= \!\!\!\!\mod{\!\!\left(\,k_{\nu}^{\mk \prime},\,m\right)}
\,+\!\!\!\!\mod{\!\!\left(n_{\nu}^{j}\,-\,n_{\nu}^{i},\,m\right)} 
    \; ,
      \label{eq:mod-m}
\end{equation}
where the difference $\bar{n}_{\nu}^j-\bar{n}_{\nu}^i$ 
is a fixed integer in the interval $[-m+1, m-1]$, must be an integer multiple
of $m$, in order to produce a nonzero value in eq.\ (\ref{eq:kvalues}).
Therefore, since $k_{\nu}$ is non-negative and smaller than $m$, we may
have\footnote{Of course, the values of $k_{\nu}$
and $\chi_{\nu}$ also depend on the (considered) indices $i,j$.
Here, however, in order to simplify the notation, we do not
make this dependence explicit.
More specifically, we could define
\begin{equation}
\chi_{\nu} \,=\, \frac{\sgn(\bar{n}_{\nu}^j - \bar{n}_{\nu}^i)\,
\left[1+\sgn(\bar{n}_{\nu}^j - \bar{n}_{\nu}^i)\right]}{2}\;m\,,
\label{eq:defdeltaofnbar}
\end{equation}
after the phases $n_\nu^j$ have been chosen, and then pick $k_\nu$ given by
eq.\ (\ref{eq:mod-m}) for every $\nu$, in order to obtain nonzero
coefficients ${\widetilde U}^{ij}_{\mu}(g;\veckpp/m)$
in eq.\ (\ref{eq:kvalues}).
In the above expression we indicate with $\sgn(x)$ the sign function, which has values $\pm 1$ or zero
according to whether $x \lesseqgtr 0$.
\label{foot:deltanu}}
\begin{equation}
\chi_{\nu}\,=\, \begin{cases}
0 & \quad\qquad \text{if $\,\bar{n}_{\nu}^j-\bar{n}_{\nu}^i \leq 0$}\,, 
\\[2mm]
m & \quad\qquad \text{if $\,\bar{n}_{\nu}^j-\bar{n}_{\nu}^i$ is positive}\,.
\end{cases}
\label{eq:deltanu}
\end{equation}
Clearly, in both cases there is only one value of 
$k_{\nu}=\chi_{\nu}-(\bar{n}_{\nu}^j - \bar{n}_{\nu}^i)$
that makes $k_{\nu}^{\mk \prime} + n_{\nu}^{j} - n_{\nu}^{i}$ an 
integer multiple of $m$, i.e.\ such that
\begin{equation}
  k_{\nu}^{\mk \prime}\,+\,
    n_{\nu}^{j}\,-\,n_{\nu}^{i} \,=\, m\,\left(\,
    K_{\nu}\,+\,
         \frac{\chi_{\nu}}{m} \,+\,
        \widetilde{n}_{\nu}^{j}\,-\,
        \widetilde{n}_{\nu}^{i} 
     \,\right) \; ,
     \label{eq:kprimeproptom}
\end{equation}
with $\chi_{\nu}/m$ equal to $0$ or $1$.
It is also evident that, for any direction $\nu$, this result does not
depend on the value of $K_{\nu}$ and we have, for any given vector 
$\vec{K}$, a set of nonzero coefficients.
In this sense, for the purpose of determining which coefficients
${\widetilde U}^{ij}_{\mu}(g;\veckpp)$ are nonzero,
see eq.\ (\ref{eq:kvalues}), we can think of $\chi_\nu$ as a ``function''
of $\bar{n}_{\nu}^j - \bar{n}_{\nu}^i$, as detailed above (see also
footnote \ref{foot:deltanu}), in such a
way that momenta $\veckpp=\veck + m\vecK$ corresponding to nonzero 
coefficients will have general $\vecK$ and specific combinations
for $\veck$, determined from (\ref{eq:mod-m}).
Thus, if we define
\begin{equation}
{\widetilde U}^{ij}_{\mu}\!\left(l;\,\frac{\veckpp}{m}\right) \,\equiv\,
{\widetilde U}^{ij}_{\mu} (l;\vec{k}, \vecK)\; ,
\end{equation}
we can collect these nonzero coefficients --- with different values
of $\vec{k}$ --- in families indexed by the vectors $\vec{K}$.
Finally, when the relation (\ref{eq:kprimeproptom}) is satisfied
(for any direction $\nu$ --- with a suitable choice for $k_{\nu}^{\mk
\prime}$ --- and with $\chi_{\nu}/m=0,1$) we can write, see eq.\ 
(\ref{eq:kvaluesx}),
\begin{equation}
{\widetilde U}^{ij}_{\mu}\!\left(l;\,\frac{\veckpp}{m}\right) \,=\,
  \exp{\!\Biggl(\!- \frac{2 \pi i \,n_{\mu}^{j}}{m\,N} \!\Biggr)} 
\sum_{\vec{x}\in\Lambda_x}
U^{ij}_{\mu}(h;\vecx) \, \exp{\!\Biggl[ - \frac{2 \pi i}{N} \,
         \sum_{\nu = 1}^d \!\left( K_{\nu}\,+\,
         \frac{\chi_{\nu}}{m} \,+\,
        \widetilde{n}_{\nu}^{j}\,-\,
        \widetilde{n}_{\nu}^{i}
         \,\right)\! x_{\nu} \Biggr]} .
    \label{eq:Ugk}
\end{equation}

Thus, considering the above result and eq.\ (\ref{eq:Utildel}), we see that,
if the Fourier transform ${\widetilde U}_{\mu}(g;\veckpp)$ of the link 
variables on the extended lattice $\Lambda_z$, evaluated for the 
wave-number vector $\veckp$, is nonzero, i.e.\ if eq.\ (\ref{eq:kprimeproptom})
is verified, then its evaluation is always reduced to a Fourier transform on the
original lattice $\Lambda_x$ for a modified wave-number vector,
with components $K_{\nu}+ \chi_{\nu}/m + \widetilde{n}_{\nu}^{j}
- \widetilde{n}_{\nu}^{i}$. 
It is important to stress again that --- while we can choose $\vec{K}$
freely --- the vector $\vec{\chi}$ depends on the considered
indices $i$ and $j$ of the coefficients.


\subsection{The diagonal elements}
\label{sec:Umu:i=j}

The results obtained in the previous section greatly simplify
when\footnote{Here, we call {\em diagonal} the coefficients
with $i=j$ --- when using the basis $\{ {\bf W}^{ij} \}$ ---
even though these coefficients do not necessarily contribute
to the diagonal elements of the corresponding matrix, given
that $( {\bf W}^{jj} )_{lm} = v_{jl}^{*} \,v_{jm}$, see eq.\
(\ref{eq:lambdaijhl}).
On the other hand, all entries of the matrix ${\bf M}^{jj}=
v\,{\bf W}^{jj}\,v^{\dagger} =
\hat{e}_j\,\hat{e}_j^{\dagger}$ are null with the exception of
the diagonal entry with indices $jj$ (which is equal to one).}
$i=j$, i.e.\ when $n_{\nu}^{j} -n_{\nu}^{i} = 0$, so that
the coefficients ${\widetilde U}^{jj}_{\mu}(g;\veckpp)$ are nonzero for,
see eq.\ (\ref{eq:mod-m}),
\begin{equation}
\chi_\nu \,=\, k_\nu \,=\, 0 \, ,
\end{equation}
yielding $k_{\nu}^{\prime} = 0,\, m,\, 2\mk m,\, \ldots,\, (N\!-\!1)\, m = K_{\nu}\, m$.
Then, we find, see eq.\ (\ref{eq:Ugk}),
\begin{equation}
{\widetilde U}^{jj}_{\mu}\!\left(l;\,\frac{\veckpp}{m}\right) =
\exp{\!\Biggl(\!\!-\frac{2 \pi i \,n_{\mu}^{j}}{m\,N}\Biggr)}
\!\!\sum_{\vec{x}\in\Lambda_x} \!\!
U^{jj}_{\mu}(h;\vecx) \,\exp{\!\Biggl(\!\!- \frac{2 \pi i}{N}
   \vecK\cdots\vecx\Biggr)} 
 = \exp{\!\Biggl(\!\!-\frac{2 \pi i \,n_{\mu}^{j}}{m\,N}\Biggr)}
   {\widetilde U}^{jj}_{\mu}(h;\vecK\!\,) ,
\label{eq:Ugtildeij}
\end{equation}
where
\begin{equation}
{\widetilde U}^{jj}_{\mu}(h;\vecK) \,\equiv\,
 \sum_{\vec{x} \in \Lambda_x} \, U^{jj}_{\mu}(h;\vecx) \,
       \exp{\left[ - \frac{2 \pi i}{N} \left(\vec{K}
       \cdots \vec{x}\right)\right]}
\label{eq:UhFourier}
\end{equation}
is the usual Fourier transform\footnote{We stress that this
is the result expected from condensed-matter physics, where
the Fourier transform of the periodic potential $U(\vecr)$
is nonzero only when considering wave-number vectors on the
reciprocal lattice (see, e.g., the second proof of Bloch's
theorem in ref.\ \cite{AM}).}
(on the original lattice $\Lambda_x$) of $U^{jj}_{\mu}(h;\vecx)$,
see eq.\ (\ref{eq:UhxFourier}).
At the same time, the components of the lattice momenta are given by
\begin{equation}
p_{\nu}(\veckpp) \, \equiv \,
  2 \, \sin{\left(\frac{\pi \, k_{\nu}^{\prime}}{m N}\right)} \, = \,
  2 \, \sin{\left(\frac{\pi \, K_{\nu}}{N}\right)} \; ,
  \label{eq:pnukrpimr=k}
\end{equation}
i.e.\ they coincide exactly\footnote{On the other hand, this result
applies only approximately when considering a generic coefficient
${\widetilde U}^{ij}_{\mu}(g;\veckpp)$ for which $k_{\nu} \neq 0$.
As a matter of fact, if $\pi k_{\nu} \ll m N$ (recall that
$k_{\nu} \in [0,m-1]$), we have
\begin{eqnarray}
p_{\nu}(\veckpp) \, &\equiv& \,
  2 \, \sin{\left(\frac{\pi \, k_{\nu}^{\prime}}{m N}\right)} \, = \,
  2 \, \sin{\left[\frac{\pi \, \left(k_{\nu} +
           K_{\nu} m\right)}{m N}\right]}
           \, \approx \,  
  2 \, \sin{\left(\frac{\pi \, K_{\nu}}{N}\right)}\,.
\end{eqnarray}
}
with the values allowed on the original
$\Lambda_x$ lattice, see eq.\ (\ref{eq:pmu}) with $k_{\nu}$ substituted
by $K_{\nu}$.

One should also note that the case $i=j$ is the only one
relevant for the evaluation of the minimizing functional --- see (in this 
order) eqs.\ (\ref{eq:EUg-Ul}), (\ref{eq:Amatrix}),
(\ref{eq:lambdatrace}),
(\ref{eq:kvaluesx})
and (\ref{eq:UhFourier}) --- since
\begin{eqnarray}
\hskip -6mm
{\cal E}_{\scriptU}[l] \,&=&\, 1-\frac{\Re \, \Tr}{N_c\,d\,V}
\sum_{\mu,\:\!\vec{x}}
 U_{\mu}(l;\vecx) \,=\, 1- \frac{\Re \, \Tr}{N_c\,d\,V} 
\sum_{\mu,\:\!\vec{x}}
\,\sum_{i,j=1}^{N_c}\,
         U^{ij}_{\mu}(l;\vecx) \, {\bf W}^{ij}\;\; \nonumber \\[2mm]
&=&\, 1- 
\sum_{\mu,\:\!\vec{x}}
\,\sum_{j=1}^{N_c} \frac{\Re \,U^{jj}_{\mu}(l;\vecx)}{N_c\,d\,V} 
=\, 1- \!\frac{\Re}{N_c\,d\,V} \sum_{\mu} \sum_{j=1}^{N_c}
\exp{\!\biggl(\!-\frac{2 \pi i \,n_{\mu}^{j}}{m\,N} \biggr)} 
{\widetilde U}^{jj}_{\mu}(h;\veczero) ,\;\;\;
\end{eqnarray}
where $\mu=1,\ldots,d\,$ and $\vec{x}\in\Lambda_x$.


\subsection{Fixed wave-number vectors}
\label{sec:fixedk}

The above results clarify for which values of $\veckp$
a given coefficient ${\widetilde U}^{ij}_{\mu}(g;\veckpp)$ is nonzero.
Now we can invert the question and try to understand which coefficients
are nonzero for a given (chosen) momentum $\veckp$.
Indeed, 
note that, in a numerical evaluation of the gluon propagator using lattice
simulations, the considered momenta $\veckp$ are usually fixed a priori.
The integers $n^i_\nu$, on the other hand, will be selected to minimize
the functional ${\cal E}_U[l]$ and we can analyze which combinations
are expected to produce a nonzero value for the propagator.
For example, if (at least) one component $k_{\nu}^{\mk\prime}$
of $\veckp$ is equal to zero, it is evident that only the
diagonal elements (i.e., $i=j$) are usually different from
zero, given that the factor, see eq.\ (\ref{eq:kvalues}),
\begin{equation}
  \sum_{y_{\nu}=0}^{m-1}
\exp{\biggl[ \, - \frac{2 \pi i}{m} \, 
 \left( \, n_{\nu}^{j} - n_{\nu}^{i}
    \, \right) y_{\nu} \, \biggr]}
\,=\,
  \sum_{y_{\nu}=0}^{m-1}
\exp{\biggl[ \, - \frac{2 \pi i}{m} \, 
 \left( \, \bar{n}_{\nu}^{j} - \bar{n}_{\nu}^{i}
    \, \right) y_{\nu} \, \biggr]}
\end{equation}
is always equal to zero for $i \neq j$, unless\footnote{Recall
that $\bar{n}_{\nu}^{j}$ and $\bar{n}_{\nu}^{i}$
take values $0,1,\ldots,m-1$, so that their difference is an
integer number in the interval $[-m+1,m-1]$.}
$\bar{n}_{\nu}^{i} = \bar{n}_{\nu}^{j}$, see again eq.\ (\ref{eq:dKronecker}).
This result is even stronger when $k_{\nu}^{\mk\prime}=0$
for more than one direction, i.e.\ it would be even more unlikely
in this case to have a nonzero coefficient when $i \neq j$.
Thus, when evaluating the zero-momentum gluon propagator, one
should recall that, except in a fortuitous event with 
$\bar{n}_{\nu}^{i} = \bar{n}_{\nu}^{j}$ for all $\nu =1,\ldots,d$,
when $i \neq j$, usually the only nonzero coefficients of
the zero-momentum link variables are the diagonal ones (i.e.\
$i=j$), given by, see eqs.\ (\ref{eq:Utildel}) and (\ref{eq:Ugtildeij})
with $\vecK = \vec{0}$,
\begin{equation}
{\widetilde U}^{jj}_{\mu}(g;\veczeropp) \, = \, m^d \,
  \exp{\biggl(- \frac{2 \pi i \,n_{\mu}^{j}}{m\,N} \, \biggr)} \,  
  {\widetilde U}^{jj}_{\mu}(h;\veczero) \; .
\end{equation}

For the same reason, if the vector $\veckpp$ has (for example) all
equal components, i.e.\
\begin{equation}
k_{\nu}^{\mk\prime} = k + K\,m
\qquad\qquad\quad \mbox{for}\;\; \nu = 1, 2, \ldots, d \; ,
\end{equation}
where $k$ and $K$ are fixed integers with values
(respectively) in $[0,m-1]$ and $[0,N\!-\!1]$, then a nondiagonal
coefficient ${\widetilde U}^{ij}_{\mu}(g;\veckpp)$
(with $i \neq j$) could be nonzero only in the unlikely event
that, for all $\nu = 1,\ldots,d$, the differences 
$\bar{n}_{\nu}^{j} - \bar{n}_{\nu}^{i}$
are either equal to $-k$ or to $m\!-\!k$, so that the
value of $\chi_{\nu}=k_{\nu}+\bar{n}_{\nu}^j-
\bar{n}_{\nu}^i=k+\bar{n}_{\nu}^j-\bar{n}_{\nu}^i$,
see eq.\ (\ref{eq:deltanu}),
is either $0$ or $m$ for
all directions $\nu$.
On the other hand, as seen in the previous section, all the diagonal
elements are (always) different from zero for $k_{\nu}^{\mk\prime}
= K\,m$ (i.e.\ $k=0$), since the factors in eq.\
(\ref{eq:kvalues}) become
\begin{equation}
  \sum_{y_{\nu}=0}^{m-1}
\exp{\biggl( \, - \frac{2 \pi i}{m} \, 
    k_{\nu}^{\mk \prime} \, y_{\nu} \, \biggr)}
\,=\,
  \sum_{y_{\nu}=0}^{m-1}
\exp{\biggl( \, - 2 \pi i \, 
    K_{\nu} \, y_{\nu} \, \biggr)}\,=\, m
    \; .
\end{equation}


\subsection{Gauge field in momentum space}
\label{sec:gluonNP}

We can now apply the outcomes obtained in the previous section
to the evaluation of the gauge field, given in terms of the 
gauge-transformed gauge link, see eqs.\ (\ref{eq:Atrac}) and
(\ref{eq:Altraceless}), as
\begin{eqnarray}
A_{\mu}(g;\vecz)\,&\equiv&\,
 \frac{1}{2 \,i}\, \left[ \,
    U_{\mu}(g;\vecz) - U_{\mu}^{\dagger}(g;\vecz)
    \, \right]_{\rm traceless} \nonumber \\[2mm]
&=&\,\frac{1}{2 \,i}\, \left[ \,
    U_{\mu}(g;\vecz) - U_{\mu}^{\dagger}(g;\vecz)
    \, \right]\,-\,\1\,\frac{\Tr}{N_c}\,
\left[ \,
    U_{\mu}(g;\vecz) - U_{\mu}^{\dagger}(g;\vecz)
    \, \right]
\label{eq:Agz}
\end{eqnarray}
--- or of its coefficients $A^{ij}_{\mu}(g;\vecz)$ --- in momentum space.
As a first step, we need to consider how eqs.\
	(\ref{eq:UFourier1})--(\ref{eq:kvaluesx})
get modified when evaluating the Fourier transform of the 
coefficients,\footnote{Note that $U_{\mu}(g;\vecz)$ is a unitary matrix,
which is written here in terms of the basis $\{ {\bf W}^{ij} \}$.}
see for example eq.\ (\ref{eq:gencoeff}),
\begin{equation}
\left[ U_{\mu}^\dagger(g;\vecz) \right]^{ij} \,=\,
w_i^{\dagger} \,U_{\mu}^\dagger(g;\vecz)\, w_j \,=\,
\left[ w_j^{\dagger}\, U_{\mu}(g;\vecz) \,w_i \right]^{*}
\,=\, \Big[{U^{ji}_{\mu}(g;\vecz)}\Big]^{*} \; .
\end{equation}
In particular, using eq.\ (\ref{eq:Ugzij}) we can write
\begin{equation}
\left[ U_{\mu}^\dagger(g;\vecz) \right]^{ij} \,=\,
    \exp{\biggl[ - \frac{2 \pi i}{m} \, \sum_{\nu = 1}^d \,
      \left( \, n_{\nu}^{j} - n_{\nu}^{i}
    \, \right) y_{\nu} \biggr]} \,
\Big[ U_{\mu}^{ji}(l;\vecx) \Big]^{*} \; ,
\label{eq:Udagij}
\end{equation}
with, see eq.\ (\ref{eq:Ulzij}),
\begin{equation}
\Big[ U^{ji}_{\mu}(l;\vecx) \Big]^{*} \, = \,
    \exp{\biggl\{ - \frac{2 \pi i}{m\,N} \, \Bigl[\,
       \sum_{\nu = 1}^d \, \left( \, n_{\nu}^{j} - n_{\nu}^{i}
    \, \right) \, x_{\nu} \, - n_{\mu}^{i}\,\Bigr] \, \biggr\}} \,
   \Big[ U^{ji}_{\mu}(h;\vecx) \Big]^{*} \, \; .
\label{eq:Ustar}
\end{equation}
Then, the difference
\begin{equation}
\left[ \, U_{\mu}(g;\vecz) \,-\,
 U_{\mu}^{\dagger}(g;\vecz) \, \right]^{ij}
\end{equation}
is simply given, see eqs.\ (\ref{eq:Ugzij}) and (\ref{eq:Udagij}), by
\begin{equation}
    \exp{\biggl[ - \frac{2 \pi i}{m} \, \sum_{\nu = 1}^d \,
      \left( \, n_{\nu}^{j} - n_{\nu}^{i}\, \right) \, y_{\nu} \biggr]} 
\, \left[ \, U^{ij}_{\mu}(l;\vecx) \,-\,
{U^{ji}_{\mu}(l;\vecx)}^{*} \, \right] \; .
\label{eq:A-Udiff}
\end{equation}
Thus, if we write, in analogy with eq.\ (\ref{eq:AFourier}),
\begin{equation}
\!\!\!\!
{\widetilde A}_{\mu}(g;\veckpp) \, \equiv \,
  \sum_{\vec{z}\in\Lambda_z} \, 
A_{\mu}(g;\vecz) \, \exp{\left[ - \frac{2 \pi i}{m\,N}
  \left(\veckp \cdots \vec{z} \,+\, \frac{k_{\mu}^{
            \mk\prime}}{2} \, \right)\right]} \; ,
\label{eq:Amugkpdef}
\end{equation}
and similarly for the coefficients ${\widetilde A}^{ij}_{\mu}(g;\veckpp)$,
we find that, see eqs.\
(\ref{eq:Agz}), (\ref{eq:gencoeff}),
(\ref{eq:kvaluesS})
and (\ref{eq:A-Udiff}),
\begin{eqnarray}
{\widetilde A}^{ij}_{\mu}(g;\veckpp) &=&
\sum_{\vec{z}\in\Lambda_z} \, 
\frac{ \left[U_{\mu}(g;\vecz)-U_{\mu}^{\dagger}(g;\vecz)\right]^{ij}
\!\!\!-\,\delta^{ij}
\frac{\Tr}{N_c} \left[ U_{\mu}(g;\vecz) - U_{\mu}^{\dagger}(g;\vecz) \right]}{2i}
\; e^{-\frac{2 \pi i}{m\,N}
  \left(\veckp \cdots \vec{z} \,+\, \frac{k_{\mu}^{
            \mk\prime}}{2} \, \right)}
\nonumber \\[2mm]
 &\propto & \sum_{\vec{y}\in\Lambda_y} \,
\exp{\biggl[ - \frac{2 \pi i}{m} \, \sum_{\nu = 1}^d \,
   \left( \, k_{\nu}^{\mk\prime} + n_{\nu}^{j} - n_{\nu}^{i}
    \, \right) \, y_{\nu} \biggr]} \; ,
\label{eq:Amugkp}
\end{eqnarray}
which is again null, see eq.\ (\ref{eq:kprimeproptom}),
unless the relation 
\begin{equation}
  k_{\nu}^{\mk \prime}\,+\,
    n_{\nu}^{j}\,-\,n_{\nu}^{i} \,=\, m\,\left(\, K_{\nu}\,+\,
         \frac{\chi_{\nu}}{m} \,+\,
        \widetilde{n}_{\nu}^{j}\,-\,
        \widetilde{n}_{\nu}^{i} 
     \,\right)
\label{eq:kprimeproptom2}
\end{equation}
is verified (for every direction $\nu$) with $\chi_{\nu}/m=0,1$
determined
by $\bar{n}_{\nu}^j-\bar{n}_{\nu}^i$, see eq.\ (\ref{eq:defdeltaofnbar}).
In this case, the r.h.s.\ in eq.\ (\ref{eq:Amugkp}) is equal to $m^d$.
Here we used the fact that the trace term is multiplied by the identity,
see eq.\ (\ref{eq:Agz}),
which has coefficients $\delta^{ij}$.
Also note that
we are writing the gauge field in
momentum space as a linear combination of the ($N_c \times N_c$)
matrices ${\bf W}^{ij} = w_i\, w_j^{\dagger}$ (with $i,j=1,\ldots,N_c$)
and that $\Tr\left(w_i\,w_j^{\dagger}\right)=\delta^{ij}$,
see eq.\ (\ref{eq:lambdatrace}).
Also, as detailed
below, the trace term does not depend on $\vecy$, in agreement with the overall
exponential factor in (\ref{eq:Amugkp}).

As for the second factor in eq.\ (\ref{eq:A-Udiff}), it is equal to
\begin{equation}
\exp{\!\Biggl[ - \frac{2 \pi i}{m\,N}
\sum_{\nu = 1}^d \left(n_{\nu}^{j} - n_{\nu}^{i}\right) x_{\nu} \Biggr]}
\,\Bigg[\,e^{-\frac{2 \pi i n_{\mu}^{j}}{m\,N}}\, U^{ij}_{\mu}(h;\vecx)
  \,-\,e^{\frac{2 \pi i \,n_{\mu}^{i}}{m\,N}}\,
    {U^{ji}_{\mu}(h;\vecx)}^{*}\, \Bigg]\, ,
      \label{eq:Azofx}
\end{equation}
where we used eqs.\ (\ref{eq:Ulzij}) and (\ref{eq:Ustar}).
At the same time, for the trace term in (\ref{eq:Agz}) and
(\ref{eq:Amugkp}) we have, 
see eqs.\ (\ref{eq:Uofg2}) and (\ref{eq:Uofl}),
\begin{eqnarray}
\frac{1}{2i}\Tr \left[ U_{\mu}(g;\vecz) - U_{\mu}^{\dagger}(g;\vecz) \right]
\,&=&\, \frac{1}{2i}
\Tr \left[ U_{\mu}(l;\vecx) - U_{\mu}^{\dagger}(l;\vecx) \right]
\nonumber \\[2mm]
& = &
\frac{1}{2i}
\Tr \left[ U_{\mu}(h;\vecx)\,e^{-i\frac{\Theta_{\mu}}{N}}\!-
     e^{i\frac{\Theta_{\mu}}{N}} U_{\mu}^{\dagger}(h;\vecx)\right] \; .
\end{eqnarray}
Hence, noting again $\Tr\left( {\bf W}^{ij} \right)=\delta^{ij}$, the above trace can be written as
\begin{equation}
\sum_{j=1}^{N_c} \frac{1}{2i} \left[
 U^{jj}_{\mu}(h;\vecx)\,e^{- \frac{2 \pi i n_{\mu}^{j}}{m N}} -
 e^{\frac{2 \pi i n_{\mu}^{j}}{m N}}\,{U^{jj}_{\mu}(h;\vecx)}^{*}
           \right]
\,=\,\sum_{j=1}^{N_c} \Im \left[U^{jj}_{\mu}(h;\vecx)
\exp{\!\left(\!\frac{- 2 \pi i \,n_{\mu}^{j}}{m N}\!\right)}\!\right]
\label{eq:trAlij}
\; ,
\end{equation}
which can also be obtained by summing eq.\ (\ref{eq:Azofx}) for $j=i$
and dividing the result by $2\,i$.
This yields
\begin{eqnarray}
A^{ij}_{\mu}(l;\vecx)\,&=&\, w_i^{\dagger}\,A_{\mu}(l;\vecx)\, w_j
\,=\, w_i^{\dagger}\,\frac{1}{2 \,i}\, \left[ \,
  U_{\mu}(l;\vecx) \, -\, U_{\mu}^{\dagger}(l;\vecx) \,
     \right]_{\rm traceless} \, w_j \nonumber \\[2mm]
 &=&\,\exp\!{\left[- \frac{2 \pi i}{m N} \sum_{\nu = 1}^d \,
                   (n_{\nu}^j\,-\,n_{\nu}^i) \, x_{\nu} \right]}\,
\Biggl\{\,\frac{1}{2 i} \,\Bigl[\,U^{ij}_{\mu}(h;\vecx)\,e^{-\frac{2\pi i n_{\mu}^{j}}{m N}} 
\nonumber \\[2mm]
&&\qquad\quad -\, e^{\frac{2\pi i n_{\mu}^{i}}{m N}} {U^{ji}_{\mu}(h;\vecx)}^{*}\,\Bigr]\,
-\, \frac{\delta^{ij}}{N_c} \,\sum_{l=1}^{N_c}\Im\! \left[
 U^{ll}_{\mu}(h;\vecx)\, e^{\frac{- 2 \pi i n_{\mu}^{l}}{m N}}\!\right] \; ,
\Biggr\}
\label{eq:Alx} 
\end{eqnarray}
where we used eqs.\ (\ref{eq:Azofx}) and (\ref{eq:trAlij}).
Then,
by recalling eq.\ (\ref{eq:Agzeta2}), it is evident that the coefficient of
proportionality in eq.\ (\ref{eq:Amugkp}) is given by
\begin{equation}
{\widetilde A}^{ij}_{\mu}\!\left(l;\,\frac{\veckpp}{m}\right)
\,=\, \sum_{\vec{x}\in\Lambda_x} \,
\exp{\left[-\frac{2 \pi i}{mN}\; \veckpp\cdots
\left(\vec{x}+\frac{\hat{e}_\mu}{2}\right)\right]}
\,A^{ij}_{\mu}(l;\vecx) \; ,
\label{eq:Aijltilde}
\end{equation}
which is the usual small-lattice definition of 
the Fourier transform of $A^{ij}_{\mu}(l;\vecx)$, 
i.e.\ eq.\ (\ref{eq:AFourier}), for the wave-number vector $\veckpp/m$.
By collecting the above results we end up with the expression\footnote{Of course,
once the nonzero coefficients ${\widetilde A}^{ij}_{\mu}(g;\veckpp)$ have been
evaluated, one can also obtain the color components ${\widetilde A}^{b}_{\mu}(g;\veckpp)$
with respect to the generators $\{ t^b \}$ using the relation
\begin{equation}
{\widetilde A}^{b}_{\mu}(g;\veckpp) \,=\, \sum_{i,j=1}^{N_c} \,
{\widetilde A}^{ij}_{\mu}(g;\veckpp)\,\frac{\Tr}{2}
\left(\,t^b\, {\bf W}^{ij}\,\right) \; .
\end{equation}
}
\begin{eqnarray}
\label{eq:Alk0}
{\widetilde A}^{ij}_{\mu}(g;\veckpp)
 \, &=& \, m^d\,{\widetilde A}^{ij}_{\mu}\!\left(l;\,\frac{\veckpp}{m}\right)
 \,=\, m^d\, \sum_{\vec{x}\in\Lambda_x} \!e^{-\frac{2 \pi i}{N}
\left(\frac{\veckpp}{m}\right)
   \,\cdots\,
\left(\vec{x}+\frac{\hat{e}_\mu}{2}\right)} \,
A^{ij}_{\mu}(l;\vecx) \; .
\label{eq:Alk}
\end{eqnarray}
Therefore, beside the factor $m^d$ and the modified wave-number
vector $\veckpp/m$ with components, see eqs.\ (\ref{eq:kprimeproptom2})
and (\ref{eq:ndecomp}),
\begin{equation}
  k_{\mu}^{\mk \prime} \,=\, m\,\left(\, K_{\mu}\,+\,
         \frac{\chi_{\mu}}{m} \,+\,
        \widetilde{n}_{\mu}^{j}\,-\,
        \widetilde{n}_{\mu}^{i} \right) \,-\,
    n_{\mu}^{j}\,+\,n_{\mu}^{i} \,=\, m\,\left(\, K_{\mu}\,+\,
         \frac{\chi_{\mu}}{m} \, \right) \,-\,
        \bar{n}_{\mu}^{j}\,+\,
        \bar{n}_{\mu}^{i} \; ,
\label{eq:kprimevalue}
\end{equation}
the only difference --- with
respect to the computation on the original lattice $\Lambda_x$, see eqs.\
(\ref{eq:AFourier}) and (\ref{eq:defAgenIm}) --- is represented by
the phase factors in eq.\ (\ref{eq:Alx}), which are a direct
consequence of the dependence of the gauge transformation on the
$\Theta_{\mu}$ matrices.

Finally, as already stressed in section \ref{sec:lambda-basis},
see comment below eq.\ (\ref{eq:lambda-ortho}), the $N_c^2$
coefficients entering the linear combination of the 
${\bf W}^{ij}=w_j\,w_i^{\dagger}$ matrices are not all independent,
when considering an element of the $su(N_c)$ Lie algebra.
Moreover, with our convention, the gauge field is Hermitian.
Then, if we write
\begin{equation}
A_{\mu}(l;\vecx)
  \, = \, \sum_{i,j=1}^{N_c}\, {\bf W}^{ij} A^{ij}_{\mu}(l;\vecx)
\end{equation}
we obtain, see eq.\ (\ref{eq:aij}), that the coefficients
$A^{ij}_{\mu}(l;\vecx)$ are complex numbers such that
\begin{equation}
{A^{ij}_{\mu}(l;\vecx)}^{*}\,=\, A^{ji}_{\mu}(l;\vecx) \; ,
\end{equation}
which can be verified directly from
eq.\ (\ref{eq:Alx}).
The above result gives, see eq.\ (\ref{eq:Alk}),
\begin{equation}
{{\widetilde A}^{ij}_{\mu}(g;\veckpp)}^{*}
 \,=\, m^d\,
{{\widetilde A}^{ij}_{\mu}(l;\veckpp/m)}^{*}
\,=\, m^d\,
{\widetilde A}^{ji}_{\mu}(l;-\veckpp/m)
\,=\, {\widetilde A}^{ji}_{\mu}(g;-\veckpp) \; .
\label{eq:Alijstar}
\end{equation}
At the same time, we have
\begin{eqnarray}
\left[\,{\widetilde A}_{\mu}(l;\veckpp/m) \right]^{\dagger} 
\,&=&\, \sum_{i,j=1}^{N_c}\, 
{{\widetilde A}^{ij}_{\mu}(l;\veckpp/m)}^{*} \;
{{\bf W}^{ij}}^{\dagger} \nonumber \\[2mm]
&=&\, \sum_{i,j=1}^{N_c}
{\widetilde A}^{ji}_{\mu}(l;-\veckpp/m)\; {\bf W}^{ji}
 \,=\, {\widetilde A}_{\mu}(l;-\veckpp/m)\qquad
\end{eqnarray}
and
\begin{equation}
\left[{\widetilde A}_{\mu}(g;\veckpp) \right]^{\dagger}
        \,=\, m^d\, 
  \left[ {\widetilde A}_{\mu}(l;\veckpp/m)
     \right]^{\dagger} \,=\, m^d\,
  {\widetilde A}_{\mu}(l;-\veckpp/m)
   \,=\, {\widetilde A}_{\mu}(g;-\veckpp) \; ,
\label{eq:Adagger}
\end{equation}
i.e.\ eq.\ (\ref{eq:Atilde-k}) is verified also on the
extended lattice $\Lambda_z$.


\subsection{Gluon propagator on the extended lattice}
\label{sec:gluon-extended}

In order to evaluate the gluon propagator on $\Lambda_z$,
it is convenient to start from eqs.\ (\ref{eq:D0defTr}) and
(\ref{eq:DkdefbisTr}), which now are written as
\begin{equation}
D(\veczeropp) \, = \, 
   \frac{\Tr}{2\,{\cal N}\,m^d}\,
   \sum_{\mu=1}^d \, \Bigl<
  \left[\, {\widetilde A}_{\mu}(g;\veczeropp)
  \,\right]^{2}
  \Bigr>
\label{eq:D0g}
\end{equation}
and
\begin{equation}
D(\veckpp) \,=\,
  \frac{\Tr}{2\,{\cal N}'\,m^d}\,
        \sum_{\mu = 1}^{d} \, \Bigl<
{\widetilde A}_{\mu}(g;\veckpp) \, {\widetilde A}_{\mu}(g;- \veckpp)
\Bigr> \; ,
\end{equation}
where the normalization factors ${\cal N}$ and ${\cal N}'$ have
been defined in section \ref{sec:gluonPBCs}.
At the same time, one can easily evaluate the trace after
expanding the gauge-field matrices in the basis ${\bf W}^{ij}
=w_i\,w_j^{\dagger}$, yielding
\begin{eqnarray}
\!\!\!\!\!\!
D(\veckpp) \, &=& \,
   \frac{1}{2\,{\cal N}'\,m^d} \,
        \sum_{\mu = 1}^{d} \, \sum_{i,j=1}^{N_c} \Bigl<
   {\widetilde A}^{ij}_{\mu}(g;\veckpp)\,
   {\widetilde A}^{ji}_{\mu}(g;-\veckpp) \Bigr> \nonumber \\[2mm]
    &=&\,\frac{m^d}{2\,{\cal N}'} \,
        \sum_{\mu = 1}^{d} \, \sum_{i,j=1}^{N_c} \Bigl<
   {\widetilde A}^{ij}_{\mu}(l;\veckpp/m)
   {\widetilde A}^{ji}_{\mu}(l;-\veckpp/m) \Bigr> \nonumber \\[2mm]\qquad
  &=&\,\frac{m^d}{2\,{\cal N}'} \,
        \sum_{\mu = 1}^{d} \, \sum_{i,j=1}^{N_c} \Bigl<\;
    \left|\, {\widetilde A}^{ij}_{\mu}(l;\veckpp/m)\,
    \right|^2 \,\Bigr>\; ,
\label{eq:DAl}
\end{eqnarray}
where we used (in this order) eqs.\ 
(\ref{eq:lambda-ortho}), (\ref{eq:Alk}) and (\ref{eq:Alijstar}).
However, as discussed above --- in order to be different from zero --- each
coefficient ${\widetilde A}^{ij}_{\mu}(l;\veckpp/m)$ requires a specific 
value for the wave-number vector $\vec{k}$ and, hence, for the wave-number
vector $\veckp=\vec{k}+m\vec{K}$, for a given $\vecK$,
see eqs.\ (\ref{eq:mod-m}), (\ref{eq:kprimeproptom2})
and (\ref{eq:kprimevalue}).
Conversely, for fixed $\vec{k}$, only some of the coefficients
entering the expression (\ref{eq:DAl}) contribute
to the gluon propagator.
On the other hand, for each choice of $\vec{k}$,
we have the freedom to choose among $N^d$ different
vectors $\vec{K}$.
In particular, as shown in section \ref{sec:fixedk}, if
we consider $k_{\nu}^{\mk\prime} = m\,K_{\nu}$, with
$K_{\nu}$ either equal to zero or to a fixed value $K$ in
the interval $[1,N\!-\!1]$, then (most likely) the gluon
propagator is given by
\begin{equation}
D(\veckpp) \, \approx \,
 \frac{m^d}{2\,(d-1)\,(N_c^2-1)\,V} \,
        \sum_{\mu = 1}^{d} \, \sum_{j=1}^{N_c} \Bigl<
    \left|\, {{\widetilde A}^{jj}_{\mu}(l;\vecK)}\,
    \right|^2 \Bigr> \; ,
\label{eq:Dkpz}
\end{equation}
i.e.\ only the diagonal elements contribute to it, with
a null vector $\vec{\chi}$, see again eq.\ (\ref{eq:mod-m}).
At the same time, from eq.\ (\ref{eq:pnukrpimr=k}), we also know that
the corresponding gluon propagator can be considered as a
function of the lattice momenta with components
\begin{equation}
p_{\nu}(\veckpp) \, = \, 2 \, \sin{\left(\frac{\pi \, K_\nu}{N}\right)}
\,=\,
\begin{cases}
   0\,,\ {\rm or}\,,   \\
  2 \, \sin{\left(\frac{\pi \, K}{N}\right)}\,.
\end{cases} \; 
\end{equation}
This observation is in agreement with our findings in ref.\
\cite{Cucchieri:2016qyc}, where indeed momenta $\veckp$
of the type $(k^{\mk\prime}, 0, 0 , \ldots, 0),\,
(k^{\mk\prime}, k^{\mk\prime}, 0, \ldots, 0), \, \ldots,
(k^{\mk\prime}, k^{\mk\prime}, k^{\mk\prime}, 
\ldots, k^{\mk\prime})$, with $k^{\mk\prime}=k+m K$,
have produced nonzero results only for $k=0$.
From eq.\ (\ref{eq:Dkpz}) it is also evident that, in order
to compare a result obtained on the extended lattice
$\Lambda_z$ with a result obtained on the original
lattice $\Lambda_x$, we have to consider $D(\veckpp) / m^d$,
which is again in agreement with the findings
presented in the same reference.

Similarly, for the case of zero momentum, we have
\begin{equation}
D(\veczero) \, = \, \frac{m^d}{2\,{\cal N}} \,
        \sum_{\mu = 1}^{d} \, \sum_{i,j=1}^{N_c} 
\Bigl< \,\left|\, {\widetilde A}^{ij}_{\mu}(l;\veczero)\, \right|^2 \, 
\Bigr>
\end{equation}
and each matrix element appearing on the r.h.s.\ is nonzero,
see eq.\ (\ref{eq:mod-m}),
only if $\bar{n}_{\nu}^{i} = \bar{n}_{\nu}^{j}$ (for any
$\nu=1,\ldots, d$).
Thus, also in this case, the main contribution to the gluon
propagator comes from the diagonal coefficients, i.e.\
\begin{equation}
D(\veczero) \, \approx \, \frac{m^d}{2\,d\,(N_c^2-1)\,V} \,
        \sum_{\mu = 1}^{d} \, \sum_{j=1}^{N_c} \Bigl<\,
   {{\widetilde A}^{jj}_{\mu}(l;\veczero)}^2\,\Bigr>\; ,
\label{eq:D0diag}
\end{equation}
where we used the result, see below eq.\ (\ref{eq:aij}), that
the coefficients $A^{jj}_{\mu}(l;\vecx)$ are real.
Note that the above approximation should become more and more
valid in the limit of $m \to +\infty$, given that the probability
of having $\bar{n}_{\nu}^{i} = \bar{n}_{\nu}^{j}$ is equal
to $1/m$, if we imagine that both $\bar{n}_{\nu}^{i}$ and
$\bar{n}_{\nu}^{j}$ have equal probability of taking one of
the possible values $0, 1, \ldots, m-1$.
Moreover, using eq.\ (\ref{eq:Aijltilde})
we can write
\begin{equation}
{\widetilde A}^{jj}_{\mu}(l;\veczero)
= \, \sum_{\vec{x}\in\Lambda_x} \, A^{jj}_{\mu}(l;\vecx)\,,
\label{eq:Ajjk=0}
\end{equation}
which are the $jj$ coefficients of the matrix $Q_{\mu}(l)$, defined
in eq.\ (\ref{eq:QmulofA}),
so that
\begin{equation}
{\widetilde A}^{jj}_{\mu}(l;\veczero) \,=\,
 w_j^{\dagger}\,Q_{\mu}(l)\,w_j
        \, = \, Q_{\mu}^{jj}(l) \; . 
\label{eq:Ajjtrace}
\end{equation}
Therefore, eq.\ (\ref{eq:aQzero}) [see also eq.\ (\ref{eq:zeromodeb})] implies that
all gauge-fixed configurations (on the extended lattice $\Lambda_z$,
for $m \to +\infty$) should be characterized by a gauge field
with almost null zero-mode coefficients ${\widetilde A}^{jj}_{\mu}(l;\veczero)$ and,
consequently, by a strongly suppressed zero-momentum gluon propagator
$D(\veczero)$.
This result was already proven in ref.\ \cite{Zwanziger:1993dh} for
the case of an absolute minimum of the minimizing functional
${\cal E}_{\scriptU}[g]$.
Here, we have shown that it applies also to any local minimum of
${\cal E}_{\scriptU}[g]$, in agreement with our numerical findings
in ref.\ \cite{Cucchieri:2016qyc}.
However, as already suggested
in the caption of fig.\ 1 of the
same reference,
this suppression is simply a
peculiar effect of the extended gauge transformations in the
limit of large $m$ --- as shown above --- and not a physically
significant result.
To further support this conclusion, we recall that null zero modes
for the gauge fields in minimal Landau gauge are also obtained on
a finite lattice with free boundary conditions (FBCs)
\cite{Schaden:1994in}.
In the present work, the BCs for the link variables $U_{\mu}(l;
\vecx)$ are given by eq.\ (\ref{eq:UlBCs}), i.e.\ they are not free
but they are
more general than the usual PBCs.
In particular, as $m \to + \infty$, we find that the toroidal BCs
(\ref{eq:Ul-toroidal}) applied to the coefficients of the link variables 
yield
\begin{eqnarray}
U^{ij}_{\mu}(l;\vec{x} + N \bar{e}_{\nu})\,&=&\,
   e^{\frac{2\pi i}{m} \left(n_{\mu}^{j}-n_{\mu}^{i}\right)}\,
   U^{ij}_{\mu}(l;\vecx) \,=\,
   e^{\frac{2\pi i}{m} \left(\bar{n}_{\mu}^{j}-\bar{n}_{\mu}^{i}\right)}\,
   U^{ij}_{\mu}(l;\vecx) \nonumber \\[2mm]
&\to&\,e^{2\pi i \left(\epsilon_{\mu}^{j}-\epsilon_{\mu}^{i}\right)}\,
  U^{ij}_{\mu}(l;\vecx) \; ,
\label{eq:Ul-toroidal2}
\end{eqnarray}
where the real parameters $\epsilon_{\mu}^{j}, \epsilon_{\mu}^{i} \in
[0,1)$ have already been defined in section
\ref{sec:minfty} and we used eq.\ (\ref{eq:ndecomp}).
Clearly, for each direction $\mu$ and for each coefficient (with
indices $i$ and $j$), there are --- in principle --- different
BCs, even though they are not completely independent of each
other.
Hence, the BCs considered for the gauge field $U_{\mu}(l;\vecx)$
are somewhat in between the PBCs for the gauge field $U_{\mu}(h;
\vecx)$ and the FBCs of ref.\ \cite{Schaden:1994in}, and it seems
reasonable to us that one finds the zero modes of the nonperiodic
gauge field $U_{\mu}(l;\vecx)$ to be (much) more suppressed than
those of the periodic gauge field $U_{\mu}(h;\vecx)$.


\section{Numerical simulations and conclusions}
\label{sec:numerical-PBCs}

Numerical simulations can be easily implemented using the Bloch setup
considered in this work (see also ref.\ \cite{Cucchieri:2016qyc}).
To this end, one just needs to generate a thermalized $d$-dimensional
link configuration $\{ U_{\mu}(x) \}$ with periodicity $N$, i.e.\
for a lattice volume $V = N^d$ with PBCs.
As for the minimization of the functional ${\cal E}_{\scriptU}[g] =
{\cal E}_{\scriptU,\scriptT}[h]$, defined in eqs.\
(\ref{eq:minimizing2}), (\ref{eq:Z}) and (\ref{eq:Uofh}), it can
be done recursively, using two alternating steps:
\begin{itemize}
\item[{\bf a)}] The matrices $\Theta_{\mu}$ are kept fixed as one
                updates the matrices $h(\vec{x})$ by sweeping through
                the lattice using a standard gauge-fixing algorithm
                \cite{Suman:1993mg,Cucchieri:1995pn,Cucchieri:1996jm,
                Cucchieri:2003fb,Leal:2022ojc}.
                In particular, one can again consider a single-site
                update (\ref{eq:htorh}), where the matrix $r(\vec{x})$
                should satisfy the inequality (\ref{eq:rineq})
                with, see eq.\ (\ref{eq:Eprime0}),
                \begin{equation}
                 w(\vec{x}) \, \equiv \, \sum_{\mu = 1}^d \,
                          \left[ \, U_{\mu}(h;\vec{x}) \,
                                   e^{- i \Theta_{\mu} / N} + \,
                   U_{\mu}^{\dagger}(h;\vec{x} - \hat{e}_{\mu})
                             \, e^{i \Theta_{\mu} / N}
                             \, \right] \; ,
                \end{equation}
                which should be compared to
                eq.\ (\ref{eq:w}).
\item[{\bf b)}] The matrices $Z_{\mu}(h)$ are kept 
                fixed in eq.\ (\ref{eq:minimizing2}) as
                one selects the matrices $\Theta_{\mu}$, belonging to
                the Cartan sub-algebra, see eq.\ (\ref{eq:tautheta}),
                in such a way that they minimize the quantities
                \begin{equation}
                  - \, \Re \, \Tr \, \frac{e^{- i \Theta_{\mu} / N}}{V}
                        \, Z_{\mu}(h)
                \end{equation}
                and satisfy the condition (\ref{eq:pbctheta}), as in
                eq.\ (\ref{eq:Thetadiag-exp}).
                We note that, for this minimization step,
                one usually does not employ a simple multiplicative update,
                as in eq.\ (\ref{eq:htorh}).
                The main problem is that the minimizing functional is
                quadratic in the matrix $v$.
                On the other hand, the dependence on the integer parameters
                $n_{\mu}^{j}$ is rather trivial.
\end{itemize}
From the above discussion is also evident that, contrary to the
situation described in section \ref{sec:numPBCs} (for the implementation
of the usual minimal-Landau-gauge condition), the organization of the
numerical algorithm is slightly more complicated, when considering the
extended lattice $\Lambda_{z}$.
Indeed, since the gauge transformation $h(\vec{x}\mkern 1mu)$ and its
update $r(\vec{x}\mkern 1mu)$ do not commute (in general) with the
$\Theta_{\mu}$ matrices, it is no longer true that we can write the
single-site update as
\begin{equation}
\exp{\left( i \sum_{\nu = 1}^d \,
          \frac{\Theta_{\nu} \, x_{\nu}}{N}
                    \right)} \, h(\vec{x}\mkern 1mu) \,=\,
l(\vec{x}\mkern 1mu)\,\to\,
   r(\vec{x}\mkern 1mu) \, l(\vec{x}\mkern 1mu) \; .
\end{equation}
Instead, we need to consider the update
\begin{equation}
l(\vec{x}\mkern 1mu)\, \to \,
\exp{\left( i \sum_{\nu = 1}^d \,
          \frac{\Theta_{\nu} \, x_{\nu}}{N}
                    \right)} \, r(\vec{x}\mkern 1mu) \,
                    h(\vec{x}\mkern 1mu) \; ,
\end{equation}
which preserves the Bloch-function structure.
Thus, we can still make use of the multiplicative updates reported
in eq.\ (\ref{eq:Uupdates}) but, beside the link configuration $\{
U_{\mu}(h;\vec{x}) \}$, we need to store (separately) the matrices
$\Theta_{\mu}$.
In fact, eq.\ (\ref{eq:minfunctUhTheta}) illustrates that it
is sufficient (and necessary) to know $\{ U_{\mu}(h;\vec{x}) \}$ and
$\{ \Theta_{\mu} \}$ in order to carry out the minimization process.

More details about the numerical implementation of this algorithm
will be discussed in a future work.
Here, we only present the numerical checks we have done to confirm
the results obtained in section \ref{sec:linkNP}.
In particular, in Figs.\ \ref{fig:spectrum1} and \ref{fig:spectrum2}
we show the ``spectrum'' of the gluon propagator or,
to be more specific, the {\em allowed} momenta, i.e.\ the
momenta for which a nonzero gluon propagator $D(\veck)$ is obtained.
To this end we recall that --- when considering $\Lambda_x$ --- the lattice momenta $p^2(\vec{k}) =
\sum_{\mu=1}^d p_{\mu}^2(\veck)$ have components $p_{\mu}(\veck) =
2 \sin{\left(\pi \, k_{\mu}/N\right)}$, see eqs.\ (\ref{eq:pmu}) and
(\ref{eq:p2def}), where $N$ is the lattice side and --- due to the
symmetry of $p^2(\vec{k})$ under the reflection $\vec{k} \to -
\vec{k}+N \hat{e}_{\mu}$ (see section \ref{sec:gluonPBCs}) ---
we just need to consider $k_{\mu}=0,1,\ldots,N/2$ (when $N$ is even).
Then, it is easy to verify that, for $N=4$ and $d=3$, there are only
$7$ different momenta (with degeneracy).
Similarly, for $N=8$ and $d=3$, there are $25$ different momenta (with
degeneracy).
These momenta --- which we call here {\em original} momenta ---
are shown (in blue) in the right column of plots a) and b) of
Figs.\ \ref{fig:spectrum1} and \ref{fig:spectrum2}, respectively
for the $N=4$ and $N=8$ case.
At the same time, for $N=128$ (and again $d=3$), there are about $45000$
different momenta (with degeneracy), which are shown (in magenta)
in the left column of plots a) and b) of both figures.\footnote{Of
course, in this case the plot resembles a ``continuum spectrum''.}
Finally, on the right column of plot b) of Figs.\ \ref{fig:spectrum1}
and \ref{fig:spectrum2} we show, in green and in red, the {\em allowed}
momenta obtained by considering two different configurations for,
respectively, the lattice $V=(4\times32)^3$ and $V=(8\times16)^3$,
using the Bloch-wave setup described above.
As one can easily see, the {\em allowed} momenta always include the
{\em original} momenta, as well as other momenta that are
configuration-dependent.
Moreover, we have considered the condition
$k_{\nu}^{\mk \prime}\,+\, n_{\nu}^{j}\,-\,n_{\nu}^{i}
\propto m$, see eq.\ (\ref{eq:kprimeproptom}), which should
be satisfied by the {\em allowed} momenta.
This has been checked using one configuration for the lattice volumes
$V=(16\times8)^3$ and $V=(32\times4)^3$, and two configurations for
each of the setups $V=(8\times16)^3$ and $V=(4\times32)^3$.
In total, for these six configurations we have found that there were
slightly more than 16,000
{\em allowed} momenta.
Of these, a little less than 6,000
are the lattice momenta that can
be considered also on the small (original) lattice.
In all cases we have checked that eq.\ (\ref{eq:kprimeproptom}) is
indeed verified for the nonzero values of the gluon propagator.

\vskip 3.5mm
\begin{figure}[t]
\includegraphics[width=7cm,height=5cm]{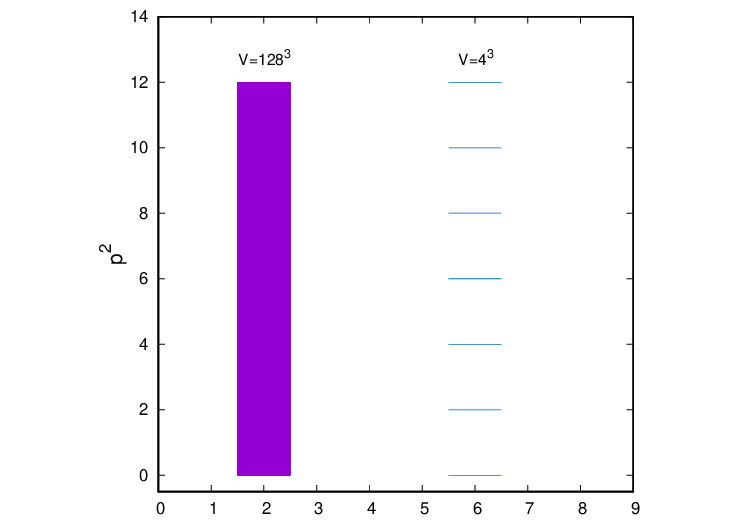}
\includegraphics[width=7cm,height=5cm]{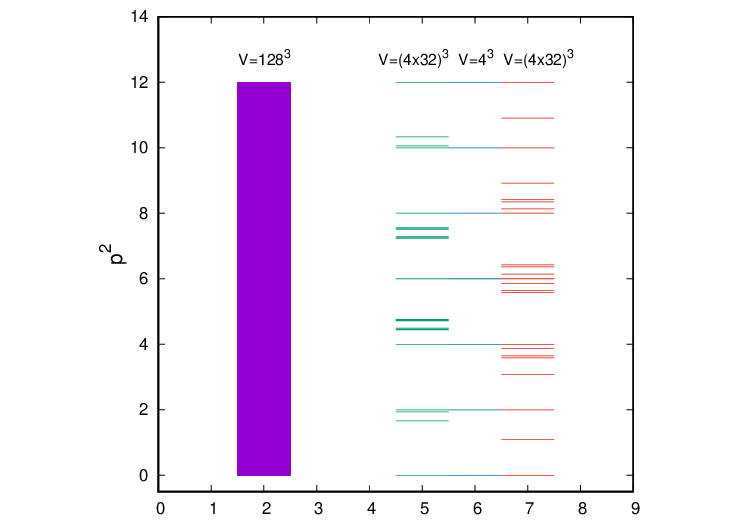}
\caption{In plot a), on the left, we show the {\em original} momenta
for the lattices $V = 128^3$ (left column) and $V = 4^3$ (right
column).
The same momenta are reported in plot b), on the right,
together with the {\em allowed} momenta, obtained by considering
two different configurations for the lattice setup $V=(4\times32)^3$,
i.e.\ with $N=4$ and $m=32$.
All simulations have been done using the SU(2) gauge group at $\beta=3.0$.
\label{fig:spectrum1}}
\end{figure} 

\vskip 3.5mm
\begin{figure}[b]
\includegraphics[width=7cm,height=5cm]{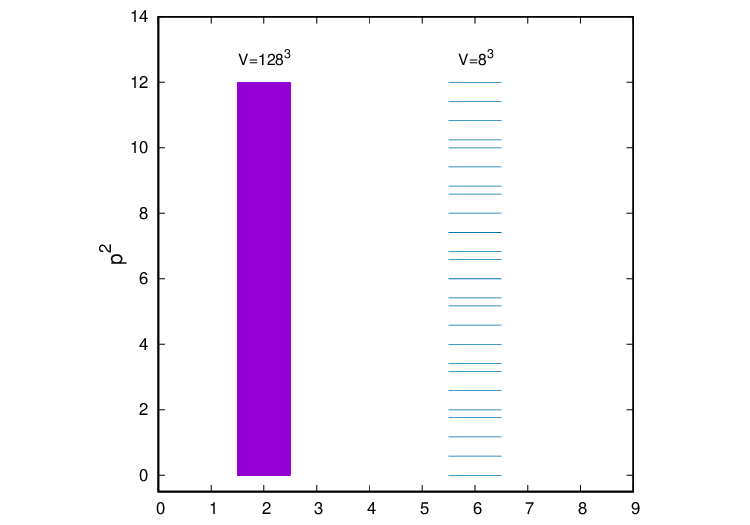}
\includegraphics[width=7cm,height=5cm]{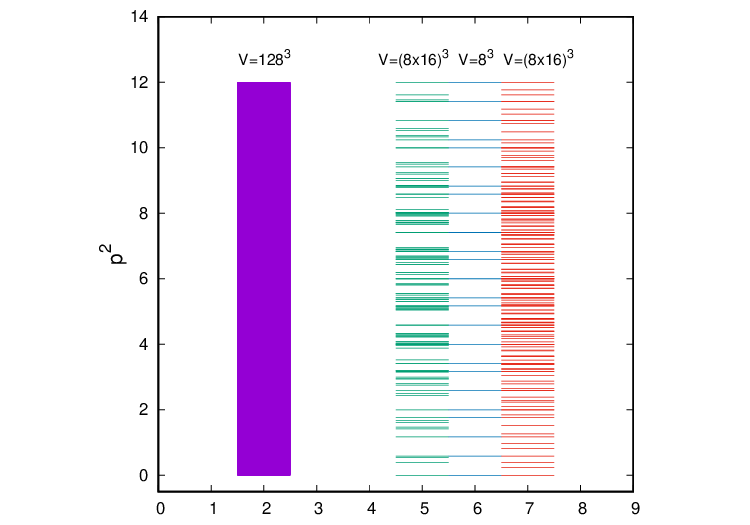}
\caption{In plot a), on the left, we show the {\em original} momenta
for the lattices $V = 128^3$ (left column) and $V = 8^3$ (right
column).
The same momenta are reported in plot b), on the right,
together with the {\em allowed} momenta, obtained by considering
two different configurations for the lattice setup $V=(8\times16)^3$,
i.e.\ with $N=8$ and $m=16$.
All simulations have been done using the SU(2) gauge group at $\beta=3.0$.
\label{fig:spectrum2}}
\end{figure}

We also stress that the explanation presented in
section \ref{sec:gluon-extended} about the suppression
of $D(\veczero)$, in the limit $m \to \infty$, is
essentially in agreement with the intuitive
argument presented in ref.\ \cite{Cucchieri:2016qyc}.
To see this, 
using eqs.\ (\ref{eq:Ajjk=0}) and (\ref{eq:Alx}), we can write
\begin{equation}
\;{\widetilde A}^{jj}_{\mu}(l;\veczero) = \sum_{\vec{x}\in\Lambda_x}\!
\Biggl\{\frac{1}{2 i}\Bigl[U^{jj}_{\mu}(h;\vecx)
e^{-\frac{2\pi i n_{\mu}^{j}}{m N}} 
\!-\, e^{\frac{2\pi i n_{\mu}^{j}}{m N}} {U^{jj}_{\mu}(h;\vecx)}^{*}\Bigr]
-\frac{1}{N_c}\!\sum_{l=1}^{N_c}\Im\! \left[
 U^{ll}_{\mu}(h;\vecx)\, e^{\frac{- 2 \pi i n_{\mu}^{l}}{m N}}\!\right]
\Biggr\} \,,\;\;
\end{equation}
i.e., we are evaluating the diagonal $jj$ coefficients of the matrix, see eq.\ (\ref{eq:Z}),
\begin{eqnarray}
  & &\, \frac{1}{2 i} \,\Bigl[\,Z_{\mu}(h) \, e^{- i \frac{\Theta_{\mu}}{N}}\,-\,
         e^{i \frac{\Theta_{\mu}}{N}} \, Z_{\mu}^{\dagger}(h)\,\Bigr]
        \, - \, \1 \, \frac{\Im\Tr}{N_c} \!\left[\,Z_{\mu}(h)
            e^{- i \frac{\Theta_{\mu}}{N}}\,\right] \nonumber \\[2mm]
  &=&\, \frac{1}{2 i} \,\Bigl[\,Z_{\mu}(h) \, v^{\dagger}\, T_{\mu}\, v -\,
         v^{\dagger}\,T_{\mu}^{\dagger}\,v \,
         Z_{\mu}^{\dagger}(h)\,\Bigr] \, - \,  \1 \,
        \frac{\Im\Tr}{N_c} \!\left[\,Z_{\mu}(h)\,v^{\dagger}\,
             T_{\mu}\, v \, \right] \, ,
\label{eq:AofZ}
\end{eqnarray}
where $T_{\mu}$ is a shorthand notation for the diagonal
matrices $T_\mu(m N;\{n_{\mu}^{j}\})$, see eqs.\ (\ref{eq:vDjj})
and (\ref{eq:Djj}).
Hence, with $w_j = v^{\dagger}\,\hat{e}_j$, we end up with the expression
\begin{eqnarray}
{\widetilde A}^{jj}_{\mu}(l;\veczero) \,&=&\,
\hat{e}_j^{\dagger}\;
    \frac{v\,Z_{\mu}(h) \, v^{\dagger}\, T_{\mu} \,-\,
         T_{\mu}^{\dagger}\,v \,
         Z_{\mu}^{\dagger}(h)v^{\dagger} }{2\,i}\, \hat{e}_j
 \, - \, \frac{\Im\Tr}{N_c} \!\left[\,Z_{\mu}(h)\,v^{\dagger}\,
             T_{\mu}\, v \, \right] \nonumber \\[2mm]
&=&\,\hat{e}_j^{\dagger}\;
    \frac{V_{\mu} \,-\, V_{\mu}^{\dagger}
          }{2\,i}\, \hat{e}_j \, - \,
        \frac{\Im\Tr}{N_c} \, V_{\mu} \; ,
\label{eq:AofZjj}
\end{eqnarray}
where we defined $V_{\mu} \equiv v\,Z_{\mu}(h) \, v^{\dagger}\, T_{\mu}$.
At the same time, in order to impose the gauge-fixing condition, we
need to maximize the quantity, see eq.\ (\ref{eq:minimizing2}),
\begin{equation}
 \Re \, \Tr \,
 \sum_{\mu = 1}^d \, Z_{\mu}(h) \,
      e^{- i \frac{\Theta_{\mu}}{N}}\,=\,
 \Re \, \Tr \, \sum_{\mu = 1}^d \,
  Z_{\mu}(h)\, v^{\dagger} \, T_{\mu}\, v
   \,=\, \Re \, \Tr \, \sum_{\mu = 1}^d \, V_{\mu} \, .
\end{equation}
Intuitively, this maximization can be easily achieved if one finds
a global rotation $v$ such that the (rotated) zero modes $v\,
Z_{\mu}(h)\,v^{\dagger}$
become close to diagonal matrices.
Then, given that in the limit $m \to \infty$ the discretized
parameters $n_{\mu}^{j}/m N$ become continuous,\footnote{By
looking at the matrix elements ${T_\mu}^{jj}$, it is clear
that the integers $n_{\mu}^{j}$ can always be limited to the
interval $[0, mN-1]$.
Then, in the limit $m \to + \infty$, the parameters $n_{\mu}^{j}/m N$ 
are real numbers belonging to the interval $[0, 1)$.}
one should be
able to use the diagonal matrices $T_{\mu}=T_\mu(m N;\{n_{\mu}^{j}\})$
--- whose elements are ${T_\mu}^{jj}=\exp{(- 2\pi i
n_{\mu}^{j}/m\,N})$ --- to bring
the matrices $V_{\mu}$ as close as
possible to real diagonal matrices.
As a consequence, both terms in eq.\ (\ref{eq:AofZjj}) should be close
to zero, implying
${\widetilde A}^{jj}_{\mu}(l;\veczero)
\approx 0$ and $D(\veczero) \approx 0$, see eqs.\ (\ref{eq:D0diag}).

As we noted in section \ref{sec:minimizing-revisited}, gauge-fixed
link configurations within each replicated lattice
$\Lambda_x^{(\vecy)}$ are rotated, transformed by global
group elements defined by the cell index ${\vec y}$, see
eq.\ (\ref{eq:Uofg2}).
The same applies to the gauge-fixed gauge-field configurations
$\{ A_{\mu}(l;\vecx) \}$,
see eq.\ (\ref{eq:Agzeta2}).
It is then natural to consider, on each replicated lattice $\Lambda_x^{(\vecy)}$,
the average color {\em magnetization} $\vec{M}_{\mu}(\vec{y})$
with (color) components\footnote{Following ref.\ \cite{Zwanziger:1990by}
one can prove that the quantity
\begin{equation}
{\cal M}\, = \, \frac{1}{d\,(N_c^2 - 1) m^d}\,
\sum_{b,\mu} \,\Bigl<\, \Bigl|\, \sum_{\vecy}
\,M_{\mu}^b(\vecy)\, \Bigr|\, \Bigr>
\end{equation}
should vanish --- in Landau gauge and in the infinite-volume limit --- at least
as fast as the inverse lattice side.
The volume dependence of ${\cal M}$ has been analyzed in detail
in two, three and four space-time dimensions in ref.\
\cite{Cucchieri:2007rg}.}
\begin{equation}
M_{\mu}^b(\vec{y}) \, = \, \frac{1}{V}\,\sum_{\vec{x}\in\Lambda_x}\,
   A_{\mu}^b(g;\vec{x} + \vec{y} N) \; ,
\end{equation}
which is related to the gluon propagator at zero momentum
since, see eq.\ (\ref{eq:Amugkpdef}),
\begin{equation}
{\widetilde A}_{\mu}^b(g;\veczeropp)\,=V\,
   \sum_{\vec{y}\in\Lambda_y}\,M_{\mu}^b(\vec{y}) \; ,
\end{equation}
so that eq.\ (\ref{eq:D0g}) implies by the expression
\begin{equation}
D(\veczeropp) \, = \, 
   \frac{\Tr}{2\,m^d\,{\cal N}}\,
   \sum_{\mu=1}^d \, \Bigl<
  \left[\, {\widetilde A}_{\mu}(g;\veczeropp)
  \,\right]^{2}
  \Bigr> \,=\,
   \frac{V^2}{m^d\,{\cal N}}\,
   \sum_{\mu=1}^d \, \sum_{b=1}^{N_c^2-1}\,
 \Biggl< \biggl[\; \sum_{\vecy \in \Lambda_y} M_{\mu}^b(\vecy)
  \;\biggr]^{2}
  \Biggr>\, ,
\end{equation}
where ${\cal N}\equiv d(N_c^2-1)V$ has been defined
in section \ref{sec:gluonPBCs}.
We show in Figs.\ \ref{fig:domains1} and \ref{fig:domains2bis}
the vectors $\vec{M}_{3}(\vec{y})$ of the color magnetization,
obtained in a simulation for the
$SU(2)$ case and with lattice volume $V = (64 \times 4)^3$
at $\beta = 3.0$. 
The vector components stand for the different values of the color index,
i.e.\ $M^b_3(\vecy)$, for $b=1,2,3$. 
One can clearly see the effect of the Bloch waves.
In particular, the average magnetization may appear ``smooth'' 
along a certain direction when moving from one cell to the next, but 
a suitably chosen projection reveals the modulated behavior, as expected.
For example, in Fig.\ \ref{fig:domains2bis}, $M_{\mu}^3(\vec{y})$ does not
change when crossing a boundary, while $M_{\mu}^1(\vec{y})$ and
$M_{\mu}^2(\vec{y})$ are rotated (see Fig.\ \ref{fig:domains3bis}).
Thus, each cell $\Lambda_x^{(\vecy)}$ may be seen as a domain, and the 
domain walls are characterized by the cell boundaries.

\vskip 3.5mm
\begin{figure}[t]
\hskip 2cm
\includegraphics[width=12cm,height=7cm]{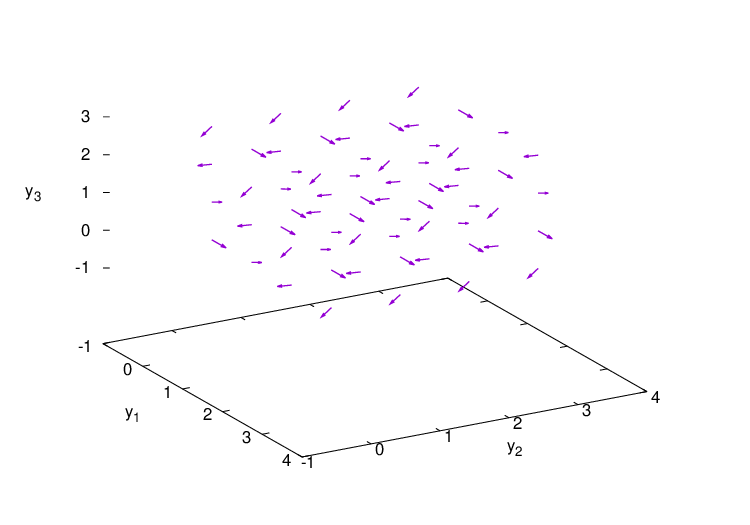}
\caption{
Average color ``magnetization'' $\vec{M}_{3}(\vec{y})$ on
each replicated lattice $\Lambda_x^{(\vecy)}$
for the pure-$SU(2)$ case and lattice volume $V = (64 \times 4)^3$,
at $\beta=3.0$.
In this case the index lattice $\Lambda_y$ is a $4^3$ lattice and $\vecy$
has components $y_{\mu}=0,1,2,3$ with $\mu =1,2,3$.
Also note that the color components $M_{3}^b(\vec{y})$ (with $b=1,2$ and $3$)
are represented along the corresponding spatial directions $\mu=1,2,3$.
\label{fig:domains1}}
\end{figure}

Finally, we present our conclusions.
Our main finding is that the gluon propagator $D(\veckpp)$
is nonzero only for the {\em allowed} momenta and, in these
cases, its value comes from some of the coefficients
${\widetilde A}^{ij}_{\mu}(g;\veckpp)$, with all the other
coefficients being equal to zero.
Hence, we now completely understand the math behind the use of Bloch
waves in minimal Landau gauge and we can perform the whole
simulation (thermalization, gauge fixing and evaluation of the
gluon propagator) in the small ``unit cell'' $\Lambda_x$.
This should permit us to produce large ensembles of data\footnote{To
this end, it may be useful to move part of the simulation from
CPUs to GPUs.}
for the infrared gluon propagator, even when we consider small unit
cells and large values of $m$.
We hope this can give us some hints about the role of the 
$\{ U_{\mu}(h,\Theta;\vec{x}) \}$ ``domains'' and of the
``magnetization'' described above.
In particular, we want to find the minimum value of
the lattice size $N$ of $\Lambda_x$ for which
the momentum-space gluon propagator $D(\veckpp)$, evaluated
on $\Lambda_z$ using the Bloch setup with a factor $m$, is
still in agreement with numerical data obtained by working
directly on a lattice of size $m N$.
More in general, we want to check the dependence of $D(\veckpp)$
on $N$, while keeping the product $m N$ fixed.
Clearly, since we know that finite-size effects in the gluon
propagator are relevant only in the infrared regime,
it is essential to consider all {\em allowed} momenta in these
numerical simulations.
Indeed, the momenta given by the discretization on the original
(small) lattice $\Lambda_x$ are insufficient to adequately probe
the infrared limit when $N$ is small.
We also plan to extend this analysis to the ghost propagator.

\vskip 3.5mm
\begin{figure}[t]
\hskip 2cm
\includegraphics[width=12cm,height=7cm]{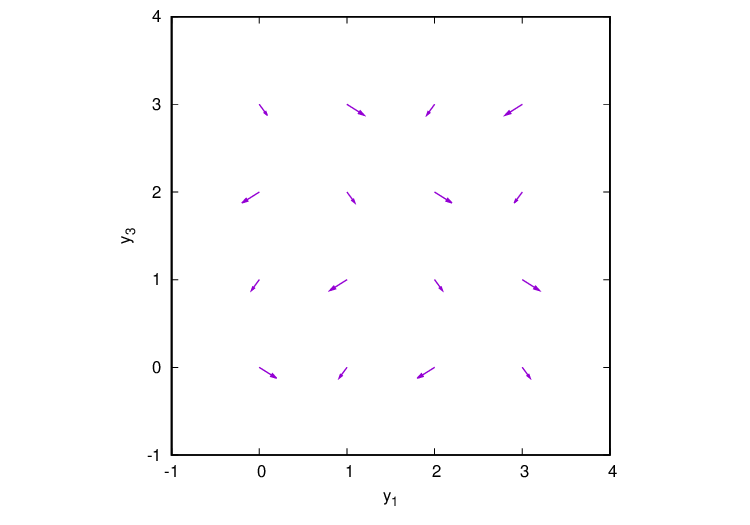}
\caption{
Average color ``magnetization'' $\vec{M}_{3}(\vec{y})$ on
each replicated lattice $\Lambda_x^{(\vecy)}$
for the pure-$SU(2)$ case and lattice volume $V = (64 \times 4)^3$,
at $\beta=3.0$.
In this case the index lattice $\Lambda_y$ is a $4^3$ lattice and $\vecy$
has components $y_{\mu}=0,1,2,3$ with $\mu =1,2,3$.
Also note that the color components $M_{3}^b(\vec{y})$ (with $b=1,2$ and $3$)
are represented along the corresponding spatial directions $\mu=1,2,3$.
Here we show the data presented in Fig.\ \ref{fig:domains1} with coordinate $y_2=0$,
projected on the $y_1\!-\!y_3$ plane.
Consequently, we are also showing only the $b=1,3$ color components.
\label{fig:domains2bis}}
\end{figure}

\vskip 3.5mm
\begin{figure}[t]
\hskip 2cm
\includegraphics[width=12cm,height=7cm]{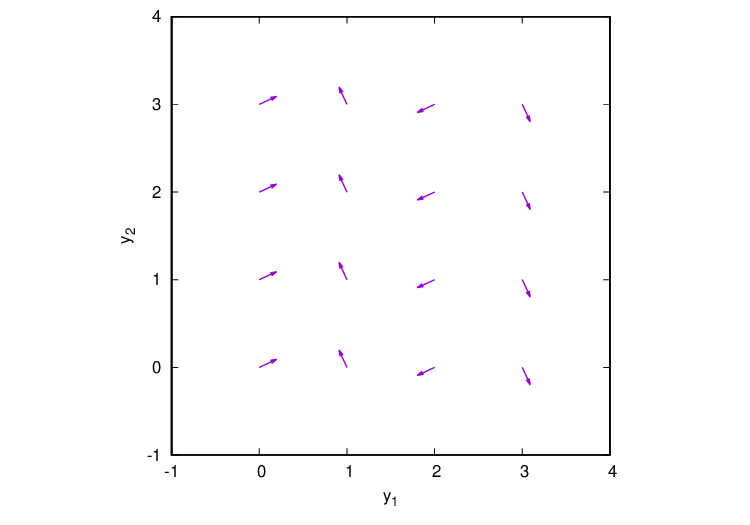}
\caption{
Average color ``magnetization'' $\vec{M}_{3}(\vec{y})$ on
each replicated lattice $\Lambda_x^{(\vecy)}$
for the pure-$SU(2)$ case and lattice volume $V = (64 \times 4)^3$,
at $\beta=3.0$.
In this case the index lattice $\Lambda_y$ is a $4^3$ lattice and $\vecy$
has components $y_{\mu}=0,1,2,3$ with $\mu =1,2,3$.
Also note that the color components $M_{3}^b(\vec{y})$ (with $b=1,2$ and $3$)
are represented along the corresponding spatial directions $\mu=1,2,3$.
Here we show the data presented in Fig.\ \ref{fig:domains1} with coordinate $y_3=0$,
projected on the $y_1\!-\!y_2$.
Consequently, we are also showing only the $b=1,2$ color components.
\label{fig:domains3bis}}
\end{figure}


\acknowledgments

The authors acknowledge partial support from FAPESP and CNPq. 
A. Cucchieri also thanks the Université Paris-Saclay (Orsay, France)
--- and in particular B. Benoit and the P\^ole Th\'eorie of the IJCLab ---
where part of this work has been done.


\appendix

\section{Cartan sub-algebra}
\label{sec:Cartan}

In this appendix we discuss properties related to the matrices 
$\Theta_{\mu}$, introduced in section \ref{sec:Bloch-theorem},
see eq.\ (\ref{eq:tautheta}). 
Recall that these matrices belong to the Cartan sub-algebra of $su(N_c)$ 
and must satisfy the periodicity condition (\ref{eq:pbctheta}), which implies
that their eigenvalues be given by $2 \pi n_{\mu} / m$, where $n_{\mu}$ is 
an integer.
We start by describing a general parametrization for the $N_c-1$ generators
of the Cartan sub-algebra
in the SU($N_c$) case 
\cite{Cornwell,vanEgmond:2021wlt,Zuber}, and then comment on possible advantages of other bases.
We also compare our setup with that considered in ref.\
\cite{Zwanziger:1993dh}.
As can be seen in section \ref{sec:linkNP}, and in particular in
subsections \ref{sec:gluonNP} and
\ref{sec:gluon-extended}, some of these properties are central to
obtain an analytic expression for the gluon propagator using the
gauge-fixed configuration
on the extended lattice $\Lambda_z$.

\vskip 3mm
We recall that we have chosen the $N_c^2 - 1$ traceless generators $t^b$ of the
$su(N_c)$ Lie algebra to be Hermitian.
Since the Cartan generators $\{ t_{\scriptC} \}$ are mutually commuting,
i.e.\ $[t^a_{\scriptC}, t^b_{\scriptC}]=0$ (for $a,b=1,\ldots,N_c-1$),
they can be simultaneously diagonalized.
For example, in the SU($N_c$) case, we can consider as
diagonal Cartan generators (in the fundamental representation) the $N_c-1$ linearly independent,
$N_c \times N_c$ Hermitian and traceless matrices $H^i$ ($i=1,\ldots,N_c-1$) defined by
\cite{Zuber}
\begin{equation}
H^i_{jk}\,=\,\xi^i\,\delta^{jk}\,\left[\delta^{ij}-\delta^{(i+1)j}\right]
\,,
\label{eq:defH}
\end{equation}
with $\xi^i$ real and $j,k=1,\ldots,N_c$.
Note that, beside being diagonal, the matrix $H^i$ only has nonzero
elements in rows/columns $i$ and $i+1$ and these two elements have
opposite signs, enforcing the tracelessness condition.
In particular, for $N_c=2$, the 
matrix $H^1$ is given by $\xi^1$ times the third Pauli matrix $\sigma_3$.
For the SU($3$) case, after setting $\xi^1=\xi^2=1$, we have
\begin{equation}
    H^1\,=\, \begin{pmatrix}
                    1 &  0 & 0 \\
                    0 & -1 & 0 \\
                    0 &  0 & 0
                 \end{pmatrix}
    \qquad \mbox{and} \qquad
    H^2\,=\, \begin{pmatrix}
                    0 & 0 & 0 \\
                    0 & 1 & 0 \\
                    0 & 0 & -1
                 \end{pmatrix}
    \; .
\label{eq:su3_original}
\end{equation}
Since they are diagonal, the above generators $H^i$ have --- as common 
eigenvectors --- the unit vectors\footnote{Let us point out that
this is the same notation as 
the one used in the main text
for the unit vectors in the $d$-dimensional Euclidean space, but clearly 
we refer here to color indices (in the fundamental representation).}
$\hat{e}_j$ [whose components are given by
$(\hat{e}_j)_k = \delta^{jk}$],
with eigenvalues 
$\lambda^i_j=\xi^i \left[ \delta^{ij} -\delta^{(i+1)j}\right]$,
where again $j=1,\ldots,N_c$.

More in general, since the matrices $H^i$ are diagonal, we may define
the Cartan generators by any combination
\begin{equation}
D^{i}\,=\, \sum_{l=1}^{N_c-1}\, R^{il}\, H^l \; ,
\label{eq:DR}
\end{equation}
where $R$ is an invertible $(N_c\!-\!1) \times (N_c\!-\!1)$ matrix.
For example, in the SU($3$) case, with the Gell-Mann choice for the 
generators of the group algebra, the Cartan sub-algebra is spanned by 
the matrices $D^{1}=H^1$
and $D^{2}=( H^1 + 2 H^2 )/\sqrt{3}$ --- usually
denoted by $\lambda_3$ and $\lambda_8$ --- instead of $H^1$ and $H^2$ given
in eq.\ (\ref{eq:su3_original}).
This corresponds to changing the basis
with the matrix
\begin{equation}
    R\,=\, \begin{pmatrix}
                    1 & 0 \\
                    \frac{1}{\sqrt{3}} & \frac{2}{\sqrt{3}}
                 \end{pmatrix} \; .
\end{equation}
In order to generalize the above bases containing Pauli and 
Gell-Mann matrices to the SU($N_c)$ case (see, e.g., appendix A1 in 
ref.\ \cite{Smit}), we may consider\footnote{Equivalently, we can use 
eq.\ (\ref{eq:DR}) with the matrix
\begin{equation}
    R\,=\, \begin{pmatrix}
                    1 & 0 & 0 & 0 & \ldots \\
                    \frac{1}{\sqrt{3}} & \frac{2}{\sqrt{3}}
                          & 0 & 0 & \ldots \\
                    \frac{1}{\sqrt{6}} & \frac{2}{\sqrt{6}}
                          & \frac{3}{\sqrt{6}} 
                          & 0 & \ldots \\
                    \frac{1}{\sqrt{10}} & \frac{2}{\sqrt{10}}
                          & \frac{3}{\sqrt{10}} 
                          & \frac{4}{\sqrt{10}} 
                          & \ldots \\
                    \ldots & \ldots & \ldots & \ldots & \ldots
                 \end{pmatrix} \; ,
\label{eq:RgeneralSUNc}
\end{equation}
i.e.\ with matrix elements $R^{il}= l \, \sqrt{\frac{2}{i(i+1)}}$ for
$l \le i$ and $R^{il}=0$ otherwise.} the matrices 
\begin{equation}
D^{i}\,=\, \sqrt{\frac{2}{i(i+1)}}
   \, \left[\, \sum_{l=1}^{i}\, l \, H^l
                     \, \right] \; ,
\label{eq:DjgenGM}
\end{equation}
with $i=1,\ldots,N_c\!-\! 1$. 
Note that, just as $H^i$, the matrices $D^{i}$ are diagonal.\footnote{For
the choice in eq.\ (\ref{eq:DjgenGM}) we have the matrix
elements $D^{i}_{jj} = \sqrt{\frac{2}{i(i+1)}}\,\xi^1$ for $j=1$,
$D^{i}_{jj} =
\sqrt{\frac{2}{i(i+1)}}\,\left[j\xi^j-(j-1)\xi^{j-1}\right]$ for 
$j=2,\ldots,i$, $D^{i}_{jj} =
- i \,\xi^i \sqrt{\frac{2}{i(i+1)}}$ for $j=i+1$, and $D^{i}_{jj} = 0$
otherwise.}
Their eigenvectors are also $\hat{e}_j$, but with eigenvalues\footnote{We
recall that each vector $\alpha_j$, with components $\alpha^{i}_j$,
corresponds to a {\em weight} (of the Cartan generators)
\cite{Cornwell,vanEgmond:2021wlt,Zuber}.}
\begin{equation}
\alpha^{i}_j\,=\,\sum_{l=1}^{N_c-1}\,
R^{il}\, \lambda^l_j
\,=\, R^{ij}\,\xi^j - R^{i(j-1)}\,\xi^{j-1} \; .
\label{eq:alpha-ij}
\end{equation}
Also, recall that, while $j$ takes values from $1$ to $N_c$, the
indices of the matrix $R^{il}$ and of the constants $\xi^i$ only
go from $1$ to $N_c\!-\!1$.
Thus, for $j=1$ we have $\alpha^{i}_1 = R^{i1}\,\xi^1$ and for
$j=N_c$ we find $\alpha^{i}_{N_c} = - R^{i(N_c-1)}\,\xi^{N_c-1}$.

Finally, it is rather evident \cite{Cornwell,Zuber} that, if $H_{
\scriptC}$ is a Cartan sub-algebra and $v$ is any element of the
Lie group, the conjugate $v^{-1} \, H_{\scriptC} \, v$ is another
Cartan sub-algebra.
Thus, for the SU($N_c)$ group we can consider as matrices
$t_{\scriptC}^b$ the set $\{ v^{\dagger} \, D^{b} \, v
\}$, with common eigenvectors $\{ w_j = v^{\dagger}\,\hat{e}_j
\;\mbox{for}\;
j = 1, \ldots, N_c \}$ --- which are orthonormal, since $w_j^{
\dagger} \,w_k = \hat{e}_j^{\dagger} \, \hat{e}_k\,=\,\delta^{
jk}$ --- and the eigenvalues $\alpha^{b}_j$ given above, where we 
switched back to the usual index $b$ for the color degrees of freedom.
This illustrates the expansion in eq.\ (\ref{eq:tautheta}).


\subsection{Comparison with reference \cite{Zwanziger:1993dh}}
\label{sec:Dan-Cartan}

We note that ref.\ \cite{Zwanziger:1993dh} defines
the $\Theta_{\mu}$ matrices, belonging to a generic Cartan
sub-algebra, as an expansion in terms of the generators $t^b$ of 
the SU($N_c$) algebra, i.e.,
\begin{equation}
    \Theta_{\mu}\,=\,\sum_{b=1}^{N_c^2-1}\,
    \theta_{\mu}^b\,t^b \,, 
\end{equation}
with real parameters $\theta_{\mu}^b$ ($\mu=1,\ldots,d$),
subject to the condition
\begin{equation}
   \left[\,\Theta_{\mu},\, \Theta_{\nu}\,\right]\,=\,
    \sum_{b,c=1}^{N_c^2-1}\,\,\theta_{\mu}^b\,\theta_{\nu}^c\,
   \left[\,t^b,\, t^c\,\right]\,=\,
    2\,i\,\sum_{a,b,c=1}^{N_c^2-1}\,\, f^{abc} 
    \,\theta_{\mu}^b\,\theta_{\nu}^c\; t^a\, \, =\, 0 \; ,
\label{eq:Thetacondition}
\end{equation}
where we denote by $f^{abc}$ the structure constants of the
$su(N_c)$ Lie algebra. 
Now, since the matrices $t^a$ are linearly independent, the above
equality implies that
\begin{equation}
   \sum_{b,c=1}^{N_c^2-1}\,f^{abc}\,\theta_{\mu}^b\,\theta_{\nu}^c
   \, =\, 0 \; ,
   \label{eq:theta-Cartan}
\end{equation}
for any $a=1,\ldots,N_c^2-1$.

In the SU($2$) case, for example, for which
the Cartan sub-algebra is one-dimensional 
and the structure constants $f^{abc}$
are given by the completely anti-symmetric tensor $\epsilon^{abc}$,
we find that the above condition is equivalent to saying that the
three-dimensional vectors $\vec{\theta}_{\mu}$ and 
$\vec{\theta}_{\nu}$ must be parallel\footnote{Indeed, in this case, 
considering the vector components $\theta^a_{\mu}$ and $\theta^a_{\nu}$, 
with $a=1,2,3$, the expression in eq.\ (\ref{eq:theta-Cartan}) corresponds to
$\vec{\theta}_{\mu}\times \vec{\theta}_{\nu} = 0$, where $\times$
indicates the usual cross product.} for any $\mu, \nu$.
This can be easily achieved \cite{Zwanziger:1993dh} with
\begin{equation}
    \Theta_{\mu}\,=\,r_{\mu}\,
    \sum_{b=1}^{3}\,\,q^b\,t^b \; ,
\label{eq:su2setup}
\end{equation}
where $r_{\mu}$ and $q^b$ are real parameters.
As a matter of fact, by factoring $\,\theta^b_\mu=\,r_\mu \,q^b$, i.e.\
by imposing that the vectors $\vec{\theta}_{\mu}$ are all proportional
to the vector $\vec{q}$, it is evident that eq.\
(\ref{eq:theta-Cartan}) is satisfied, since
$\sum_{b,c=1}^{3} f^{abc} q^b q^c = 0$.
Note that matrices $\Theta_{\mu}$ defined
in this way are not necessarily diagonal.
On the other hand, they are mutually commuting since they are
proportional to the same matrix $\sum_{b=1}^{3}\,\,q^b\,t^b$.
One can also write
\begin{equation}
   \Theta_{\mu}\,=\,
     r_{\mu} \, v^{\dagger} \, \sigma_3 \,v \; ,
\label{eq:su2setup2}
\end{equation}
where $v \in $ SU(2) and $\sigma_3$ is the third Pauli matrix,
which is diagonal.
Indeed, eqs.\ (\ref{eq:su2setup}) and (\ref{eq:su2setup2})
are completely equivalent.\footnote{This is a general
result: any element of the $su(N_c)$ Lie algebra is conjugate
to an element of a Cartan sub-algebra (see, for example,
\cite{Zuber} and references therein).
In the case of the SU(2) group one can check this directly if
$t^c$ are the three Pauli matrices $\sigma_c$.
Indeed, by writing $v$ as $v_0 \1 + i \vec{\sigma} \cdot \vec{v}$,
where $\1$ is the $2 \times 2$ identity matrix and $v_0^2 + \vec{v}^2
= 1$, one recovers eq.\ (\ref{eq:su2setup}) --- starting from
eq.\ (\ref{eq:su2setup2}) --- by using the relation
\begin{equation}
\sigma_i \sigma_j = \1\, \delta^{ij}+i\sum_{k=1}^3 \epsilon^{ijk}\sigma_k \; .
\end{equation}
}
Hence, the above parametrization (\ref{eq:su2setup2}) is clearly in
agreement with the previously discussed setup and we can say that the
expansion used in ref.\ \cite{Zwanziger:1993dh} corresponds to a transformation
$v^{-1} \, H_{\scriptC} \, v$ of the Cartan sub-algebra given by $\sigma_3$.
Note that, using eq.\ (\ref{eq:su2setup2}), the matrices $\Theta_{\mu}$
trivially have eigenvectors
\begin{equation}
v^{\dagger} \begin{pmatrix} 1 \\ 0 \end{pmatrix}
\label{eq:w1su2}
\end{equation}
and
\begin{equation}
v^{\dagger} \begin{pmatrix} 0 \\ 1 \end{pmatrix}
\label{eq:vdagger}
\end{equation}
with eigenvalues $\pm\,r_{\mu}$.

In like manner, in the SU(3) case, which has rank two, we can write
\cite{Zwanziger:1993dh}
\begin{equation}
    \Theta_{\mu}\,=\,r_{\mu,3}\,
    \sum_{b=1}^{8}\,\,q^b_3\,t^b \,+\,
                     r_{\mu,8}\,
    \sum_{b=1}^{8}\,\,q^b_8\,t^b
\end{equation}
with real parameters $r_{\mu,3},\, r_{\mu,8},\, q^b_3$ and $q^b_8$, 
i.e.\ we now factored $\,\theta^b_\mu=r_{\mu,3}\,q^b_3 + r_{\mu,8}\,q^b_8$.
This yields, see eq.\ (\ref{eq:theta-Cartan}),
\begin{equation}
   \sum_{b,c=1}^{8}\,f^{abc}\,q^b_3\,q^c_8\,
   \left(\, r_{\mu,3}\,r_{\nu,8} \,-\,
   r_{\mu,8}\,r_{\nu,3} \,\right) \,=\, 0\; ,
\end{equation}
where we used the (obvious) relation $f^{abc}= -f^{acb}$.
Clearly, since the above expression must be valid for any values of
the parameters $r_{\mu,3}$ and $r_{\mu,8}$, we must
select two commuting matrices
\begin{equation}
   \widetilde{\lambda}_3\,=\,
    \sum_{b=1}^{8}\,\,q^b_3\,t^b
      \qquad \qquad \mbox{and} \qquad \qquad 
   \widetilde{\lambda}_8\,=\,
    \sum_{c=1}^{8}\,\,q^c_8\,t^c
\end{equation}
to parametrize the expansion of $\Theta_{\mu}$, so that
\begin{equation}
\frac{1}{2}\, \Tr\,\left\{\, t^a \,
\left[\,\widetilde{\lambda}_3,\,\widetilde{\lambda}_8\,\right]
\right\}\,=\,
\sum_{b,c=1}^{8}\,f^{abc}\,q^b_3\,q^c_8
\,=\, 0 \; .
\end{equation}
Then, we recover again our definition for $\Theta_{\mu}$ --- in terms
of diagonal matrices and the transformation
$v^{-1} \, H_{\scriptC} \, v$ --- if we consider
\begin{equation}
    \Theta_{\mu}\,=\,
      r_{\mu,3}\, \widetilde{\lambda}_3 \,+\,
      r_{\mu,8}\, \widetilde{\lambda}_8 \,=\,
    \,v^{\dagger} \, \left(\,
      r_{\mu,3}\, \lambda_3 \,+\,
      r_{\mu,8}\, \lambda_8 \,\right)\, v  \; ,
\end{equation}
where $\lambda_3$ and $\lambda_8$ are the two diagonal
Gell-Mann matrices and $v \in $ SU(3).


\subsection{New basis for the Lie algebra}
\label{sec:lambda-basis}

As already noted above, the matrices
$\Theta_{\mu}$ --- which belong to the Cartan sub-algebra 
and are written in terms of the basis 
$\,t_{\scriptC}^{b} = v^{\dagger} D^{b} v$ ---
have eigenvectors 
$w_j = v^{\dagger}\,\hat{e}_j$ (for $j=1,\ldots,N_c$), with eigenvalues
given by
\begin{equation}
\Theta_{\mu} \, w_j \, = \, \left[\,\sum_{b=1}^{N_c-1}\,
 \theta_{\mu}^{b}\, \alpha^{b}_j\,\right]\, w_j\,\equiv\,
               \beta_{\mu}^{j} \, w_j\,=\,
 \frac{2 \pi \,n_{\mu}^{j}}{m} \, w_j \; ,
\label{eq:Theta-wi}
\end{equation}
where the parameters $\theta_{\mu}^{b}$ refer to the expansion in
eq.\ (\ref{eq:tautheta}),
$\alpha^{b}_j$ is defined in eq.\ (\ref{eq:alpha-ij}) and,
in the last step, we imposed the constraint (\ref{eq:pbctheta}), i.e.\ 
that $n_{\mu}^{j}$ be integers.

Then,
it is natural to consider a new basis for $\Theta_{\mu}$, with matrices 
defined as an outer product of these eigenvectors, i.e.,
\begin{equation}
{\bf W}^{ij}\equiv w_i\,w_j^{\dagger}
\,=\,v^{\dagger}\,\hat{e}_i\,\hat{e}_j^{\dagger}\,v
\,\equiv\,v^{\dagger}\,{\bf M}^{ij}\,v
\,,
\label{eq:defW}
\end{equation}
where the matrix element $\,lm\,$ of ${\bf W}^{ij}$ is given by 
$(w_i)_l\,(w_j^{\dagger})_m$ (for $l,m=1,\ldots,N_c$).
Similarly, we have also defined the $N_c \times N_c$ matrix 
${\bf M}^{ij}=\hat{e}_i\,\hat{e}_j^{\dagger}$, 
whose elements are simply\footnote{In other words,
all the entries of ${\bf M}^{ij}$ are null with the exception of
the entry with indices $i,j$, which is equal to 1.}
$({\bf M}^{ij})_{lm}=\delta^{il}\delta^{jm}$.
In this way, considering $\,\Theta_\mu = \sum_{i,j} c_\mu^{ij}\, {\bf W}^{ij}$, 
we can left multiply eq.\ (\ref{eq:Theta-wi}) by $w_i^{\dagger}$ to obtain 
the expansion parameters
\begin{equation}
    c_{\mu}^{ij}\,=\,
    w_i^{\dagger}\, \Theta_{\mu} \, w_j \, = \, \beta_{\mu}^{i} \,
    \delta^{ij}
    \,=\, \frac{2 \pi \,n_{\mu}^{i}}{m} \, \delta^{ij}
\label{eq:betamuk}
\end{equation}
in the ${\bf W}^{ij}$ basis.
As expected, they are nonzero only for $i=j$,
since the eigenvectors $w_i$ form an orthonormal set.
Thus, we can write
\begin{equation}
\Theta_{\mu} \, = \, \sum_{j=1}^{N_c} \beta_{\mu}^{j} \; {\bf W}^{jj}
\,=\, \sum_{j=1}^{N_c}\,
   \frac{2 \pi\, n_{\mu}^{j}}{m} \, v^{\dagger}\,{\bf M}^{jj}\,v
   \label{eq:Thetadiag}
\end{equation}
and
\begin{equation}
  \exp{\left( i \,\frac{\Theta_{\mu}}{N} \right)} \,=\, v^{\dagger}\,
  \exp{\left( \sum_{j=1}^{N_c}\,
   \frac{2 \pi i \,n_{\mu}^{j}}{mN} \,{\bf M}^{jj} \,\right)} \,v \; ,
   \label{eq:Thetadiag-exp}
\end{equation}
which are important results for our analysis in sections
\ref{sec:proof} and \ref{sec:minimizing-revisited}.
Of course, when $v=\1$ --- or, equivalently, when considering the
basis
${\bf M}^{ij}$ --- the matrices in eqs.\ (\ref{eq:Thetadiag}) and
(\ref{eq:Thetadiag-exp}) are diagonal.

Let us stress that the matrices ${\bf W}^{ij}$ trivially satisfy the
trace condition\footnote{Of course, similar expressions apply to the
basis ${\bf M}^{ij}=\hat{e}_i\,\hat{e}_j^{\dagger}$.}
\begin{equation}
\Tr\left({\bf W}^{ij}\right)\,=\,
\Tr\left(w_i\,w_j^{\dagger}\right)\,=\, w_j^{\dagger}\,w_i
\,=\, \delta^{ij}
\label{eq:lambdatrace}
\end{equation}
and the orthonormality relations
\begin{equation}
\Tr\left({\bf W}^{ij}\, {{\bf W}^{lm}}^{\dagger} \,\right) =\,
\Tr\left({\bf W}^{ij}\, {\bf W}^{ml} \right) =
\Tr\left(w_i\,w_j^{\dagger}\,
w_m\,w_l^{\dagger}\,\right) \,=\, \delta^{jm}\,\Tr\left(
w_i\,w_l^{\dagger}\,\right)=\, \delta^{jm}\,\delta^{il} \; ,
\label{eq:lambda-ortho}
\end{equation}
where we used\footnote{The property
\begin{equation}
{{\bf W}^{ij}}^{\dagger}={\bf W}^{ji}
\end{equation}
can be seen directly from the definition (\ref{eq:defW}).
Thus, we see 
that each matrix ${\bf W}^{ij}$ is {\em not} Hermitian, unless $i=j$.
\label{WnHerm}}
${{\bf W}^{lm}}^{\dagger}={\bf W}^{ml}$,
the orthonormality of the $w$'s and (\ref{eq:lambdatrace}).
Hence, the $N_c^2$ matrices ${\bf W}^{ij}$ are indeed a
basis for any $N_c \times N_c$ matrix, which is --- as seen
in section \ref{sec:linkNP} --- the most natural one to consider when
analyzing the impact of the index lattice on the
evaluation of the gluon propagator.
On the other hand, elements of the (real) SU($N_c$) Lie group --- as
well as of the corresponding $su(N_c)$ Lie algebra --- are written
in terms of $N_c^2\!-\!1$ real independent parameters.
Therefore, when using this basis, the $N_c^2$ coefficients entering
the linear combination of the ${\bf W}^{ij}$ matrices are not all
independent.
As a matter of fact, a generic matrix
\begin{equation}
A \, = \, \sum_{i,j=1}^{N_c} A^{\mk ij}\, {\bf W}^{ij}\,,
\label{eq:Amatrix}
\end{equation}
where the coefficients $A^{ij}$ are given, see eq.\ (\ref{eq:betamuk}), by
\begin{equation}
    A^{ij}\,=\,
    w_i^{\dagger}\, A \, w_j \; , 
\label{eq:gencoeff}
\end{equation}
is (in general) not traceless, due to eq.\ (\ref{eq:lambdatrace}).
Thus, for the $su(N_c)$ Lie algebra we have to enforce the
constraint
\begin{equation}
    0\,=\,\Tr A\,=\,\sum_{j=1}^{N_c} \, A^{\mk jj} \; ,
    \label{eq:TrA=0}
\end{equation}
yielding the relation, see eq.\ (\ref{eq:Thetadiag}),
\begin{equation}
0 \, = \, \sum_{j=1}^{N_c} \, \beta_{\mu}^{j}
  \, = \, \frac{2 \pi}{m}\,
     \sum_{j=1}^{N_c} \, n_{\mu}^{j}
     \label{eq:tracewith-n}
\end{equation}
for the $\Theta_{\mu}$ matrices.
Of course, this condition is automatically satisfied if the $\beta_{
\mu}^{j}$ eigenvalues are given, see eqs.\ (\ref{eq:Theta-wi})
and (\ref{eq:alpha-ij}), by
\begin{equation}
\beta_{\mu}^{j}\,=\,\sum_{b=1}^{N_c-1}\,
 \theta_{\mu}^{b}\, \alpha^{b}_j \,=\,
\sum_{b=1}^{N_c-1}\, \theta_{\mu}^{b}\, \left[\,
 R^{bj}\,\xi^j - R^{b(j-1)}\,\xi^{j-1} \,\right] \; ,
\end{equation}
which implies
\begin{equation}
 \sum_{j=1}^{N_c} \, \beta_{\mu}^{j}\,=\, \sum_{j=1}^{N_c-1} \, \sum_{b=1}^{N_c-1}\, \theta_{\mu}^{b}\,
 R^{bj}\,\xi^j\, -\,
 \sum_{j=2}^{N_c} \, \sum_{b=1}^{N_c-1}\, \theta_{\mu}^{b}\,
 R^{b(j-1)}\,\xi^{j-1}  \,=\, 0 \; .
\end{equation}
Indeed, when written in terms of the coefficients $\theta_{\mu}^{b}$,
see eq.\ (\ref{eq:tautheta}), the matrices $\Theta_{\mu}$ depend
on $d\,(N_c\!-\!1)$ free parameters; on the other hand, when they
are written using the $n_{\mu}^{j}$ coefficients, see eq.\
(\ref{eq:Thetadiag}), we have $d\,N_c$ free parameters, subject
to the $d$ constraints (\ref{eq:tracewith-n}).

Beside being traceless, an element of the SU($N_c)$ Lie algebra
should also be (with our convention) Hermitian.
Hence, if we impose $A^{\dagger}=A$ in eq.\ (\ref{eq:Amatrix}),
we find
\begin{equation}
  \left(A^{\mk ij}\right)^{*} \,=\, A^{\mk ji} \; ,
\label{eq:aij}
\end{equation}
given that $\left({\bf W}^{ij}\right)^{\dagger} = {\bf W}^{ji}$,
see footnote \ref{WnHerm}.
(Here, $^{*}$ denotes complex conjugation.)
The last result, together with eq.\ (\ref{eq:TrA=0}), implies that
the diagonal coefficients $A^{\mk jj}$ are real and that only
$N_c\!-\!1$ of them are independent.
At the same time, from eq.\ (\ref{eq:aij}) we find that there are
only $N_c\,(N_c\!-\!1)/2$ independent complex off-diagonal elements,
yielding a total of $(N_c-1)+(N_c^2-N_c)=N_c^2\!-\!1$ free real
parameters (as expected).
We stress that the coefficients $A^{\mk ij}$ are {\em not} the matrix
elements of $A$, which are given by the expression
\begin{equation}
A_{lm} \, = \, \sum_{i,j=1}^{N_c} \left({\bf W}^{ij}\right)_{lm}
                  \; A^{\mk ij}
\end{equation}
with
\begin{equation}
\left({\bf W}^{ij}\right)_{lm}\,=\,\sum_{k,n=1}^{N_c}
\left(v^{\dagger}\right)_{lk} \,\left({\bf M}^{ij}\right)_{kn} \,v_{nm} \,=\,
\left(v^{\dagger}\right)_{li} \,v_{jm} \,=\,
v_{il}^{*} \,v_{jm} \; ,
\label{eq:lambdaijhl}
\end{equation}
so that one has (as always for a Hermitian matrix)
\begin{eqnarray}
A_{lm}^{*}\,&=&\,\sum_{i,j=1}^{N_c}\,\left({\bf W}^{ij}\right)_{lm}^{*}\,
   \left(A^{\mk ij}\right)^{*}\,=\,\sum_{i,j=1}^{N_c}\,
\left[\,v_{il}^{*} \,v_{jm}\,\right]^{*}\,\left(A^{\mk ij}\right)^{*}
 \nonumber \\[2mm]
  &=&\, \sum_{i,j=1}^{N_c}\,v_{jm}^{*}\,v_{il}\,A^{\mk ji}\,=\,
\sum_{j,i=1}^{N_c}\,\left({\bf W}^{ji}\right)_{ml}\,A^{\mk ji}\,=\,
A_{ml} \; .
\end{eqnarray}

Finally, eq.\ (\ref{eq:Thetadiag}) tells us that we can easily
relate the Cartan sub-algebra, defined by the diagonal matrices
in eqs.\ (\ref{eq:defH}) and (\ref{eq:DR})
above, with the matrices ${\bf M}^{jj}$ (or the matrices ${\bf W}^{jj}$).
Indeed, given that $({\bf M}^{jj})_{lm}=\delta^{jl} \delta^{jm}$,
we can write (for $i=1,\ldots,N_c\!-\!1$)
\begin{equation}
 H^{i} \,=\, \xi^i\,
   \left[\, {\bf M}^{ii} \,-\, {\bf M}^{(i+1)(i+1)}\,\right]
\end{equation}
so that
\begin{equation}
  D^{i} \,=\, \sum_{j=1}^{N_c-1} R^{ij}\; \xi^j\, 
          \left[\, {\bf M}^{jj} \,-\, {\bf M}^{(j+1)(j+1)}
              \,\right]
\label{eq:Digen}
\end{equation}
and
\begin{equation}
 t_{\scriptC}^i \,=\,
 v^{\dagger}\,D^{i}\,v
 \,=\, \sum_{j=1}^{N_c-1} R^{ij}\, \xi^j\,
          \left[\, {\bf W}^{jj} \,-\,
       {\bf W}^{(j+1)(j+1)} \, \right] \; .
\end{equation}
In particular, if we set $\xi^j=1$ in eq.\ (\ref{eq:Digen}) and
we use the matrix $R$ defined in eq.\ (\ref{eq:RgeneralSUNc}), the
matrices $D^{i}$ recover the generalized diagonal Gell-Matrices
matrices, see eq.\ (\ref{eq:DjgenGM}).
It is interesting that, using the matrices ${\bf M}^{ij}$, we can easily
define also the generalized nondiagonal Gell-Matrices matrices
(see again appendix A1 in ref.\ \cite{Smit}):
\begin{equation}
\label{eq:lambdabsym}
t^b \,=\, {\bf M}^{ij}\,+\,{\bf M}^{ji} 
\end{equation}
and
\begin{equation}
t^b \,=\, -i\,\left(\, {\bf M}^{ij}\,-\,{\bf M}^{ji}\,\right) \; ,
\label{eq:lambdabantisym}
\end{equation}
with $i,j=1,\ldots,N_c$ and $i<j$.
Note that there are $N_c(N_c\!-\!1)/2$ symmetric matrices
(\ref{eq:lambdabsym}), $N_c(N_c\!-\!1)/2$ anti-symmetric matrices
(\ref{eq:lambdabantisym}) and $N_c\!-\!1$ diagonal matrices $t_{\scriptC}^i
=D^{i}$,
for a total of $N_c^2-1$ (Hermitian and traceless) generators.

The above results imply that the (generic) matrix
\begin{equation}
M_{\scriptC}\,\equiv\, \sum_{i=1}^{N_c-1}\, m^i\,
   t_{\scriptC}^i
 \,=\, \sum_{i=1}^{N_c-1}\, m^i\,
    \sum_{j=1}^{N_c-1} R^{ij}\; \xi^j\,
          \left[\, {\bf W}^{jj} \,-\,
       {\bf W}^{(j+1)(j+1)} \, \right] \; ,
\label{eq:M_Cgen}
\end{equation}
which is in the Cartan sub-algebra, can also be written as
\begin{equation}
M_{\scriptC}\,=\, \sum_{j=1}^{N_c}\, a^{jj}\, {\bf W}^{jj}
\label{eq:M_C}
\end{equation}
with\footnote{A linear relation among the coefficients
$a^{jj}$ and $m^i$ is, of course, expected for any
change of basis in the Cartan sub-algebra.}
\begin{equation}
a^{jj}\,=\,\sum_{i=1}^{N_c-1}\,m^i\,\left[\,R^{ij}\,
    \xi^j \,-\,R^{i(j-1)}\,\xi^{j-1}\right] \; .
\label{eq:all}
\end{equation}
However, one should stress that, on the l.h.s.\ of the above
equation, the index $j$ takes (integer) values in the interval
$[1,N_c]$,
while, on the r.h.s., the indices $j$ and $j\!-\!1$ of $R$ and of $\xi$
are always restricted to the interval
$[1,N_c-1]$.
This implies the relations
\begin{eqnarray}
a^{11}\,&=&\,\sum_{i=1}^{N_c-1}\,m^i\,R^{i1}\,\xi^1\\[2mm]
a^{22}\,&=&\,\sum_{i=1}^{N_c-1}\,m^i\,\left(\,R^{i2}\,\xi^2
    \,-\,R^{i1}\,\xi^1\right) \\[2mm]
a^{33}\,&=&\,\sum_{i=1}^{N_c-1}\,m^i\,\left(\,R^{i3}
    \,\xi^3\,-\,R^{i2}\,\xi^2\,\right) \\[2mm]
  &\ldots& \\[2mm]
a^{\scriptscriptstyle{(N_c-1)(N_c-1)}}\,&=&\,
      \sum_{i=1}^{N_c-1}\,m^i\,\left[\,R^{i(N_c-1)}
          \,\xi^{N_c-1} \,-\,
      R^{i(N_c-2)}\,\xi^{N_c-2}\,\right] \\[2mm]
a^{\scriptscriptstyle{N_c N_c}}\,&=&\,-\,
  \sum_{i=1}^{N_c-1}\,m^i\, R^{i(N_c-1)}\,\xi^{N_c-1}\; ,
\end{eqnarray}
which trivially ensure the constraint (\ref{eq:TrA=0}).



\end{document}